\documentclass{article}%
\usepackage{graphicx}
\usepackage{amsfonts}
\usepackage{geometry}
\usepackage{amsmath}
\usepackage{authblk}

\usepackage{amssymb}%
\providecommand{\U}[1]{\protect\rule{.1in}{.1in}}
\numberwithin{equation}{section}
\begin{document}
\title{Constraint Algebra in LQG Reloaded: Toy Model of a $\mathrm{U}(1)^{3}$ Gauge
Theory I}

\author[a]{Adam Henderson}
\author[a]{Alok Laddha}
\author[a,b]{Casey Tomlin}
\affil[a]{Institute for Gravitation and the Cosmos\\Pennsylvania State University, University Park, PA 16802-6300, U.S.A}
\affil[b]{Raman Research Institute\\Bangalore-560 080, India}

\maketitle

\begin{abstract}
We analyze the issue of anomaly-free representations of the constraint algebra
in Loop Quantum Gravity (LQG) in the context of a diffeomorphism-invariant
$\mathrm{U}(1)^{3}$ theory in three spacetime dimensions. We construct a
Hamiltonian constraint operator whose commutator matches with a quantization
of the classical Poisson bracket involving structure functions. Our
quantization scheme is based on a geometric interpretation of the Hamiltonian
constraint as a generator of phase space-dependent diffeomorphisms. The
resulting Hamiltonian constraint at finite triangulation has a conceptual
similarity with the $\bar{\mu}$-scheme in loop quantum cosmology and highly
intricate action on the spin-network states of the theory. We construct a
subspace of non-normalizable states (distributions) on which the continuum
Hamiltonian constraint is defined which leads to an anomaly-free
representation of the Poisson bracket of two Hamiltonian constraints in loop
quantized framework.\newline Our work, along with the work done in
\cite{madtom} suggests a new approach to the construction of
anomaly-free quantum dynamics in Euclidean LQG.

\end{abstract}

\section{Introduction}
Loop Quantum gravity (LQG) started out as an approach to non-perturbative quantization of the gravitational field using a classical canonical formulation of gravity as a starting point \cite{newvar,barbero}. The (spatial) diffeomorphism invariance of the
theory guaranteed a pretty kinematical framework with a tight analytic control rarely seen in four-dimensional quantum field theories \cite{almmt}. The initial attempts at a formulation of the dynamics (via implementation of the Hamiltonian constraint) were very promising. In the mid-nineties Thiemann  proposed a quantization of the Euclidean as well as the Lorentzian Hamiltonian constraint in a series of remarkable papers titled Quantum Spin Dynamics \cite{qsd1,qsd2,qsd3}. Thiemann's Hamiltonian constraint had some rather intriguing properties like UV finiteness. However, despite this initial promise 
the Hamiltonian constraint program in canonical Loop quantum gravity (LQG) has reached a strange impasse. Several issues still remain open and in our opinion it is hard to argue against an assertion that there is no satisfactory definition of Hamiltonian constraint in canonical LQG. [For some very interesting recent progress in this direction we refer the reader to \cite{val1,perezham}.] This impasse has in turn led to new avenues to analyze the dynamics in LQG, e.g. the Master constraint program \cite{master1,master2}, the covariant spin-foam models (for reviews, see \cite{perezsfreview,alexandrov}), and the deparametrized dust models \cite{giesel1,giesel2}. 
There are two primary reasons for the above
assertion. On one hand there is no unique definition of the Hamiltonian
constraint. Quantization of Hamiltonian constraint in LQG involves (just as
for any composite operator in any quantum field theory) an intermediate choice
of regularization. This regularization amounts to choosing a family of loops,
edges, surfaces and certain discrete representation theoretic labels from
which a regularized Hamiltonian constraint operator is built. There are an
infinite number of choices for each of the regulating structures involved and
in principle each such choice can give rise to a distinct operator which is
well defined on the kinematical Hilbert space of LQG, $\mathcal{H}%
_{\mathrm{kin}}$. This would not be a problem if the continuum limit of the
regularized constraint was independent of the regulating structures involved.
However this is not the case and even the continuum Hamiltonian has an
infinite dimensional parameter worth of ambiguity.\newline As the Hamiltonian
constraint in canonical gravity is a generator of the so called Dirac algebra,
a priori one might expect that as we require the quantum Hamiltonian
constraint to be anomaly-free (in the sense that there exist a representation
of the Dirac algebra in quantum theory), there will be certain non-trivial
restrictions on the quantization choices mentioned above. However as it turns
out, this is not quite true. Anomaly freedom is achieved in LQG through what
is known in gauge theories as an on-shell closure. (That is we require both
the left and the right hand sides of constraint algebra to trivialize on
states that are solution to (at least some of) the constraints) and this
condition barely places any restrictions on the quantization ambiguities which
infect the definition of the constraint. Even more worryingly there are some
very concrete signs that this trivialization of the right hand side and left
hand side of Dirac algebra might even extend to states which in a precise
sense are not on-shell, thus indicating some serious problems with the
definition of the quantum constraint.\newline In this paper, we analyze the
issue in a simple three-dimensional diffeomorphism-invariant gauge theory with
Abelian gauge group $\mathrm{U}(1)^{3}$ (To the best of our knowledge this model was first conceived by Smolin in four space-time dimensions \cite{Smolin}) .
Our aim is to quantize the
Hamiltonian constraint of the theory (in loop formulation) such that it has a
chance to generate an anomaly-free Dirac algebra in the quantum theory. In the
next section, we outline the problem with Thiemann's Hamiltonian constraint in
more detail and outline the work done in this paper.

\section{Motivation and Outline of the paper}

\subsection{The issue}
In this section we explain the problems with Thiemann's Hamiltonian constraint
that we referred to in the introduction. The underlying issues are rather
involved and we may not be able to do justice to different points of view
which exist in the literature. We refer the reader to (\cite{ttbook}%
,\cite{alreview}, \cite{abreview}) for more details.\newline

Traditionally the continuum Hamiltonian constraint operator is a densely
defined operator on $\mathcal{H}_{\mathrm{kin}}$. This is accomplished by
placing a rather unusual topology (referred to as the URS topology) on the space of operators in which
convergence of the one parameter family of finite triangulation Hamiltonian
Constraint operators tuned out to be an operator on $\mathcal{H}%
_{\mathrm{kin}}$. Roughly speaking this topology was such that limit
points of any two operator sequences (indexed by finite triangulation $T$)
$\hat{O}_{1T}$ and $\hat{O}_{2T}$ which are such that
\begin{equation}
\hat{O}_{1T}-\hat{O}_{2T}=\left(  \hat{U}(\phi)-1\right)  \hat{O}%
_{3T}\label{property1}%
\end{equation}
for some $\phi\in\text{Diff}(\Sigma),$ ($\hat{U}$ being the unitary
representation of diffeomorphism group on $\mathcal{H}_{kin}$) and for some
$\hat{O}_{3\ T}$, are identified. That is in this topology $\lim
_{T\rightarrow\infty}(\hat{O}_{1T}-\hat{O}_{2T})=0$.\newline In this topology
the commutator of two (continuum) Hamiltonian constraints $\hat{H}[N],\hat
{H}[M]$ on $\mathcal{H}_{\mathrm{kin}}$ vanishes. As shown in \cite{qsd3},
there exists a quantization (at finite triangulation) of the right hand side
of the Poisson bracket relation (see Section \ref{classytheory})%
\begin{equation}
\{H[N],H[M]\}=V[\vec{\omega}]\label{eq:hhbracket}%
\end{equation}
which is of the form $\left(  \hat{U}(\phi)-1\right)  \hat{O}_{3T}$, whence
its continuum limit in the URS topology is zero. Thus the quantization of both
the left and right hand sides of (\ref{eq:hhbracket}) vanish. Although this
was originally taken to be a sign of internal consistency and anomaly freedom,
a closer look at the structural aspects of the computation was performed in a
series of remarkable papers by Gambini, Lewandowski, Marolf, and Pullin
(\cite{lm1},\cite{lm2}). They looked at the convergence of the
finite-triangulation commutator sequence not on $\mathcal{H}_{\mathrm{kin}},$
but on certain distributional spaces (which are extensions of the space of
diffeomorphism-invariant distributions) referred to as habitats. As the
habitat consists of (distributional) states which are not
diffeomorphism-invariant, a priori neither the quantization of the LHS or RHS
of (\ref{eq:hhbracket}) is expected to be a trivial operator. However, it
turned out that $\widehat{\{H[N],H[M]\}}$ is the zero operator on the habitat,
and there exists a quantization of $V[\vec{\omega}]$ at finite triangulation
on $\mathcal{H}_{\mathrm{kin}}$ whose continuum limit on the habitat is
trivial. However, this vanishing of the quantization of both the RHS and LHS
of (\ref{eq:hhbracket}) on diffeomorphism non-invariant states is rather
unsatisfactory. Even more worrysome, the reasons for the vanishing of the LHS
and RHS are entirely different. The most transparent way to see this is given
in \cite{lm2}, where the authors argue that, if instead of working with
density weight one constraints, one works with higher density constraints, but
keeps the quantization choices essentially the same, then on their habitat,
the LHS will continue to vanish, but the RHS will not, whence suggesting the
presence of an anomaly in the whole scheme (we come back to this point in more
detail below).

\subsection{Our Goal}

Success of the canonical loop quantum gravity program is defined by the
following: Starting with the (unique) diffeomorphism-covariant representation
of the holonomy-flux algebra, does there exist a vector space $\mathcal{V}$
(whose elements are linear combinations of spin network states) that is a
representation space for the Dirac algebra, in the sense that
\begin{align}
\hat{U}(\phi_{1})\cdot\hat{U}(\phi_{2})\Psi &  =\hat{U}(\phi_{1}\circ\phi
_{2})\Psi\label{eq:qdirac}\\
\left[  \hat{H}[N],\hat{H}[M]\right]  \Psi &  =\hat{V}[\vec{\omega}%
]\Psi\nonumber\\
\hat{U}(\phi)\hat{H}[N]\hat{U}(\phi^{-1})\Psi &  =\hat{H}[\phi^{\ast}%
N]\Psi\nonumber
\end{align}
$\forall\Psi\in\mathcal{V}$, where

\begin{enumerate}
\item[(i)] $\phi$ is a spatial diffeomorphism (usually taken to be in the
semi-analytic category) and $\hat{U}(\phi)$ is a representation of the
diffeomorphism group on $\mathcal{V}$ induced via its unitary representation
on $\mathcal{H}_{\mathrm{kin}}$.\footnote{In light of work done in
\cite{amdiff}, one could ask for a genuine representation of Dirac algebra
involving the diffeomorphism constraint operator $\hat{V}[\vec{N}]$ instead of
$\hat{U}(\phi)$.}

\item[(ii)] $\hat{H}[N]$ is a continuum quantum Hamiltonian constraint
operator obtained as a limit point of a net of finite-triangulation operators
defined on $\mathcal{H}_{\mathrm{kin}}$.
\end{enumerate}

We refer to (\ref{eq:qdirac}), and in particular to the second equation in
(\ref{eq:qdirac}), as the off-shell closure condition for $\hat{H}[N]$
\cite{nicolaietal}.

Once a quantization of the Hamiltonian constraint is found which meets the
above criteria, its kernel (in $\mathcal{V}$) is expected to give rise to the
physical Hibert space of the theory. This is a rather ambitious aim, and it is
instructive to accomplish it in models which are, on one hand,
diffeomorphism-invariant field theories with a gauge algebra being the Dirac
algebra, and on the other hand, are simpler and more tractable than gravity.
Models where this aim has been accomplished include (\textbf{a}) a
two-dimensional parametrized field theory (PFT) \cite{amhamcons,ttpft}, and
(\textbf{b}) the Husain-Kucha\v{r} (HK) model\footnote{The HK model is
essentially canonical gravity without the Hamiltonian constraint.}
\cite{amdiff}. However, the main reasons that the constraint algebra in these
models could be represented in an anomaly-free manner are the following:

\begin{enumerate}
\item[(1)] Two dimensions are special in the sense that the Dirac algebra in
two-dimensional PFT is a true Lie algebra which is isomorphic to the Witt
algebra ($\text{diff}(S^{1})\oplus\text{diff}(S^{1})$). In the HK model, the
only constraint (apart from the Gauss constraint) is the spatial
diffeomorphism constraint whose algebra is isomorphic to the Lie algebra of
(spatial) vector fields.

\item[(2)] The Poisson action of constraint functionals on classical fields in
these models has a clear geometric interpretation which provided key insights
into the possible quantization choices.
\end{enumerate}

Neither of the above are true in the case of canonical gravity, and so as
appealing as the results obtained in PFT and the HK model are, we need to see
if the lessons learned there can be applied to models which were more closely
related to canonical gravity. In this paper we propose just such a model. It
is a diffeomorphism-invariant (in fact topological) field theory in three
dimensions which can be thought of as a weak coupling limit of
three-dimensional Euclidean gravity.\footnote{We are indebted to Miguel
Campiglia for pointing this out to us.} As a weak coupling limit amounts to
switching off self-interactions, in this case it amounts to switching the
gauge group from $\mathrm{SU}(2)$ to $\mathrm{U}(1)^{3}$. The canonical
formulation of the theory corresponds to the phase space of a $\mathrm{U}%
(1)^{3}$ Yang-Mills theory in $2+1$ dimensions with Hamiltonian,
diffeomorphism, and Gau\ss \ constraints. On the (Gau\ss ) gauge-invariant
sector of the phase space, the remaining constraints (Hamiltonian and
diffeomorphism) generate the Dirac algebra.

Our aim in this paper is to loop quantize this system such that the
(continuum) Hamiltonian constraint satisfies the second equation in
(\ref{eq:qdirac}).\footnote{The remaining relations which are related to the
diffeomorphism covariance properties of the Hamiltonian constraint will be
analyzed in \cite{hat2}.} \footnote{In the terminology of \cite{nicolaietal} we
are aiming towards a quantization of the Hamiltonian constraint which
satisfies the off-shell closure condition.} However, blindly looking for a
possible quantization of the Hamiltonian constraint which will lead to
off-shell closure is a hopeless task, so we draw upon the lessons learned in
\cite{amhamcons,amdiff} to achieve our goal. In order to familiarize the
reader with these lessons, we briefly recall them below.

\subsubsection{Determination of the Correct Density Weight}

As explained rather beautifully in \cite{lm1}, if one chooses to work with the
density one Hamiltonian constraint, then no matter what domain one chooses to
take the limit of finite-triangulation Hamiltonian constraint on, the
resulting operator will always have a vanishing commutator with itself (as
long as limits of the finite-triangulation commutator are well-defined). In
particular, the commutator of two density one constraints can never give rise
to an operator which could resemble a quantization of the RHS. The reason for
this is rather simple. Consider a smeared Hamiltonian constraint in $d$
spatial ($d\geq2$) dimensions,
\begin{equation}
H[N]=\int_{\Sigma}\mathrm{d}^{d}x~N^{(1-k)}(x)H^{(k)}(x),
\end{equation}
where the superscripts indicate the density weights of the
various fields. The smearing function $N$ is a scalar density of weight
$\left(  1-k\right)  $ while the local Hamiltonian density $H^{(k)}%
(x)=q^{\left(  k-2\right)  /2}FEE(x)$ has density weight $k$. In LQG, the
quantization of $H[N]$ proceeds by first approximating the integral by a
Riemann sum over simplices (which constitute a triangulation $T$ of $\Sigma$)
and then approximating each term in the sum by a function of appropriate
holonomies and fluxes. Typically in LQG the \textquotedblleft
fineness\textquotedblright\ of the triangulation is measured by a parameter
$\delta$ (usually associated with coordinate volume of a simplex in $T$), and
a simple dimensional analysis shows that
\begin{align}
H_{T(\delta)}[N] &  =\sum_{\triangle\in T(\delta)}\delta^{d-(2+2d-2)+d(2-k)}%
N^{(1-k)}(v(\triangle))O_{\triangle}(v(\triangle))\nonumber\\
&  =\sum_{\triangle\in T(\delta)}\delta^{d\left(  1-k\right)  }N^{(1-k)}%
(v(\triangle))O_{\triangle}(v(\triangle))
\end{align}
where $O_{\triangle}(v(\triangle))$ is a function of holonomies and fluxes,
constructed out of loops and surfaces associated to $\triangle$. Quantization
choices involved in the definition of $O_{\triangle}(v(\triangle))$ are such
that it has no explicit dependence on $\delta$. It is easy to see that for any
$d$, if the lapse $N$ is a scalar with density weight zero ($k=1$), then
$H_{T(\delta)}[N]$ has no explicit dependence on $\delta$. Along with a
special property of Thiemann's quantization that the Hamiltonian constraint
operator does not act on the vertices (of a spin network) that it creates,
this ensures that commutator of two density one operators vanish. Note that as
there are no explicit factors of $\delta$ left in the definition of $\hat
{H}_{T(\delta)}[N]$ and so also in the commutator, a quantization of the RHS
can never arise, as it involves derivatives of the lapse functions,
which\ themselves require at least one factor of $\delta^{-1}$. This
observation led the authors of \cite{lm2} to conclude that one must quantize
higher density constraints in order to have explicit factors of $\delta^{-1}$
which in the continuum limit could give rise to terms like $\left(
N\partial_{a}M-M\partial_{a}N\right)  $. \emph{The lesson we draw from these
arguments is that even though density one constraints can be quantized on
$\mathcal{H}_{\mathrm{kin}}$, if we are interested in seeking an anomaly-free
representation of the constraint algebra, then one needs to work with higher
density constraints.} [For an interesting counter point to this argument, see
\cite{ttpft}.] It was also argued in \cite{lm2} that, if one chose to work
with higher density constraints such that one has enough factors of
$\delta^{-1}$ to obtain a non-trivial quantization of $V[\vec{\omega}]$ on the
habitat, then the LHS will continue to vanish unless the Hamiltonian
constraint acts non-trivially on the vertices that it creates. This ought to
be rather worrysome.

In \cite{amhamcons} these observations were taken seriously and applied to
two-dimensional PFT. It was shown that one could obtain an anomaly-free
representation of the Dirac algebra if one quantized density-two constraints.
The key point was to work with an appropriate density weight such that at
finite triangulation, the Hamiltonian constraint, when written in terms of
holonomies and fluxes should have precisely one explicit factor of
$\delta^{-1}$. In this case, the commutator will have a factor of
$(\delta\delta^{\prime})^{-1}$, and this has the correct dimensionality to
yield a quantization of $V[\vec{\omega}]$ in continuum limit.\footnote{One
factor of $\delta^{-1}$ can, in the continuum limit, give one derivative, and
hence can yield terms like $N\partial_{a}M-M\partial_{a}N,$ and the other
factor is precisely the factor one needs to obtain a quantum diffeomorphism
constraint, which we expect to be linked with Lie derivatives.}

In the model considered in this paper, where the classical Hamiltonian
constraint is
\begin{equation}
H[N]=\int_{\Sigma}\mathrm{d}^{2}x\ Nq^{-k/2}\epsilon^{ijk}F_{ab}^{i}E_{j}%
^{a}E_{k}^{b}(x),
\end{equation}
it is rather easy to see that, in order to quantize the constraint such that
at finite triangulation it has an explicit factor of $\delta^{-1}$, we need
$k=\frac{1}{2}$ whence $N$ needs to be a scalar density of weight $-\frac
{1}{2}$.

\subsubsection{What Should the Constraint Operators Do?}

As we mentioned above, one of the aspects which distinguish two-dimensional
PFT and the HK model from canonical gravity (or for that matter the model of
this paper) is that Poisson action of the constraints on the phase space data
has a transparent geometric interpretation. In the first case, the Hamiltonian
constraint, being a generator of the Witt algebra, is intricately linked to
spatial diffeomorphisms on $S^{1}$. In the HK model, as the only constraint is
the diffeomorphism constraint, its Poisson action on $(A,E)$ is nothing but
the Lie derivative by the shift field. These interpretations were key inputs
in pinning down the quantization choices for these constraints. The connection
between the geometric interpretation and quantization can be encoded in the
following schematic equation. Given a spin network state $\Psi$, let the
corresponding classical cylindrical function be denoted $\Psi(A)$. Then
\begin{equation}
\hat{O}_{T(\delta)}[V]\Psi\equiv\frac{1}{\delta}\left[  \text{Finite action
generated by }O[V]\text{, parametrized by $\delta$}-1\right]  \Psi(A)
\end{equation}
For example, in the case of the HK model, the quantization choices made in
\cite{amdiff} to construct the diffeomorphism constraint operator were such
that this operator at finite triangulation equalled $D[\vec{N}]=\frac
{1}{\delta}(\hat{U}(\phi_{\delta}^{\vec{N}})-1)$.

If we were to follow this route to find out what quantization choices are to
be made to construct ${\hat{H}}[N]$, we are forced to look for a geometric
interpretation of Poisson action of the Hamiltonian constraint. As we show in
Section \ref{classint}, there indeed does exist such an
interpretation\footnote{Rather remarkably this interpretation also holds for
the SU(2) case, and is likely to be important in extending this program to
Euclidean quantum gravity.}
\begin{equation}
X_{H[N]}A_{a}^{i}\approx\epsilon^{ijk}\mathcal{L}_{q^{-1/4}N\vec{E}_{j}}%
A_{a}^{k}\cong\epsilon^{ijk}\frac{(\phi_{\delta}^{q^{-1/4}N\vec{E}_{j}}%
)^{\ast}-1}{\delta}A_{a}^{k} \label{eq:twistdiff}%
\end{equation}
where $\approx$ refers to equality modulo Gau\ss \ law. Thus the change in,
say, $A^{1}$ under the action of the Hamiltonian vector field of $H[N]$ equals
Lie derivative of $A^{3}$ with respect to the vector field $q^{-1/4}N\vec
{E}_{2}$ minus the Lie derivative of $A^{2}$ with respect to the vector field
$q^{-1/4}N\vec{E}_{3}$. The second approximation is a discrete approximant to
the Hamiltonian vector field. We will seek a quantization of $H[N]$ at finite
triangulation which mimics this action on spin-network states.

\subsubsection{Where Should the Continuum Limit be Taken?}

As the finite-triangulation Hamiltonian constraint has an explicit factor of
$\delta^{-1}$, it cannot admit a continuum limit on $\mathcal{H}%
_{\mathrm{kin}}$ (in any operator topology), and whence the obvious question
is, is there any admissible topology on the space of operators, and are there
any subspaces of the space of distributions on which the continuum limit can
be taken? Once again the case of two-dimensional PFT and the HK model provide
important clues. One needs to build spaces (or habitats as termed in
\cite{lm1}) by studying the specific deformations of a spin network that the
finite triangulation constraints generate. The topology on the space of
operators can then come by looking at seminorms defined by (generalized)
matrix elements of operators between these habitat states and states in
$\mathcal{H}_{\mathrm{kin}}$. A priori, there can be infinitely-many habitats
which can function as a home for the quantum constraints. In this paper, we
consider the simplest possible habitat on which the continuum limit of the
Hamiltonian constraint exists and on which the off-shell closure relation be
checked. We do not know if this habitat is physically interesting in the sense
that it is\textbf{ }a representation space for Dirac observables. However, as
our modest aim in this paper is to see if there is an anomaly free
representation of the Hamiltonian constraint on \emph{some} space, we leave a
detailed analysis of the construction of \textquotedblleft physically
interesting\textquotedblright\ habitats for future research.

In the remainder of this section, we outline how we implement the above
lessons in our model and arrive at a quantum Hamiltonian constraint which
satisfies the off-shell closure condition.


\subsection{Outline}



\subsubsection{The Idea of the Quantum Shift and the Role of Inverse Volume}

As we want to quantize the Hamiltonian constraint at finite triangulation such
that it mimics the action in (\ref{eq:twistdiff}) on charge network
states,\footnote{We refer to the spin network states of the $\mathrm{U}%
(1)^{3}$ theory as charge networks \cite{rfock}, as the edge labels in this
case are $\mathrm{U}(1)$ charges.} we need to define the quantum counterparts
of the classical vector fields $q^{-1/4}N\vec{E}_{i}$. This is where the loop
representation throws its first surprise. Although classically the triad
fields are smooth, quantum mechanically they turn out to be operator-valued
distributions. More precisely, given any charge network state $|c\rangle$, the
graph of this charge network is the \textquotedblleft locus of
discontinuity\textquotedblright\ of the $\hat{q}^{-1/4}N\widehat{\vec{E}}_{i}$
operator. Nonetheless, as we show below, one can quantize $q^{-1/4}N\vec
{E}_{i}$ in such a way that each charge network is an eigenstate, with its
spectrum belonging to $T_{x}\Sigma$. Given any charge network state
$|c\rangle$ based on the graph $\gamma,$ the expectation value of $\hat
{q}^{-1/4}N\widehat{\vec{E}}_{i}(x)$ is non-vanishing only if $x\in
V(\gamma)$ and the resulting \textquotedblleft vectors\textquotedblright%
\ at the vertices of $\gamma$ will be referred to as the \emph{quantum shift}
associated to that state.

\subsubsection{Image of the Regularized Hamiltonian Constraint and the Birth
of Extraordinary Vertices}

Given a charge network state $|c\rangle$, the action of the regularized
Hamiltonian constraint operator produces two generic effects:

\begin{enumerate}
\item[(i)] The change in the edge labels is state dependent; and

\item[(ii)] a new degenerate vertex (by degenerate we mean that the inverse
volume operator acting at that vertex vanishes) is created whose location
depends on the quantum shift $\langle\hat{E}_{i}(v)\rangle$.
\end{enumerate}

We call such vertices \emph{extraordinary vertices}. The restriction of a
charge network in the neighborhood of an extraordinary vertex has certain
invariant properties that we enumerate below, and these help us isolate all
the charge network states which lie in the image of the regularized
Hamiltonian constraint operator.

\subsubsection{Geometric Interpretation of the Poisson Action of the Product
of Hamiltonian Constraints}

As the extraordinary vertices which are created by the Hamiltonian constraint
are degenerate, naively one would expect the (regularized) Hamiltonian
constraint to act trivially at such vertices. However, this result relies upon
specific quantization choices and would lead to an anomaly in the constraint
algebra. We cure this problem by once again going back to the analogous
classical computation.

The classical discrete approximant to the Hamiltonian vector field of $H[N]$
on a cylindrical function $f_{c}(A)$ creates a linear combination of
cylindrical functions with exact analogs of the extraordinary vertices
mentioned above. These vertices are located along the integral curves of a
triad-dependent vector field, so the action of a second Hamiltonian constraint
on a cylindrical function containing such vertices moves them (as such an
action will have a non-trivial effect on integral curves of phase
space-dependent vector fields). This observation helps us in \emph{modifying}
the action of the Hamiltonian constraint on extraordinary vertices, by making
use of the quantization ambiguities that are available to us due to the
structure of the loop representation.

\subsubsection{Proposal for the Habitat}

Finally by studying the precise nature of extraordinary vertices (that is, the
deformations in a charge network that the Hamiltonian constraint operator
creates), we propose a definition of a habitat. It is a subspace of
distributions, where each distribution is a linear combination of an infinite
number of a specific class of charge networks with coefficients being
dependent on the vertex set of the charge network. The underlying idea of this
habitat is precisely the same as that proposed in \cite{lm1} (though the
habitat itself is completely different) whence we call it a
Lewandowski-Marolf-Inspired (LMI) habitat. As we show in Appendix \ref{A3}, on
this habitat, the regularized Hamiltonian constraint admits a continuum limit.

\subsubsection{Quantization of the RHS and the Off-Shell Closure Condition}

We finally demonstrate that there exists a quantization of the RHS, $\hat
{V}[\vec{\omega}]$ on the LMI habitat such that the continuum limit of
regularized commutators between the two Hamiltonian constraints equals
$\hat{V}[\vec{\omega}]$. This is our main result.

We finally end with conclusions and highlight the open issues and some of the
unsatisfactory aspects of our construction. Some of these issues will be
analyzed in the sequel \cite{hat2}.


\section{Classical Theory}

\label{classytheory}

In this section we describe the constrained Hamiltonian system that we aim to
quantize in this paper. The algebra of constraints that we eventually arrive
at can be obtained simply by replacing the internal gauge group SU(2) of
general relativity in connection variables with a direct product of three
commuting copies of U(1), but there is another way, due to Smolin
\cite{Smolin}, in which one takes Newton's gravitational constant
$G_{\mathrm{N}}\rightarrow0$ at the level of the action, and analyzes the
resulting canonical theory. We follow this second route.

Our starting point is the Palatini action for general relativity in three
dimensions:%
\begin{equation}
S_{\mathrm{P}}[e,\omega]=\frac{1}{16\pi G_{\mathrm{N}}}\int_{M}\mathrm{d}%
^{3}x~\eta^{\mu\nu\rho}\Omega_{\mu\nu}^{i}e_{\rho i},
\end{equation}
where the basic variables are a (dimensionless) co-triad $e_{\mu}^{i},$ and a
connection $\omega_{\mu}^{i}$ (with dimensions of inverse length) with
curvature $\Omega_{\mu\nu}^{i}$. $G_{\mathrm{N}}$ has units of inverse
momentum (in $c=1$ units), and the Planck length is defined as $l_{\mathrm{P}%
}=\hbar G_{\mathrm{N}}$ (where $\hbar$ as usual has units of angular
momentum). Here $\mu,\nu,\dots=0,1,2$ are spacetime indexes (while below we
will use $a,b,\dots=1,2$ as spatial indexes), and $i,j,\dots$ label the
generators of a group $G$ in whose Lie algebra both $e_{\mu}$ and $\omega
_{\mu}$ take values. $\eta^{\mu\nu\rho}$ is the ($e$- and $A$-independent)
Levi-Civita tensor density of weight $+1$ on the manifold $M.$ We take the
manifold $M$ to be topologically $\Sigma\times%
\mathbb{R}
$ with $\Sigma$ a closed two-dimensional Riemann surface. For $G=\mathrm{SU}%
(2),\ $the action is equivalent to that of Euclidean-signature general
relativity, while $G=\mathrm{SU}(1,1)$ corresponds to Lorentzian general
relativity. In the SU(2) case,
\begin{equation}
\Omega_{\mu\nu}^{i}=2\partial_{\lbrack\mu}\omega_{\nu]}^{i}+\epsilon
^{ijk}\omega_{\mu}^{j}\omega_{\nu}^{k},
\end{equation}
and setting $A_{\mu}^{i}:=(8\pi G_{\mathrm{N}})^{-1}\omega_{\mu}^{i}$ (which
has units of momentum (or mass) per length), one can rewrite the action as%
\begin{equation}
S_{\mathrm{P}}[e,\omega]=\tfrac{1}{2}\int_{M}\mathrm{d}^{3}x~\eta^{\mu\nu\rho
}\left(  2\partial_{\lbrack\mu}A_{\nu]}^{i}+8\pi G_{\mathrm{N}}\epsilon
^{ijk}A_{\mu}^{j}A_{\nu}^{k}\right)  e_{\rho i}.
\end{equation}
In the limit $G_{\mathrm{N}}\rightarrow0,$ we obtain the following action%
\begin{equation}
S[e,A]=\tfrac{1}{2}\int_{M}\mathrm{d}^{3}x~\eta^{\mu\nu\rho}F_{\mu\nu}%
^{i}e_{\rho i},\qquad F_{\mu\nu}^{i}:=2\partial_{\lbrack\mu}A_{\nu]}^{i},
\end{equation}
in which the $\mathrm{SU}(2)$ gauge symmetry
\begin{equation}
\omega_{\mu}\rightarrow g\omega_{\mu}g^{-1}-\frac{1}{8\pi G_{\mathrm{N}}}\left(
\partial_{\mu}g\right)  g^{-1}%
\end{equation}
has become a $\mathrm{U}(1)^{3}$ gauge symmetry:%
\begin{equation}
A_{\mu}^{i}\rightarrow A_{\mu}^{i}-\partial_{\mu}\theta^{i}%
\end{equation}
This is the action we wish to study.

\subsection{Constraints}

In this section we essentially follow \cite{qsd4}.\\

Canonical analysis of the theory defined by $S$ reveals $E_{i}^{a}:=\eta
^{ab}e_{b}^{i}$ as the momentum conjugate to $A_{a}^{i}$, where $\eta^{ab}$ is
the Levi-Civita density on $\Sigma,$ a symplectic structure given by
\begin{equation}
\{A_{a}^{i}(x),E_{j}^{b}(y)\}=\delta_{a}^{b}\delta_{j}^{i}\delta^{(2)}(x,y),
\end{equation}
and first class constraints%
\begin{equation}
G[\Lambda]:=\int\mathrm{d}^{2}x~\Lambda^{i}\partial_{a}E_{i}^{a},\qquad
F[N]:=\tfrac{1}{2}\int\mathrm{d}^{2}x~N_{i}\eta^{ab}F_{ab}^{i},
\end{equation}
where $\Lambda^{i},N_{i}$ are Lagrange multipliers. $G[\Lambda]$ constitutes
three U(1) Gauss\ constraints, and $F[N]$ is referred to as the curvature constraint.

Considering $F[N]$, when the 2-metric $q_{ab}:=e_{a}^{i}e_{b}^{j}\delta_{ij}$
has non-zero determinant $\det q\equiv q$, one may perform an invertible phase
space-dependent transformation on the Lagrange multipliers $N_{i}$ 
and arrive at an alternative set of constraints that more closely resemble
those that arise in 3+1 dimensions (for Euclidean signature and
Barbero-Immirzi parameter equal to 1). Namely, one may define a vector field
$N^{a}$ and a scalar density $N$ of weight $-\frac{1}{2}$ such that%
\begin{equation}
N^{i}=N^{a}\eta_{ab}E_{i}^{b}+Nq^{-1/4}E^{i},
\end{equation}
where $E^{i}:=\frac{1}{2}\epsilon^{ijk}\eta_{ab}E_{i}^{a}E_{j}^{b}$ is the
called degeneracy vector in \cite{qsd4}, which satisfies $E^{i}E^{i}=q$ and $E^{i}%
E_{i}^{a}=0.$ With this decomposition of $N^{i},$ the single curvature
constraint can be written as the sum of two constraints%
\begin{equation}
F[N]=V[\vec{N}]+H[N],
\end{equation}
where%
\begin{equation}
V[\vec{N}]:=\int\mathrm{d}^{2}x~N^{a}F_{ab}^{i}E_{i}^{b},\qquad H[N]:=\tfrac
{1}{2}\int\mathrm{d}^{2}x~Nq^{-1/4}\epsilon^{ijk}F_{ab}^{i}E_{j}^{a}E_{k}^{b}
\label{hammndiff}%
\end{equation}
By subtracting a multiple of the Gau\ss \ constraint from $V[\vec{N}]$ one
obtains the generator of diffeomorphisms:%
\begin{equation}
D[\vec{N}]:=V[\vec{N}]-G[A\cdot\vec{N}]=\int\mathrm{d}^{2}x~E_{i}%
^{a}\pounds _{\vec{N}}A_{a}^{i}=-\int\mathrm{d}^{2}x~A_{a}^{i}\pounds _{\vec
{N}}E_{i}^{a}.
\end{equation}
The Poisson algebra of constraints $G[\Lambda],D[\vec{N}],H[N]$ is first
class:%
\begin{align}
\{G[\Lambda],G[\Lambda^{\prime}]\}  &  =\{G[\Lambda],H[N]\}=0\\
\{D[\vec{N}],G[\Lambda]\}  &  =G[\pounds _{\vec{N}}\Lambda]\\
\{D[\vec{N}],D[\vec{N}^{\prime}]\}  &  =D[\pounds _{\vec{N}}\vec{N}^{\prime
}]\\
\{D[\vec{N}],H[N]\}  &  =H[\pounds _{\vec{N}}N]\\
\{H[N],H[M]\}  &  =D[\vec{\omega}]+G[A\cdot\vec{\omega}]=V[\vec{\omega
}],\qquad\omega^{a}:=q^{-1/2}E_{i}^{a}E_{i}^{b}\left(  M\partial
_{b}N-N\partial_{b}M\right)
\end{align}

\section{Quantum Kinematics}

Here we briefly review the Hilbert space on which the basic kinematical
operators, the holonomies and fluxes, are defined. It is in complete analogy
with the SU(2) case so we direct the reader to \cite{ttbook} for further
details. The kinematical Hilbert space can be defined by specifying a complete
orthonormal basis, as follows. We refer to the basis states as charge
networks, which, like the SU(2) spin networks, are specified by a graph with
representation labels. These states are written as%
\begin{equation}\label{eq:fourpointone}
|c\rangle=%
{\textstyle\bigotimes\nolimits_{i=1}^{3}}
{\textstyle\bigotimes\nolimits_{I=1}^{N}}
h_{e_{I}}^{n_{e_{I}}^{i}}.
\end{equation}
$c$ denotes the compound label $c=\{\gamma,\{\vec{n}_{I}\}_{I=1}^{N}\}$ where
$\gamma$ is a finite, piecewise-analytic graph embedded in $\Sigma$ consisting
of $N$ oriented analytic edges $e_{I}$ meeting at vertices $v$ (technically,
since the usual GNS construction would provide states based on three distinct
graphs $\gamma_{i},$ $i=1,2,3$, where the charge labels on each $\gamma_{i}$
are all non-zero, we consider the graph $\gamma$ to be the finest possible
graph associated with the union $\cup_{i}\gamma_{i}$). Each edge $e_{I}$ is
colored by a triplet $n_{e_{I}}^{i}$ ($i=1,2,3$ labeling the different U(1)
copies) of integers, which we denote in vector notation by $\vec{n}%
_{I}=(n_{e_{I}}^{1},n_{e_{I}}^{2},n_{e_{I}}^{3}).$ By splitting edges in their
interior at `trivial vertices' (points at which $\gamma$ remains analytic), we
arrange that at each non-trivial vertex all edges are outgoing by reversing
the orientation of appropriate segments; if a segment's orientation is
reversed by this procedure, the corresponding charges undergo a change of
sign. On the right side of \ref{eq:fourpointone},
\begin{equation}
h_{e_{I}}^{i}[A]:=\mathrm{e}^{\mathrm{i}\kappa\int_{e_{I}}A_{a}^{i}%
\mathrm{d}x^{a}}%
\end{equation}
is the holonomy of the U(1)$_{i}$ connection $A^{i}$ along the oriented edge
$e_{I}$ in the fundamental representation ($n_{e_{I}}^{i}=1$), and $h_{e_{I}%
}^{n_{e_{I}}^{i}}[A]$ is the holonomy in the $n_{e_{I}}^{i}$-representation
(the factor of $\kappa$ has the same units as $G_{\mathrm{N}}$ and is needed
to make the exponent dimensionless). $|c\rangle$ will be gauge-invariant with
respect to U(1)$^{3}$ gauge transformations only if it is gauge-invariant with
respect to each U(1)$_{i}$ separately, and $|c\rangle$ is U(1)$_{i}$
gauge-invariant if, at each non-trivial vertex $v,$ the sum of the charges
$\sum_{I}n_{e_{I}}^{i}$ on (outgoing) edges $e_{I}$ at $v$ vanishes. The set
of all gauge-invariant charge networks provides a complete orthonormal basis
(with respect to the Ashtekar-Lewandowski measure built from the normalized
U(1) Haar measure) for the kinematical U(1)$^{3}$ gauge-invariant Hilbert
space $\mathcal{H}_{\text{kin}}$.

In the connection representation, holonomies act on charge network functions
$c(A)$ by multiplication, and the densitized triads as%
\begin{equation}
\hat{E}_{i}^{a}(x)c(A)=\mathrm{i}\hbar\{E_{i}^{a}(x),c(A)\}=-\mathrm{i}%
\kappa\hbar\frac{\delta c(A)}{\delta A_{a}^{i}(x)}=\kappa\hbar\sum_{I}\left(
{\textstyle\int_{0}^{1}}
\mathrm{d}t_{I}~\delta^{(2)}(x,e_{I}(t_{I}))\dot{e}_{I}^{a}(t_{I})\right)
n_{e_{I}}^{i}c(A)
\end{equation}
where each edge $e_{I}$ is parameterized by $t_{I}\in\lbrack0,1],$ and
$\dot{e}_{I}^{a}$ is its tangent. Given a one-dimensional oriented surface
$L$, parameterized by $s\in\lbrack0,1]$ with tangent $\dot{L}^{a}$ one can
define a flux operator%
\begin{equation}
\hat{E}_{i}(L):=\int_{0}^{1}\mathrm{d}s~\eta_{ab}\hat{E}_{i}^{a}(L(s))\dot
{L}^{b}(s).
\end{equation}
Its action on a holonomy functional based on an edge which emanates from $L$
is given by%
\begin{equation}
\hat{E}_{i}(L)h_{e}^{n^{i}}[A]=\kappa\hbar\int_{0}^{1}\mathrm{d}s~\eta
_{ab}\dot{L}^{b}(s)\int_{0}^{1}\mathrm{d}t~\delta^{(2)}(L(s),e(t))\dot{e}%
^{a}(t)n^{i}h_{e}^{n^{i}}[A]=\tfrac{1}{2}\kappa\hbar\epsilon(L,e)n^{i}%
h_{e}^{n^{i}}[A]
\end{equation}
where $\epsilon(L,e)=\pm1,0$ is the relative orientation $L$ and $e.$ The
factor of $\frac{1}{2}$ appears because we have assumed that $e$ has an
endpoint on $L$ and evaluated one of the $\delta$-functions at the boundary of
integration. If $e$ has an endpoint on the boundary of $L,$ then an additional
factor of $\frac{1}{2}$ appears. $\hat{E}_{i}(L)$ extends by Leibniz' rule to
charge networks, and we observe that it is diagonal in the charge network basis.

\section{The Action of Thiemann's Hamiltonian Constraint}

In this section we describe Thiemann's seminal construction of the Euclidean
Hamiltonian constraint in LQG. As we are working in 2+1 dimensions, we will
summarize the construction as given in \cite{qsd4}, and will restrict ourself
to the $\mathrm{U}(1)^{3}$ case.\footnote{Although Thiemann has defined a
quantum Hamiltonian constraint in 2+1 as well as 3+1 dimensions for the
$\mathrm{SU}(2)$ theory, his constructs can be trivially generalized to any
compact group, and in particular $\mathrm{U}(1)^{3}$} We will only focus on
the salient features of his construction which are most relevant to us. This
will help us bring out the contrast between quantization choices we make and
the choices made in \cite{qsd4}.

Given a graph $\gamma$ with a vertex set $V(\gamma)$, Thiemann's construction
involves a choice of the following ingredients:

\begin{enumerate}
\item[(1)] A one parameter family of triangulations $T(\gamma)$ adapted to the
graph $\gamma$.

\item[(2)] An approximation of the classical smeared Hamiltonian constraint by
a suitable Riemann sum over the simplices of $T(\gamma)$ such that, in the
limit of shrinking triangulation, one recovers the continuum expression.

\item[(3)] Associated to each simplex in the triangulation which contributes
to the Riemann sum, the choice of a loop (to approximate the curvature) and a
choice of a collection of edge segments (to approximate the inverse metric determinant).

\item[(4)] The approximant to the continuum curvature is a holonomy around the
pre-chosen (based) loop in a representation which Thiemann selects to be the
fundamental representation (or at least this representation is chosen to be
fixed once and for all and is not considered to be state dependent).
\end{enumerate}

\textbf{Choice of Triangulation}: Given a vertex $v\in V(\gamma),$ let there
be $n$ (outgoing) edges emanating from $v$, $\{e_{1},\dots,e_{n}\}$. Let us
assume that these edges are such that $(e_{i},e_{i+1})$ are right oriented
$\forall i\in\{1,\dots,n\}$ with $n+1:=1$.\footnote{The notion of orientation
is given in (definition 4.1) \cite{qsd4}. Roughly speaking, it means that,
given a pair of edges $e,e^{\prime},$ if upon starting at $e$ and moving
counterclockwise one encounters $e^{\prime}$ before encountering the analytic
extension of $e,$ then the order pair $(e,e^{\prime})$ is said to be right
oriented.} Now assign to each pair $(e_{i},e_{i+1})$ a two simplex
$\triangle_{i}(v)|_{i=1,\dots,n}$ which has one of the vertices as $v$ and
whose boundary is traversed by two segments $s,s^{\prime}$ in $e,e^{\prime}$
respectively, and an (analytic) arc $a_{s,s^{\prime}}$ between end points of
$s$ and $s^{\prime}$. Note that Thiemann's choice of triangulation is such
that the Riemann sum which approximates $H[N]$ is given by
\begin{equation}
H_{T(\gamma)}[N]=\sum_{v\in V(\gamma)}\frac{4}{n}\sum_{i=1}^{n}\mathcal{F}%
_{\triangle_{i}(v)}(A,E,N)+\ \text{Sum over simplices which do not contain
vertices of $\gamma$}%
\end{equation}
Here $\mathcal{F}_{\triangle_{i}(v)}(A,E)$ is a suitable approximant to
$\int_{\triangle_{i}(v)}\mathrm{d}^{3}x~N\sqrt{q}^{-k}\epsilon^{ijk}F_{ab}%
^{i}E_{j}^{a}E_{k}^{b}$ written in terms of various holonomies and
fluxes.\footnote{Usually $k$ is chosen to be 1, but at least as far as the
finite-triangulation operator is concerned, one can be more general. The
density weight of the lapse depends on $k.$}

This nice split of the Riemann sum means that upon quantization, the
s$\text{um over simplices which do not contain vertices of $\gamma$ }$gives
the zero operator. Thus%
\begin{equation}
\hat{H}_{T(\gamma)}[N]=\sum_{v\in V(\gamma)}\frac{4}{n}\sum_{i=1}%
^{n}N(v)\widehat{\mathcal{F}_{\triangle_{i}(v)}(A,E)}%
\end{equation}
where $\widehat{\mathcal{F}_{\triangle_{i}(v)}(A,E)}$ is a composite operator
built out of $\hat{h}_{\partial(\triangle_{i}(v))},\hat{h}_{s_{i}},\hat
{h}_{s_{i+1}}\ \text{and}\ \hat{V}(v)$. Schematically, it looks like
\begin{equation}
\widehat{\mathcal{F}_{\triangle_{i}(v)}(A,E)}=\frac{1}{\delta^{m}}O(\hat
{h}_{\partial(\triangle_{i}(v))}\hat{h}_{s_{i}}\hat{h}_{s_{i+1}},\hat{V}(v)),
\end{equation}
where the parameter $\delta$ is such that the coordinate area of
$\triangle_{i}(v)$ is $O(\delta^{2})$ and $m$ depends on what density weight
constraint we are working with.\footnote{Thiemann's quantization is specific
to the density one constraint. This is because it is only for density weight
one that one can take the continuum limit of $\hat{H}_{T(\gamma)}[N]$ on
$\mathcal{H}_{\mathrm{kin}}$. But if one were to work with higher density
constraints (whose continuum limit will not be a well-defined operator on
$\mathcal{H}_{\mathrm{kin}}$ but on some distributional space, the germ of
Thiemann's construction would essentially go through except for extra factors
on $\frac{1}{\delta^{m}}$ floating around.}

In the above equation $s_{i}$ and $s_{i+1}$ are arc segments in $e_{i}%
,e_{i+1}$. We emphasize three of the four features mentioned at the beginning
of this section once again, as it will help us illustrate the key difference
between Thiemann's regularization and the one we adapt in this paper.

\begin{enumerate}
\item[(1)] All the holonomies in the construction are typically in the
fundamental (or at least a state-independent) representation.

\item[(2)] The action of this (finite triangulation) Hamiltonian constraint on
a state based on $\gamma$ results in the addition of two new vertices and one
new edge. \emph{The location of these vertices and edges is independent of the
colorings of the state and only depend on the graph.}

\item[(3)] The Hamiltonian constraint has a trivial action on the newly
created vertices.\footnote{This statement is less obvious in two dimensions
than in three, however in \cite{qsd4} it is ensured by constraining the
tangent space structure of the graph at these vertices.}
\end{enumerate}

In Section \ref{HatFiniteTriangulation}, we will see how the quantization of
the Hamiltonian constraint performed in this paper differs in these three
aspects from Thiemann's construction.

\section{A Classical Computation}

\label{classint}

\label{classical}

In this section we exhibit a classical computation which motivates our
proposal for the action of the quantum Hamiltonian constraint operator. We
work with the density-two Hamiltonian constraint; i.e., with no power of the
metric determinant $q$ appearing. This simplifies the calculation
considerably, and allows for a geometric interpretation of the action of the
constraint. Moreover, in the U(1)$^{3}$ quantum theory, $\hat{q}$ is just a
multiple of the identity on charge networks (and hence any power of it is
also), so we expect this simplified calculation to capture the most important
ingredients of the classical theory that we want to retain in quantization.

Consider the action of the density two Hamiltonian constraint
\begin{equation}
H[N]=\tfrac{1}{2}\int\mathrm{d}^{2}x~N\epsilon^{ijk}E_{i}^{a}E_{j}^{b}%
F_{ab}^{k} \label{density2}%
\end{equation}
on cylindrical functions (here $N$ is a scalar density of weight $-1$). First
observe that the action of the corresponding Hamiltonian vector field on the
connection can be written as%
\begin{equation}
X_{H[N]}A_{a}^{i}(x):=\{H[N],A_{a}^{i}(x)\}=-\epsilon^{ijk}NE_{j}^{b}%
F_{ab}^{k}(x)=\epsilon^{ijk}\mathcal{L}_{V_{j}}A_{a}^{k}(x)-\epsilon
^{ijk}\partial_{a}(NE_{j}^{b}A_{b}^{k}(x)). \label{eq:hamvec1}%
\end{equation}
Here $V_{i}^{a}:=NE_{i}^{a}$ is a phase space-dependent vector field (of
density weight zero) for each value of $i.$ The second term can be seen as the
result of a U(1) gauge transformation, and since we work at the U(1)$^{3}$
gauge-invariant level, it will not contribute to the analysis. We see that the
action thus results in a linear combination of (phase space-dependent)
infinitesimal\ `diffeomorphisms.' Of course these are not diffeomorphisms in
the usual sense since the U(1)$^{3}$ indexes get reshuffled.

\begin{figure}
\begin{center}
\includegraphics[scale=0.675]{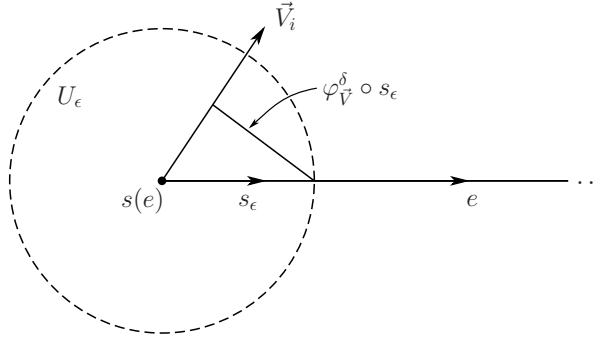}
\caption{The action of a diffeomorphism generated by the vector field $\vec
{V}_i$ (which is compactly supported in $U_{\epsilon}%
$) on a holonomy functional.}
\label{schmigure}
\end{center}
\end{figure}

Consider now the action of $X_{H[N]}$ on a holonomy functional associated with
an edge $e,$%
\begin{equation}
h_{e}^{\vec{n}_{e}}\equiv h_{e}^{n_{e}^{1}}h_{e}^{n_{e}^{2}}h_{e}^{n_{e}^{3}%
}:\mathcal{A}\rightarrow\mathrm{U}(1)^{3},\qquad A=(A^{1},A^{2},A^{3})\mapsto
h_{e}^{n_{e}^{1}}[A^{1}]h_{e}^{n_{e}^{2}}[A^{2}]h_{e}^{n_{e}^{3}}[A^{3}].
\end{equation}
where ${\cal A}$ is the space of smooth $U(1)^{3}$ connections. 
We suppose that the vector fields $V_{i}^{a}$ have support only in some
$\epsilon$-neighborhood $U_{\epsilon}$ of $s(e),$ the source of $e$ (as
mentioned above, in the quantum theory, which features a non-trivial power of
$\hat{q}$, this is the only relevant situation, in fact with $\epsilon$ as
small as one pleases, since $\hat{q}$ acts only at the vertices of charge
networks). Using (\ref{eq:hamvec1}), we find (discarding the terms coming form
the total derivative)%
\begin{equation}
X_{H[N]}h_{e}^{\vec{n}_{e}}[A]=\kappa\left(  \mathrm{i}n_{e}^{2}%
{\textstyle\int_{e}}
\mathcal{L}_{V_{3}}A^{1}-\mathrm{i}n_{e}^{3}%
{\textstyle\int_{e}}
\mathcal{L}_{V_{2}}A^{1}+\mathrm{cyclic}\right)  h_{e}^{n_{e}^{1}}[A^{1}%
]h_{e}^{n_{e}^{2}}[A^{2}]h_{e}^{n_{e}^{3}}[A^{3}] \label{shithorse2}%
\end{equation}
We approximate the Lie derivatives by
\begin{equation}
\mathcal{L}_{V}A=\frac{1}{\delta}((\varphi_{V}^{\delta})^{\ast}A-A)+O(\delta),
\end{equation}
where $\varphi_{V}^{\delta}$ is a one-parameter family (parameterized by
$\delta$) of finite transformations generated by the vector field $V$,
$(\varphi_{V}^{\delta})^{\ast}$ being the pullback map. Since
supp$(V)=U_{\epsilon},$ we have%
\begin{align}
\mathrm{i}\kappa n_{e}%
{\textstyle\int_{e}}
\mathcal{L}_{V}A  &  =\mathrm{i}n_{e}%
{\textstyle\int_{s_{\epsilon}}}
\mathcal{L}_{V}A\nonumber\\
&  =\frac{1}{\delta}\mathrm{i}\kappa n_{e}%
{\textstyle\int_{s_{\epsilon}}}
((\varphi_{V}^{\delta})^{\ast}A-A)+O(\delta)\nonumber\\
&  =\frac{1}{\delta}\kappa\left(  \mathrm{i}n_{e}%
{\textstyle\int_{\varphi_{V}^{\delta}\circ s_{\epsilon}}}
A-\mathrm{i}n_{e}%
{\textstyle\int_{s_{\epsilon}}}
A\right)  +O(\delta)\nonumber\\
&  =\frac{1}{\delta}\left(  h_{\varphi_{V}^{\delta}\circ s_{\epsilon}}^{n_{e}%
}[A]-h_{s_{\epsilon}}^{n_{e}}[A]\right)  +O(\delta)
\end{align}
where $s_{\epsilon}=e\cap U_{\epsilon}$ is a small segment of $e$ lying in
$U_{\epsilon}$ (see Figure (\ref{schmigure})%
) Substituting in (\ref{shithorse2}), we obtain%
\begin{align}
X_{H[N]}h_{e}^{\vec{n}_{e}}[A]  &  =\frac{1}{\delta}\left(  \left(
h_{\varphi_{V_{3}}^{\delta}\circ s_{\epsilon}}^{n_{e}^{2}}[A^{1}%
]-h_{s_{\epsilon}}^{n_{e}^{2}}[A^{1}]\right)  -\left(  h_{\varphi_{V_{2}%
}^{\delta}\circ s_{\epsilon}}^{n_{e}^{3}}[A^{1}]-h_{s_{\epsilon}}^{n_{e}^{3}%
}[A^{1}]\right)  +\mathrm{cyclic}\right)  h_{e}^{n_{e}^{1}}[A^{1}]h_{e}%
^{n_{e}^{2}}[A^{2}]h_{e}^{n_{e}^{3}}[A^{3}]+O(\delta)\nonumber\\
&  =\frac{1}{\delta}\left(  h_{\varphi_{V_{3}}^{\delta}\circ s_{\epsilon}%
}^{n_{e}^{2}}[A^{1}]h_{s_{\epsilon}}^{-n_{e}^{2}}[A^{1}]-1\right)
h_{s_{\epsilon}}^{n_{e}^{2}}[A^{1}]h_{e}^{n_{e}^{1}}[A^{1}]h_{e}^{n_{e}^{2}%
}[A^{2}]h_{e}^{n_{e}^{3}}[A^{3}]\nonumber\\
&  \qquad-\frac{1}{\delta}\left(  h_{\varphi_{V_{2}}^{\delta}\circ
s_{\epsilon}}^{n_{e}^{3}}[A^{1}]h_{s_{\epsilon}}^{-n_{e}^{3}}[A^{1}]-1\right)
h_{s_{\epsilon}}^{n_{e}^{3}}[A^{1}]h_{e}^{n_{e}^{1}}[A^{1}]h_{e}^{n_{e}^{2}%
}[A^{2}]h_{e}^{n_{e}^{3}}[A^{3}]+\mathrm{cyclic}+O(\delta)
\end{align}
Approximating $h_{s_{\epsilon}}^{n_{e}^{i}}[A^{1}]$ outside the parentheses as
$1+O(\epsilon)$, we have finally%
\begin{align}
X_{H[N]}h_{e}^{\vec{n}_{e}}[A]  &  =\frac{1}{\delta}\left(  h_{\varphi_{V_{3}%
}^{\delta}\circ s_{\epsilon}}^{n_{e}^{2}}[A^{1}]h_{s_{\epsilon}}^{-n_{e}^{2}%
}[A^{1}]-1\right)  h_{e}^{n_{e}^{1}}[A^{1}]h_{e}^{n_{e}^{2}}[A^{2}%
]h_{e}^{n_{e}^{3}}[A^{3}]\nonumber\\
&  \qquad-\frac{1}{\delta}\left(  h_{\varphi_{V_{2}}^{\delta}\circ
s_{\epsilon}}^{n_{e}^{3}}[A^{1}]h_{s_{\epsilon}}^{-n_{e}^{3}}[A^{1}]-1\right)
h_{e}^{n_{e}^{1}}[A^{1}]h_{e}^{n_{e}^{2}}[A^{2}]h_{e}^{n_{e}^{3}}%
[A^{3}]+\mathrm{cyclic}+O(\epsilon,\delta). \label{HAMVECK4}%
\end{align}

We can extend this calculation to charge networks. Consider a charge network
$c$ based on a graph containing an $N$-valent vertex $v\in\mathrm{supp}%
(V_{i})$ (with $\mathrm{supp}(V_{i})$ an $\epsilon$-neighborhood of $v$) and
suppose no other vertex of $c$ lies in $\mathrm{supp}(V_{i}).$ Then a simple
Leibniz rule application of (\ref{HAMVECK4}) yields%
\begin{equation}
X_{H[N]}c(A)=c(A)\frac{1}{\delta}\sum_{I=1}^{N}\left[  \left(  h_{\varphi
_{V_{3}}^{\delta}\circ s_{\epsilon}^{I}}^{n_{e_{I}}^{2}}[A^{1}]h_{s_{\epsilon
}^{I}}^{-n_{e_{I}}^{2}}[A^{1}]-1\right)  -\left(  h_{\varphi_{V_{2}}^{\delta
}\circ s_{\epsilon}^{I}}^{n_{e_{I}}^{3}}[A^{1}]h_{s_{\epsilon}^{I}}%
^{-n_{e_{I}}^{3}}[A^{1}]-1\right)  \right]  +\mathrm{cyclic}+O(\epsilon
,\delta). \label{glue}%
\end{equation}
We can rewrite this result in terms of a product over $I$ by noting that,
given some $\epsilon$-dependent quantities $f_{I}(\epsilon)=1+\epsilon g_{I}$
(short holonomies being an example),%
\begin{equation}%
{\textstyle\sum\nolimits_{I}}
(f_{I}(\epsilon)-1)=%
{\textstyle\prod\nolimits_{I}}
f_{I}(\epsilon)-1+O(\epsilon^{2}). \label{shitface}%
\end{equation}
Using (\ref{shitface}), (\ref{glue}) becomes%
\begin{equation}
X_{H[N]}c(A)=c(A)\frac{1}{\delta}\left[
{\textstyle\prod\nolimits_{I=1}^{N}}
h_{\varphi_{V_{3}}^{\delta}\circ s_{\epsilon}^{I}}^{n_{e_{I}}^{2}}%
[A^{1}]h_{s_{\epsilon}^{I}}^{-n_{e_{I}}^{2}}[A^{1}]-%
{\textstyle\prod\nolimits_{I=1}^{N}}
h_{\varphi_{V_{2}}^{\delta}\circ s_{\epsilon}^{I}}^{n_{e_{I}}^{3}}%
[A^{1}]h_{s_{\epsilon}^{I}}^{-n_{e_{I}}^{3}}[A^{1}]\right]  +\mathrm{cyclic}%
+O(\epsilon,\delta) \label{HAMM6}%
\end{equation}
It is easy to check that if $c(A)$ is gauge-invariant at $v$ then the
$X_{H[N]}c(A)$ derived in (\ref{HAMM6}) will be gauge-invariant as well.

We can now restate our goal: Our aim is to quantize the Hamiltonian constraint
(\ref{hammndiff}) at finite triangulation in such a way that its action on
charge networks gives the linear combination in (\ref{HAMM6}) (up to factors
coming from the quantization of the non-trivial power of $q$ appearing in
(\ref{hammndiff})). Of course (\ref{HAMM6}) is not the only approximant one
can obtain starting with the geometric action of $X_{H[N]}$. The justification
of the choices we have made lies in the fact that the off-shell closure
condition is satisfied.

\section{Definition of $\hat{H}[N]$}

The classical Hamiltonian constraint (\ref{hammndiff}) is written in terms of
the local connection and densitized triad fields, but neither of these objects
is a well-defined operator in our quantum theory, so we cannot immediately
write down an operator $\hat{H}[N]$ corresponding to (\ref{hammndiff}). The
strategy \cite{ttbook} is to first derive a classical approximant $H_{T}[N]$
to $H[N]$, where $H_{T}[N]$ is written solely in terms of holonomies and
fluxes, and then quantize it as an operator on $\mathcal{H}_{\text{kin}}.$
Since there are an uncountably-infinite number of different ways that the
connection and triad can be approximated using holonomies and fluxes
(generally leading to inequivalent quantum operators) we tune our choices to
the end we seek, which is to mimic the classical action found above. There are
many choices to be made, and in the following subsections, we motivate and
specify each.

\subsection{Choice of Triangulation}

All subsequent regularization choices are based first on a one parameter
family of \textit{triangulations} $T(\delta)$ of $\Sigma$, by which we mean,
for a fixed value of $\delta,$ a tesselation or cover of $\Sigma$ by subsets
$\triangle\subset\Sigma$.\footnote{We use the term triangulation rather
loosely here. It does not mean a triangulation in the sense of algebraic
topology where an $n$-dimensional triangulation of a space implies covering
the space with $n$-dimensional simplices which themselves intersect only in lower
dimensional simplices.} We proceed to spell out what is required of
$T(\delta).$

We fix once and for all a volume form $\omega$ on $\Sigma$ used to assign
areas in subsequent constructions. Then, the argument $\delta$ of $T(\delta)$
is a parameter which roughly measures the (square root of the) area of each
$\triangle\in T(\delta)$. All we require of $T(\delta)$ is that the Riemann
sum $\sum_{\triangle\in T(\delta)}H_{\triangle}[N]$ converge to $H[N]$ as
$\delta\rightarrow0$ (the triangulation becoming infinitely fine), where
$H_{\triangle}[N]$ is some approximant to $H[N]$ in $\triangle.$ The allowed
values of $\delta$, and hence the class of admissible triangulations, will be
fixed by the charge network state $c$ that $\hat{H}_{T}$ will act on. We
emphasize that any one parameter family $T(\delta)$ satisfying this
requirement is an admissible family of triangulations, in the sense that the
Riemann sum correctly approximates the classical Hamiltonian constraint, and
we use this freedom to choose $T(\delta)$ which are `adapted' to charge
networks, in a way we describe below.

\begin{figure}
\begin{center}
\includegraphics[scale=0.9]{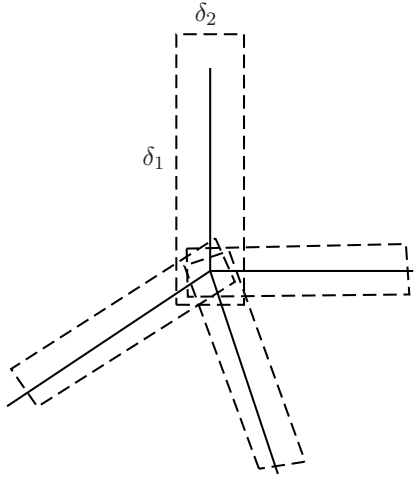}
\caption
{Portion of an admissible triangulation adapted to a graph near a vertex. Each plaquette $\triangle
^{I}_{v}$ containing the vertex overlaps the others in an area $\sim
\delta_2^2$.}
\label{plaq}
\end{center}
\end{figure}

Let $c$ be a charge network with an underlying graph $\gamma$. We construct a
one parameter family of triangulations $T(\delta,\gamma)$ adapted to the graph
$\gamma,$ satisfying the following criteria:

\begin{enumerate}
\item[(i)] The coordinate areas $|\triangle|~=\int_{\triangle}\omega$ of all
$\triangle\in T(\gamma,\delta)$ containing a vertex of $\gamma$ satisfy
\begin{equation}
|\triangle|~=\delta_{1}\delta_{2}=\delta^{2}, \label{eq:area1}%
\end{equation}
where $\delta_{1}\gg\delta\gg\delta_{2}$. That is, the plaquettes are chosen
to be `long rectangles' in some local coordinates.

\item[(ii)] We tailor $T(\delta,\gamma)$ so that each $N$-valent vertex $v$ in
the vertex set $V(\gamma)$ of $\gamma$ is in the interior of precisely $N$
plaquettes $\{\triangle_{v}^{I}\}_{I=1}^{N},$ moreover, we require that
$\triangle_{v}^{I}$ is aligned along the edge $e_{I}$ emanating from $v$ as
shown in Figure (\ref{plaq}). This requirement ensures that the overlap of any
plaquettes is a region of area $\sim\delta_{2}^{2},$ and hence the
contribution to the Riemann sum of these regions will be sub-leading in
$\delta$ in the sense that it will vanish in the continuum limit.

\item[(iii)] The triangulation of the space $\Sigma-\bigcup_{v\in V(\gamma
)}\bigcup_{I}\triangle_{v}^{I}$ is only subject to the requirements that no
plaquettes overlap (except in their boundaries), and that their areas scale
with $\delta^{2}$.
\end{enumerate}

The existence of such $T(\delta)$ in which the contribution to a Riemann sum
from overlapping cuboids which vanishes in the continuum limit was shown in
\cite{amdiff} for three dimensions. We assume here a precisely similar
construct for two dimensional case.%

\subsection{Riemann Sum}

Given an admissible triangulation $T(\delta,\gamma)$ adapted to $\gamma,$ our
next task is to construct an approximant $H_{\triangle}[N]$ to $H[N]$ in each
$\triangle\in T(\delta,\gamma).$ First we expand the classical Hamiltonian
$H[N]$ into terms labeled by the curvature's U(1) index:%
\begin{equation}
H[N]=\tfrac{1}{2}\int\mathrm{d}^{2}x~Nq^{-1/4}\left(  \epsilon^{1jk}F_{ab}%
^{1}E_{j}^{a}E_{k}^{b}+\epsilon^{2jk}F_{ab}^{2}E_{j}^{a}E_{k}^{b}%
+\epsilon^{3jk}F_{ab}^{3}E_{j}^{a}E_{k}^{b}\right)  =:\sum_{i=1}^{3}H^{(i)}[N]
\end{equation}
Let us focus on $H^{(1)}$; $H^{(2)}$ and $H^{(3)}$ can be obtained by cyclic
permutations of the U(1)$_{i}$ indexes. For an admissible $T(\delta,\gamma)$,
the following expression converges to $H^{(1)}[N]$ as $\delta\rightarrow0$:%
\begin{equation}
H_{T(\delta,\gamma)}^{(1)}[N]=\tfrac{1}{2}\sum_{\triangle\in T(\delta,\gamma
)}\left\vert \triangle\right\vert N(v_{\triangle})q^{-1/4}(v_{\triangle
})\epsilon^{1jk}F_{ab}^{1}(v_{\triangle})E_{j}^{a}(v_{\triangle})E_{k}%
^{b}(v_{\triangle}),
\end{equation}
where $v_{\triangle}$ is a point in $\triangle$, which we specify after
splitting the sum in the following way: The sum over $\triangle$ is split into
those $\triangle_{v}$ that contain a vertex $v\in V(\gamma),$ and those
$\bar{\triangle}$ that do not:%

\begin{equation}
2H_{T(\delta,\gamma)}^{(1)}[N]=\sum_{\triangle_{v}|v\in V(\gamma)}\left\vert
\triangle_{v}\right\vert Nq^{-1/4}\epsilon^{1jk}F_{ab}^{1}E_{j}^{a}E_{k}%
^{b}(v)+\sum_{\bar{\triangle}}\left\vert \bar{\triangle}\right\vert
Nq^{-1/4}\epsilon^{1jk}F_{ab}^{1}E_{j}^{a}E_{k}^{b}(v_{\bar{\triangle}%
}),\label{bling}%
\end{equation}
and in the first sum, $v_{\triangle}$ is chosen to be $v,$ the vertex
contained in $\triangle,$ while in the second sum, $v_{\bar{\triangle}}$ is a
basepoint of $\bar{\triangle},$ chosen once and for all. When we quantize
(\ref{bling}) as an operator acting on charge networks based on the graph
$\gamma$ with $\hat{q}^{-1/4}$ acting right-most, the latter sum will not
contribute, since as shown in Appendix \ref{A1}, $\hat{q}^{-1/4}$ acts
non-trivially only at charge network vertices.

Next we need to approximate the various local fields in the first sum of
(\ref{bling}) by holonomies and fluxes. This consists of choosing surfaces
over which the triads are smeared, holonomies around loops that feature in the
curvature approximant, and holonomies along short paths that feature in the
inverse metric determinant approximant. We take a cue from the treatment of
the diffeomorphism constraint \cite{amdiff}, and tailor the curvature
approximant\ to the underlying state. In this case however, there is no fixed
shift vector field that one can use to define a small loop. To stand in its
place, we introduce a key ingredient in our construction, the \textit{quantum
shift.}

\subsection{The Quantum Shift}

Note that classically $Nq^{-1/4}E_{j}^{a}$ is a vector field (of density
weight zero) for each $j$. The rough idea is to quantize this operator on
charge networks and use its eigenvalues as shift vector components, which then
feed into the definition of the small loop used to approximate the curvature.
It turns out that in the U(1)$^{3}$ theory, our quantization of $Nq^{-1/4}%
E_{j}^{a}$ yields an operator diagonal in the charge network basis, so we
define the (regularized) quantum shift components by%
\begin{equation}
V_{j}^{a}(x)|_{\epsilon}:=\langle c|N\hat{E}_{j}^{a}(x)|_{\epsilon}\hat
{q}_{\epsilon}^{-1/4}(x)|c\rangle.
\end{equation}
where $\hat{E}_{j}^{a}|_{\epsilon}$ and $\hat{q}_{\epsilon}^{-1/4}$ denote
some $\epsilon$-regularized $\hat{E}_{j}^{a}$ and $\hat{q}^{-1/4},$ which we
construct below. As suggested by the notation, we will quantize $E_{j}^{a}$
and $q^{-1/4}$ seperately.

As explained in Appendix (\ref{A1}), the regulated operator $\hat{q}%
_{\epsilon}^{-1/4}$ we employ is proportional to $(\kappa\hbar)^{-1},$ as well
as the small parameter $\epsilon$ used to construct the classical identity
that is quantized to define $\hat{q}_{\epsilon}^{-1/4}$. We leave these
factors explicit, and write the eigenvalues as
\begin{equation}
\hat{q}_{\epsilon}^{-1/4}(v)|c\rangle=:\frac{\epsilon}{\kappa\hbar}%
\lambda(\vec{n}_{v}^{c})|c\rangle,
\end{equation}
where $v\in V(c)$ (otherwise the right-hand side is zero), and the
$\lambda(\vec{n}_{v}^{c})$ are dimensionless numbers depending on relations
amongst the tangents of the edges emanating from $v,$ their charges, as well
as additional regularization choices.

As for $\hat{E}_{j}^{a}|_{\epsilon},$ we require some extra structure: At each
vertex $v\in c,$ we fix, once and for all, an $\epsilon$-neighborhood
$U_{\epsilon}(\gamma,v)$ with a coordinate chart $\{x_{v}\}$ with origin at
$v,$ and a coordinate ball $B_{x}(v,\epsilon)\subset U_{\epsilon}(\gamma,v)$
of radius $\epsilon$ centered at $v.$ Using this structure, we regularize the
$\delta$-function appearing in the action of $\hat{E}_{j}^{a}$, resulting in a
regularized operator $\hat{E}_{j}^{a}|_{\epsilon}$ which acts as%
\begin{equation}
\hat{E}_{j}^{a}|_{\epsilon}(v_{x})|c\rangle:=\kappa\hbar\sum_{e_{I}\cap
v}\left(
{\textstyle\int_{0}^{1}}
\mathrm{d}t~\frac{\chi_{B_{x}(v,\epsilon)}(e_{I}(t))}{\pi\epsilon^{2}}\dot
{e}_{I}^{a}(t)\right)  n_{I}^{j}|c\rangle,
\end{equation}
where $\chi_{S}$ is the characteristic function on $S.$ $v_{x}$ reminds us
that the coordinate chart $\{x_{v}\}$ is required extra structure. This
evaluates to
\begin{equation}
\hat{E}_{j}^{a}|_{\epsilon}(v_{x})|c\rangle=\frac{\kappa\hbar}{\pi\epsilon
^{2}}\sum_{e_{I}\cap v}\left(
{\textstyle\int_{e_{I}\cap B_{x}(v,\epsilon)}}
\mathrm{d}e_{I}^{a}\right)  n_{I}^{j}|c\rangle=\frac{\kappa\hbar}{\pi\epsilon
}\sum_{e_{I}\cap v}\hat{e}_{I}^{a}n_{I}^{j}|c\rangle+O(1)
\end{equation}
where $\hat{e}_{I}^{a}$ is a unit vector in $\{x_{v}\}$ which passes through
$e_{I}\cap B_{x}(v,\epsilon).$ Thus we have
\begin{equation}
\hat{V}_{j}^{a}(v_{x})|_{\epsilon}|c\rangle=N(v_{x})\hat{E}_{j}^{a}%
|_{\epsilon}(v)\hat{q}_{\epsilon}^{-1/4}(v)|c\rangle=\frac{1}{\pi}%
N(v_{x})\lambda(\vec{n}_{v}^{c})\sum_{e_{I}\cap v}\hat{e}_{I}^{a}n_{I}%
^{j}|c\rangle=:V_{j}^{a}(v_{x},\epsilon,c)|c\rangle
\end{equation}
This is a heavily coordinate-dependent construction (note that it also depends
on choices made in the classical identity used in the construction of $\hat
{q}_{\epsilon}^{-1/4}$). We place a bound $\delta_{0}:=\delta_{0}(c,\epsilon)$
on the parameter $\delta$ (associated with $T(\delta,\gamma)$) by requiring
that for all $\delta\leq\delta_{0},$ the endpoint of the arc $\delta E_{j}%
^{a}$ is in the ball $U_{\epsilon}(\gamma,v)$.

\subsection{The Hamiltonian Constraint Operator at Finite Triangulation}

\label{HatFiniteTriangulation}

In this subsection we lay out our proposal for the Hamiltonian constraint
operator at finite triangulation. Let us order $\hat{H}_{T(\delta,\gamma
)}^{(1)}[N]$ in the following way:
\begin{equation}
2\hat{H}_{T(\gamma,\delta)}^{(1)}[N]=\sum_{v\in V(\gamma)}\sum_{\triangle
_{v}|v\in V(\gamma)}\left\vert \triangle_{v}\right\vert \epsilon
^{1jk}(\widehat{F_{ab}^{1}E_{k}^{b}})_{\delta}\hat{V}_{j}^{a}(v_{x})|_{\delta
}+\sum_{\bar{\triangle}}~\left\vert \bar{\triangle}\right\vert \epsilon
^{1jk}(\widehat{F_{ab}^{1}E_{k}^{b}})_{\delta}\hat{V}_{j}^{a}(v_{\bar
{\triangle}})|_{\delta}%
\end{equation}
Since $\hat{q}_{\delta}^{-1/4}$ vanishes everywhere except at the vertices of
$\gamma,$ the second sum gives no contribution, leaving%
\begin{equation}
2\hat{H}_{T(\gamma,\delta)}^{(1)}[N]|c\rangle=\sum_{v\in V(\gamma)}%
\sum_{\triangle_{v}|v\in V(\gamma)}\left\vert \triangle_{v}\right\vert \left(
V_{2}^{a}(v_{x},\delta,c)(\widehat{F_{ab}^{1}E_{3}^{b}}(\triangle
_{v}))_{\delta}-V_{3}^{a}(v_{x},\delta,c)(\widehat{F_{ab}^{1}E_{2}^{b}%
}(\triangle_{v}))_{\delta}\right)  |c\rangle
\end{equation}
We now use the eigenvalues $V_{i}^{a}$ to specify the loops used to define the
curvature operator. Specifically, at a given vertex $v$, we associate one loop
with each edge emanating from $v$. $T(\gamma,\delta)$ is chosen such that for
an $N$-valent vertex $v,$ there are $N$ plaquettes $\{\triangle_{v}%
^{I}\}_{I=1}^{N}$ containing $v,$ and with each a loop is associated. We now
construct these loops.

For a given edge $e_{I},$ one segment of the loop is formed by a
coordinate-length $\delta$ segment of $e_{I}$ itself, and another by a segment
of length $\delta|E_{i}^{a}|$ in the direction of $V_{i}^{a}$. Note here that
\begin{equation}
\left\vert E_{i}^{a}\right\vert :=\frac{\left\vert V_{i}^{a}(v_{x}%
,\delta,c)\right\vert }{\left\vert N(v_{x})\lambda(\vec{n}_{v}^{c})\right\vert
}%
\end{equation}
is the norm of the quantum shift eigenvalue, apart from the inverse volume
eigenvalue and the value of the lapse; that is, we do not use the entire
quantum shift for the loop specification (if $N(v_{x})\lambda(\vec{n}_{v}%
^{c})=0,$ then the quantum shift is zero, and the Hamiltonian operator is
defined to act trivially). First we describe the generic case, where the
endpoint of the arc $\delta E_{i}^{a}$ does not lie on $\gamma$, and later
describe the special case when this endpoint lies on $\gamma.$

\begin{figure}
\begin{center}
\includegraphics[scale=0.8]{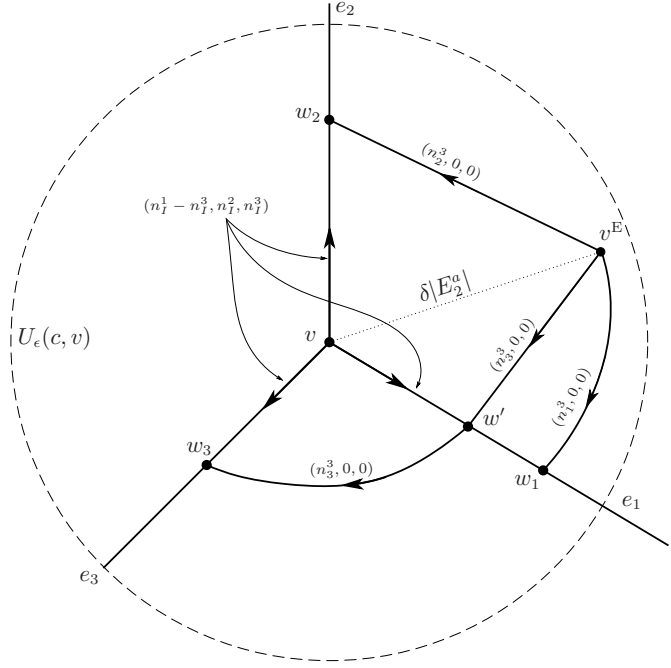}
\caption{The state $|c_1\cup\alpha^{\delta}_v(\langle\hat{E}_2\rangle
,n^3),c_2,c_3\rangle$ as produced by the action of $\hat{H}^{(1)}%
_{T(\gamma, \delta)}[N]$ in the generic case where $v^{\mathrm{E}%
}$ does not lie on $\gamma
$. Each segment $e_I$ now leaving $v$ has had its U(1)$_1$ charge shifted by $-n_I^3$, and the segments which leave the extraordinary vertex $v^{\mathrm
{E}}$ are only charged in U(1)$_1$. The dotted segment $\delta
|E^a_2|$ shared by all $\beta_{2,I}%
^v$ is totally uncharged as a result of gauge invariance. The tri-valent vertices $w_I$ are such that all edge tangents are here parallel or antiparallel, and the 4-valent vertex $w'$ is such that there are two pairs of edges which are analytic extensions of each other.}
\label{Haction}
\end{center}
\end{figure}

The final segment is an arc connecting the ends of these two segments which is
tangent to the edges of $\gamma$ at its endpoints (this is a consequence of
the fact that the quantum shift direction determined by $E_{i}^{a}$ is
tangential to each edge at the arc position, and it ensures that the operator
$\hat{q}^{-1/4},$ and hence the Hamiltonian, acts trivially at these newly
created trivalent vertices). We postpone specifying further properties of
these arcs, as they are irrelevant, except that they do not create any
spurious new intersections, and that their areas satisfy a property spelt out
below. Let us denote the full loops by $\beta_{i,I}^{v}.$ By convention, they
are oriented such that the segment which overlaps $e_{I}$ is ingoing at the
vertex. Note that the segment $\delta E_{i}^{a}$ is shared by all $\beta
_{i,I}^{v}$ (as $I$ varies). Now consider the following classical approximant:%
\begin{equation}
V_{i}^{a}(F_{ab}^{j}E_{k}^{b}(\triangle_{v}^{I}))_{\delta}=N(v_{x}%
)\frac{(h_{\beta_{i,I}^{v}}^{j})^{n_{e_{I}}^{k}}-1}{\mathrm{i}\kappa n_{e_{I}%
}^{k}|\beta_{i,I}^{v}|}\frac{E_{k}(L_{I})}{|L_{I}|}q(v_{x})^{-1/4}%
\end{equation}
Here $(h_{\beta_{i,I}^{v}}^{j})^{n_{e_{I}}^{k}}$ is the U(1)$_{j}$ holonomy
around the loop $\beta_{i,I}^{v}$ in the $n_{e_{I}}^{k}$-representation, and
$|\beta_{i,I}^{v}|$ is the coordinate area of $\beta_{i,I}^{v},$ and $L_{I}$
is a flux surface transverse to $e_{I}$ of area $\left\vert L_{I}\right\vert
.$ This converges to $V_{i}^{a}F_{ab}^{j}E_{k}^{b}(v)$ classically as
$|\beta_{i,I}^{v}|,\left\vert L_{I}\right\vert \rightarrow0.$ Making this
replacement as an operator in $\hat{H}_{T(\gamma, \delta)}^{(1)}[N],$ we
obtain%
\begin{align}
2\hat{H}_{T(\gamma, \delta)}^{(1)}[N]|c\rangle &  =\sum_{v\in V(\gamma
)}N(v_{x})\lambda(\vec{n}_{v}^{c})\sum_{\triangle_{v}|v\in V(\gamma
)}\left\vert \triangle_{v}\right\vert \left(  \frac{(h_{\beta_{2,I}^{v}}%
^{1})^{n_{e_{I}}^{3}}-1}{\mathrm{i}\kappa n_{e_{I}}^{3}|\beta_{2,I}^{v}|}%
\frac{\hat{E}_{3}(L_{I})}{\left\vert L_{I}\right\vert }-\frac{(h_{\beta
_{3,I}^{v}}^{1})^{n_{e_{I}}^{2}}-1}{\mathrm{i}\kappa n_{e_{I}}^{2}|\beta
_{3,I}^{v}|}\frac{\hat{E}_{2}(L_{I})}{\left\vert L_{I}\right\vert }\right)
|c\rangle\nonumber\\
&  =\frac{\hbar}{\mathrm{i}}\sum_{v\in V(\gamma)}N(v_{x})\lambda(\vec{n}%
_{v}^{c})\sum_{I}\frac{\left\vert \triangle_{I}\right\vert }{\left\vert
L_{I}\right\vert }\left(  \frac{(h_{\beta_{2,I}^{v}}^{1})^{n_{e_{I}}^{3}}%
-1}{|\beta_{2,I}^{v}|}-\frac{(h_{\beta_{3,I}}^{1})^{n_{e_{I}}^{2}}-1}%
{|\beta_{3,I}^{v}|}\right)  |c\rangle
\end{align}
where the sum over $I$ extends over the valence of the vertex $v$ and we have
chosen flux surfaces $L_{I}$ such that $\epsilon(L_{I},e_{I})=+1.$ The charges
$n_{e_{I}}^{2},n_{e_{I}}^{3}$ are chosen to be those coloring the edge $e_{I}$
of $c.$ If either $n_{e_{I}}^{2},n_{e_{I}}^{3}$ is zero, then we choose the
holonomy to be in the fundamental representation. We have the freedom of
tuning the loop, flux, and plaquette areas so as to arrive at an overall
factor of $\delta^{-1}$:%
\begin{equation}
2\hat{H}_{T(\gamma, \delta)}^{(1)}[N]|c\rangle=\frac{\hbar}{\mathrm{i}}%
\sum_{v\in V(\gamma)}N(v_{x})\lambda(\vec{n}_{v}^{c})\frac{1}{\delta}\sum
_{I}\left(  \left(  (h_{\beta_{2,I}^{v}}^{1})^{n_{e_{I}}^{3}}-1\right)
-\left(  (h_{\beta_{3,I}^{v}}^{1})^{n_{e_{I}}^{2}}-1\right)  \right)
|c\rangle
\end{equation}
We may again pass to the product form (discarding terms which vanish
classically as $\delta\rightarrow0$)\footnote{The reader may wonder about the physical motivation in switching to the product form. Although the anomaly freedom of the continuum Hamiltonian constraint achieved in this paper is intricately tied to the structure of the operator and in particular to this product form, we believe that off-shell closure condition could be satisfied even without switching to the product form, if the key ideas developed here are followed closely. However the real reason to pass to the product form lies in keeping an eye on the $SU(2)$ theory, which is our main goal. In that case, only the product rule will ensure that the newly created vertex will be non-degenerate and second Hamiltonian constraint will have a non-trivial action on it.}

\begin{align}
\hat{H}_{T(\gamma, \delta)}^{(1)}[N]|c\rangle &  =\frac{\hbar}{2\mathrm{i}%
}\sum_{v\in V(\gamma)}N(v_{x})\lambda(\vec{n}_{v}^{c})\left(  \frac{%
{\textstyle\prod\nolimits_{I}}
(h_{\beta_{2,I}^{v}}^{1})^{n_{e_{I}}^{3}}-1}{\delta}-\frac{%
{\textstyle\prod\nolimits_{I}}
(h_{\beta_{3,I}^{v}}^{1})^{n_{e_{I}}^{2}}-1}{\delta}\right)  |c\rangle
\nonumber\\
&  =:\frac{\hbar}{2\mathrm{i}\delta}\sum_{v\in V(\gamma)}N(v_{x})\lambda
(\vec{n}_{v}^{c})\left(  |c_{1}\cup\alpha_{v}^{\delta}(\langle\hat{E}%
_{2}\rangle,n^{3}),c_{2},c_{3}\rangle-|c_{1}\cup\alpha_{v}^{\delta}%
(\langle\hat{E}_{3}\rangle,n^{2}),c_{2},c_{3}\rangle\right)  \label{dinky}%
\end{align}
We have introduced the following notation:%
\begin{equation}
|c\rangle\equiv|c_{1},c_{2},c_{3}\rangle,\qquad|c_{1}\cup\alpha_{v}^{\delta
}(\langle\hat{E}_{2}\rangle,n^{3}),c_{2},c_{3}\rangle:=%
{\textstyle\prod\nolimits_{I}}
(h_{\beta_{2,I}^{v}}^{1})^{n_{e_{I}}^{3}}|c\rangle,
\end{equation}
suggestive of the fact that only $c_{i=1}$ has been altered by the action of
$\hat{H}_{T(\gamma, \delta)}^{(i=1)},$ and this deformation is performed near
the vertex $v,$ is of `size' $\delta,$ and depends on the  (vector-valued) eigenvalue
$\langle\hat{E}_{2}\rangle$ (in the state $c$) as well as the values of the
$n^{i=3}$ charges of $c.$ If $c$ is not charged in the $i=2$ or $i=3$ factors
of U(1), then $\hat{H}_{T(\delta,\gamma)}^{(1)}$ annihilates $|c\rangle,$ with
similar statements for $\hat{H}_{T(\gamma, \delta)}^{(2)}$ and $\hat
{H}_{T(\gamma, \delta)}^{(3)}.$ To see this, note that if $c$ is not charged
in U(1)$_{2}$ for example, then $n_{e_{I}}^{2}=0,$ and hence $\beta_{2,I}^{v}$
collapses to a retraced curve (so that the corresponding holonomy equals 1),
and $\hat{E}_{2}$ annihilates $|c\rangle.$ The term $|c_{1}\cup\alpha
_{v}^{\delta}(\langle\hat{E}_{2}\rangle,n^{3}),c_{2},c_{3}\rangle$ produced by
$\hat{H}_{T(\gamma, \delta)}^{(1)}[N]$ is depicted in Figure (\ref{Haction}).

\begin{figure}
\begin{center}
\includegraphics[scale=0.8]{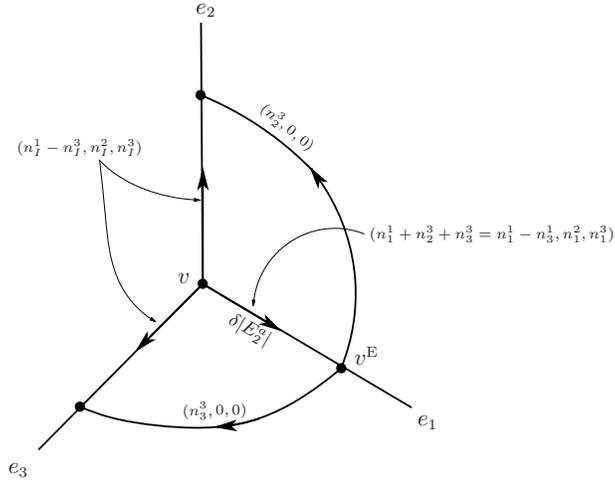}
\caption{The state $|c_1\cup\alpha^{\delta}_v(\langle\hat{E}_2\rangle
,n^3),c_2,c_3\rangle$ as produced by the action of $\hat{H}^{(1)}%
_{T(\gamma, \delta)}[N]$ in the special case where $v^{\mathrm{E}%
}$ lies on $\gamma$.}
\label{HactionB}
\end{center}
\end{figure}

In the case that the endpoint of $\delta E_{i}^{a}$ lies on $\gamma,$ the
construction proceeds just as above, however we observe that the quantum shift
actually points along some edge in this case, and hence the resulting state
has all edge tangents parallel or antiparallel at this point, as shown in
Figure (\ref{HactionB}).%

At each $N$-valent vertex $v$, $\hat{H}_{T(\gamma,\delta)}^{(i)}[N]$ acts by
attaching at most $N$ loops $\beta_{j,I}^{v}$ charged in U(1)$_{i}$ only with
charge $n_{e_{I}}^{k},$ the charge on the edge that $\beta_{j,I}^{v}$
partially overlaps. Our construction is such that at most two loops do not
intersect any other edges except the ones they overlap, and remaining loops
will have non-trivial intersection with the edges apart from the ones they
overlap (this trivial observation, unavoidable in two dimensions, will be
important later).%

Recall that all the attached loops have precisely one common segment which is
given by the straight line $\delta E_{i}^{a}$. Gauge invariance ensures that
this segment is (as part of the resulting state $|c_{1}\cup\alpha_{v}^{\delta
}(\langle\hat{E}_{2}\rangle,n^{3}),c_{2},c_{3}\rangle$ for instance)
uncharged, and thus its endpoint (the beginning point being $v$) will be an
$N_{v}\leq N$-valent vertex charged only in U(1)$_{i}$. Whence the action of
$\hat{H}_{T(\gamma, \delta)}^{(1)}[N]$ at a charge network vertex creates
precisely two charge network states, each of which have an additional $N_{v}%
$-valent vertex, and precisely $N_{v}$ additional tri-valent vertices. We will
refer to the $N_{v}$-valent vertex created by the action of $H_{T(\gamma,
\delta)}^{(1)}[N]$ an \emph{extraordinary vertex} $v^{\mathrm{E}}$. Note that
this extraordinary vertex can lie off the original graph or be in the interior
of one of the edges of the original graph depending on the quantum shift. By
construction, $v^{\mathrm{E}}$ lies in the interior of $U_{\epsilon}(\gamma,
v)$. In order to specify the action of the Hamiltonian constraint on arbitrary
charge networks we need a classification scheme given in the following section.

To summarize, the action of Hamiltonian constraint at finite triangulation
creates three kinds of vertices. The extraordinary vertices, whose location
depends on the quantum shift, a set of tri-valent vertices which by
construction are such that $\hat{q}^{-1/4}$ vanishes at such vertices, and the
four-valent vertices which have very specific charge configurations and
analyticity properties. As we will see later, these tri- and four-valent
vertices will play no role in proving the off-shell closure condition, and
hence we will refer to them as\emph{ irrelevant vertices.}

\subsection{Classification of Extraordinary Vertices}

As we saw above, the action of the Hamiltonian constraint $\hat{H}%
_{T(\gamma,\delta)}^{(i)}[N]$ on a charge network state $|c\rangle$ results in
the creation of what we called \emph{extraordinary} (EO) vertices. In this
section, we analyze the structure of these vertices in more detail. Our aim is
to argue that, given a charge-network state $|\bar{c}\rangle$ with its vertex
set $V(\bar{c}_{1}\cup\bar{c}_{2}\cup\bar{c}_{3})$, we can uniquely determine

\begin{enumerate}
\item[(a)] Which of the vertices are EO;

\item[(b)] If $v^{\mathrm{E}}$ is EO, then there exists a \emph{unique} charge
network state $|c\rangle$ such that the action of $\hat{H}_{T(\gamma,\delta
)}^{(i)}(v)$ on $|c\rangle$ (for a unique $v$ in $V(c_{1}\cup c_{2}\cup
c_{3})$) results in $|\bar{c}\rangle$ with $v^{\mathrm{E}}$ (where $\hat
{H}_{T(\gamma,\delta)}^{(i)}(v)$ is defined via $\hat{H}_{T(\gamma,\delta
)}^{(i)}[N]|c\rangle=\sum_{v\in V(\gamma(c))}N(v)\hat{H}_{T(\gamma,\delta
)}^{(i)}(v)|c\rangle$).
\end{enumerate}

We will first give a classification scheme, which help us isolate EO vertices
inside any charge network $\bar{c}$ unambiguously. We then show that the
removal of an EO vertex $v^{\mathrm{E}}$ along with all the edges incident on
it and appropriate shifts in charges on the remaining edges of the graph
results in $\vec{c}$ with a vertex $v$ such that the action of $\hat
{H}_{T(\gamma,\delta)}^{(i)}(v)$ for some $\delta$ and $i$ results in $\bar
{c}$.

As we saw earlier, if we act on a state $|c\rangle$ by $\hat{H}_{T(\gamma
,\delta)}^{(i)}[N]$, EO vertices are endpoints of a straight line arc
determined by quantum shift vectors. Generically these vertices will lie off
the graph $\gamma(c)$; however, there can be states in which the EO vertices
will lie on some edge which was already present in the original graph. We will
distinguish these two types of vertices as type A and type B vertices, respectively.

Given a charge-network $c$ with a vertex $v^{\mathrm{E}}$, we give a minimal
set of independent conditions which, if satisfied, determine that
$v^{\mathrm{E}}$ is an EO vertex. The conditions characterizing type A
vertices are given below. The set of conditions characterizing an EO vertex of
type B are given in Appendix (\ref{AtypeB}). We caution the reader that the
conditions as listed here are rather technical and not too illuminating. The
most efficient way to understand them is to consult Figure (\ref{Haction}) simultaneously.

Let $\bar{c}=:(\bar{c}_{1},\bar{c}_{2},\bar{c}_{3})$ be a charge network with
$\gamma(\bar{c})$ the coarsest graph associated underlying it. Let
$v^{\mathrm{E}}$ be a vertex of $\gamma(\bar{c})$. We will call $v^{\mathrm{E}%
}$ an EO vertex of type $(\mathrm{A},M\in\{\mathrm{12,3}\},j\in\{1,2,3\})$
or type $(\mathrm{B},M\in\{\mathrm{1,2,3}\},j\in\{1,2,3\})$ \emph{iff} it
satisfies the set of conditions A or B, respectively.

\textit{Remark on notation}: Sometimes we will indicate the type of EO
vertices only by omitting one or two of the labels. E.g., when the analysis
only depends on the fact that the EO vertex is type $(M=\mathrm{I},j=2),$ we
will omit the label $\mathrm{A/B}$.

\textbf{Set A}:

\begin{enumerate}
\item[(1)] All the edges beginning at $v^{\mathrm{E}}$ are charged in the
$M^{\text{th}}$ copy.

\item[(2)] If the valence of $v^{\mathrm{E}}$ is $N_{v}$,\footnote{The
subscript $v$ in $N_{v}$ may seem out of place, however when we list down all
the conditions in A, its relevance will become clear.} then we will denote the
$N_{v}$ vertices which are the end points of the $N_{v}$ edges beginning at
$v^{\mathrm{E}}$ by the set $\mathcal{S}_{v^{\mathrm{E}}}:=\{v_{(1)}%
^{\mathrm{E}},\dots,v_{(N_{v})}^{\mathrm{E}}\}$. The valence of all these
vertices is bounded between 3 and 4.

\begin{enumerate}
\item[(2a)] At most two vertices in $\mathcal{S}_{v^{\mathrm{E}}}$ are tri-valent.
\end{enumerate}

\item[(3)] The tri-valent vertices are such that the edges which are not
incident on $v^{\mathrm{E}}$ are analytic extensions of each other and the
four-valent vertices are such that two of the edges which are not incident on
$v^{\mathrm{E}}$ are analytic extensions of each other, and fourth edge is the
analytic extension of the edge which is incident at $v^{\mathrm{E}}$.

\begin{enumerate}
\item[(3a)] Any four-valent vertex defined in {(3)} is such that, if the four
edges $(e_{1},e_{2},e_{3},e_{4})$ incident on it are such that $e_{1}\circ
e_{2}$ is entire-analytic and $e_{3}\circ e_{4}$ is entire-analytic then
$\vec{n}_{e_{1}}=\vec{n}_{e_{2}}$, $\vec{n}_{e_{3}}=\vec{n}_{e_{4}}$.
\end{enumerate}

\item[(4)] Let $e_{v^{\mathrm{E}}}$ be an edge beginning at $v^{\mathrm{E}}$
which ends in a four-valent vertex $f(e_{v^{\mathrm{E}}})$. By {(3)}, there
exists an edge $e_{v^{\mathrm{E}}}^{\prime}$ beginning at $f(e_{v^{\mathrm{E}%
}})$ such that $e_{v^{\mathrm{E}}}\circ e_{v^{\mathrm{E}}}^{\prime}=:\tilde
{e}_{v^{\mathrm{E}}}$ is the analytic extension of $e_{v^{\mathrm{E}}}$ in
$E(\bar{c})$ beginning at $v^{\mathrm{E}}$. The final vertex $f(\tilde
{e}_{v^{\mathrm{E}}})$ of $\tilde{e}_{v^{\mathrm{E}}}$ is always tri-valent.
Thus, restricting attention to analytic extensions of each of the edges
beginning at $v^{\mathrm{E}}$, all such edges end in tri-valent vertices, and
all of these tri-valent vertices are such that the remaining two edges
incident on them are analytic extensions of each other (see Figure
\ref{Haction}). The set of these $N_{v}$ three-valent vertices
\textquotedblleft associated" to $v^{\mathrm{E}}$ is $\mathcal{\bar{S}%
}_{v^{\mathrm{E}}}:=\{\bar{v}_{1}^{\mathrm{E}},...,\bar{v}_{N_{v}}%
^{\mathrm{E}}\}$.\footnote{Note that $\mathcal{S}_{v^{\mathrm{E}}}%
\cap\ \mathcal{\bar{S}}_{v^{\mathrm{E}}}\ =\ \text{3-valent vertices in
}\mathcal{S}_{v^{\mathrm{E}}}$.}

\begin{enumerate}
\item[(4a)] All three edges incident on any element in $\mathcal{\bar{S}%
}_{v^{\mathrm{E}}}$ have parallel (or anti-parallel) tangents.
\end{enumerate}

\item[(5)] Let us denote these (maximally analytic inside $E(\gamma)$) edges
beginning at $v^{\mathrm{E}}$ by $\{\tilde{e}_{v^{\mathrm{E}}}^{(1)}%
,...\tilde{e}_{v^{\mathrm{E}}}^{N_{v}}\}$.
Without loss of generality, consider the case when all the edges incident at
$v^{\mathrm{E}}$ are charged in first copy.\footnote{In this case we will say
that $v^{\mathrm{E}}$ is of type $(\mathrm{A},M=\mathrm{I},j\ \in\ \{2,3\})$.}
Let the charges on these edges be $\{(n_{\tilde{e}_{v^{\mathrm{E}}}^{(1)}%
},0,0),...,(n_{\tilde{e}_{v^{\mathrm{E}}}^{(N_{v})}},0,0)\}$.\newline If
$\tilde{e}_{v^{\mathrm{E}}}^{(k)}$ ($k\in\{1,...,N_{v}\}$) ends in a
tri-valent vertex $f(\tilde{e}_{v^{\mathrm{E}}}^{(k)})$ and if the charges on
the remaining two (analytically related) edges $e_{v^{\mathrm{E}}}^{(k)\prime
},e_{v^{\mathrm{E}}}^{(k)\prime\prime}$ incident on $f(\tilde{e}%
_{v^{\mathrm{E}}}^{(k)})$ are $(n_{e_{v^{\mathrm{E}}}^{(k)\prime}}%
^{(1)},n_{e_{v^{\mathrm{E}}}^{(k)\prime}}^{(2)},n_{e_{v^{\mathrm{E}}%
}^{(k)\prime}}^{(3)})$ and $(n_{e_{v^{\mathrm{E}}}^{(k)\prime\prime}}%
^{(1)},n_{e_{v^{\mathrm{E}}}^{(k)\prime\prime}}^{(2)}=n_{e_{v^{\mathrm{E}}%
}^{(k)\prime}}^{(2)},n_{e_{v^{\mathrm{E}}}^{(k)\prime\prime}}^{(3)}%
=n_{e_{v^{\mathrm{E}}}^{(k)\prime}}^{(3)}$, then either

\begin{enumerate}
\item[(a)] $n_{\tilde{e}_{v^{\mathrm{E}}}^{(k)}}^{(1)}=n_{e_{v^{\mathrm{E}}%
}^{(k)\prime}}^{(2)}$, or

\item[(b)] $n_{\tilde{e}_{v^{\mathrm{E}}}^{(k)}}^{(1)}=n_{e_{v^{\mathrm{E}}%
}^{(k)\prime}}^{(3)}$.
\end{enumerate}

\item[(6)] Now consider the set $\mathcal{\bar{S}}_{v^{\mathrm{E}}}$. Recall
that each element in this set is a tri-valent vertex. Consider the vertex
$f(\tilde{e}_{v^{\mathrm{E}}})$ whose three incident edges are $\tilde
{e}_{v^{\mathrm{E}}},e_{v^{\mathrm{E}}}^{\prime},\ \text{and}%
\ e_{v^{\mathrm{E}}}^{\prime\prime}$. Recall that $\tilde{e}_{v^{\mathrm{E}}%
}^{\prime},\tilde{e}_{v^{\mathrm{E}}}^{\prime\prime}$ are analytic
continuations of each other. Depending on whether $n_{\tilde{e}_{v^{\mathrm{E}%
}}}\lessgtr0$, choose one out of the two edges, $\tilde{e}_{v^{\mathrm{E}}%
}^{\prime},\tilde{e}_{v^{\mathrm{E}}}^{\prime\prime}$ which has lesser or
greater charge in the first copy than the other edge. Consider the set of all
such chosen edges for each vertex in $\mathcal{\bar{S}}_{v^{\mathrm{E}}}$. We
refer to this set as $\mathcal{T}_{v^{\mathrm{E}}}$.

\begin{enumerate}
\item[(6a)] If all these edges meet in a vertex $v$ which is such that, if the
number of edges incident on $v$ is greater then $N_{v}$ and \emph{if the
charges $\{\tilde{e}_{v^{\mathrm{E}}}^{(k)}\}_{k=1,...,N_{v}}$ are the
$\mathrm{U}(1)_{j}$ charges on the edges in $\mathcal{T}_{v^{\mathrm{E}}},$
then the $\mathrm{U}(1)_{j}$ charge on edges incident at $v$ which are not in
$\mathcal{T}_{v^{\mathrm{E}}}$ is zero.} As shown in Appendix \ref{A3}, $v,$
if it exists, is unique.
\end{enumerate}

\item[(7)] Finally consider the graph $\gamma:=\gamma(\bar{c})-\{\tilde
{e}_{v^{\mathrm{E}}}^{(1)},...,\tilde{e}_{v^{\mathrm{E}}}^{(N_{v})}\}$ and a
charge-network $c$ based on $\gamma$ obtained by deleting $\{\tilde
{e}_{v^{\mathrm{E}}}^{(1)},...,\tilde{e}_{v^{\mathrm{E}}}^{(N_{v})}\}$ along
with the charges on them, and also deleting exactly the same amount of charge
from the edges in $\mathcal{T}_{v^{\mathrm{E}}}$. Note that by construction,
$v$ belongs to $\gamma$. Now consider $U_{\epsilon}(\gamma,v)$. The final and
key feature of an EO vertex $v^{\mathrm{E}}$ is,
\end{enumerate}

$\bigskip$

$v^{\mathrm{E}}\in U_{\epsilon}(\gamma,v)$ and $v^{\mathrm{E}}$ is the
endpoint of the \textquotedblleft straight line curve\textquotedblright%
\ $\delta\langle\hat{E}_{j}^{a}\rangle_{c}$ for some $\delta$, where $j=2$ if
in ({6}) condition (a) is satisfied, and $j=3$ if in ({5}), (b) is satisfied.

\bigskip

It is easy to see that the conditions listed above are independent of each
other, as one could easily conceive of a charge network state which satisfies
all but one of the conditions. If all the conditions given in \textbf{Set A}
above, or \textbf{Set B} in the Appendix, are satisfied, then we call the pair
$(v,v^{\mathrm{E}})$ \emph{extraordinary}. For the benefit of the reader we
emphasize once again that the type of extraordinariness of $v^{\mathrm{E}}$ is
labeled by the triple $(\mathrm{A/B},M\in\{\mathrm{1,2,3}\},j\in\{{1,2,3}\})$.
For example, $M=\mathrm{1}$ when all the edges incident at $v^{\mathrm{E}}$
are only charged under $\mathrm{U}(1)_{1}$ and $j\in\{2,3\}$ if these charges
equal the charges in $\mathrm{U}(1)_{j}$ on edges in $\mathcal{T}%
_{v^{\mathrm{E}}}$.\newline

We now prove a lemma which shows that EO vertices are always associated to the
action of some $\hat{H}_{T(\gamma,\delta)}^{(i)}[N]$. This will imply that any
charge network which has an EO vertex is always in the image of $\hat
{H}_{T(\gamma,\delta)}^{(i)}[N]$ for some $i\ \text{and}\ N$.

\textbf{Claim}: Let $\cup_{K\in\{\mathrm{A,B}\}}\cup_{j\in\{2,3\}}\{\bar
{c}_{K}^{j}\}=\mathcal{C}$ be the set of charge network states such that
$(v,v^{\mathrm{E}})$ is an EO pair for each charge network in this set and
$v^{\mathrm{E}}$ is an EO vertex of type $(\mathrm{1},K,j)$.\footnote{We are
restricting attention to the $M=\mathrm{I}$ case in this lemma. The proof is
exactly analogous for $M\ \in\ \{\mathrm{2,3}\}$.} Let $c$ be a charge network
obtained by performing the surgery described in condition {(7)}
above.\footnote{It is easy to see that under this surgery, one ends up with
the same charge network $c$ no matter which $\bar{c}_{K}^{j}\ \in
\ \mathcal{C}$ one starts with.} Also let $N$ be a lapse function such that it
has support in a neighborhood of $v$ (which, as we saw above, belongs to both
$c$ and $\bar{c}_{K}^{j}\forall K,j$. If the vertex $v$ is non-degenerate
(i.e. $\langle\hat{q}^{-1/4}\rangle_{c}\neq0$), and if
\begin{equation}
\hat{H}_{T(\gamma,\delta)}^{(1)}[N]|c\rangle=\frac{1}{\delta}N(v)\langle
\hat{q}^{-1/4}\rangle_{c}(v)\left[  \alpha|c^{\prime}\rangle+\beta
|c^{\prime\prime}\rangle\right]
\end{equation}
where $\alpha,\beta\ \in\{\pm1\}$, then

\begin{enumerate}
\item[(a)] Both $c^{\prime},c^{\prime\prime}$ belong to the set $\cup
_{K,j}\{\bar{c}_{K}^{j}\}$.

\item[(b)] Conversely, given any $\bar{c}$ that is obtained from $c$
(containing a non-degenerate vertex $v$ which is not EO by adding an EO vertex
$v^{\mathrm{E}}$ of type $(\mathrm{1},K,j)$ for some $K,j$, then $\bar{c}$ is
always one of the two charge networks one gets by letting $\hat{H}%
_{T(\gamma,\delta)}^{(1)}(v)$ act on $|c\rangle$ for some $\delta$.
\end{enumerate}

\textbf{Proof}: {(a)} follows by construction. That is, it is straightforward
to verify that both $c^{\prime}$ and $c^{\prime\prime}$ satisfy all the
conditions listed in \textbf{Set A} or \textbf{Set B}.

For {(b)}, consider a charge network $\bar{c}$ obtained by adding an EO vertex
$v^{\mathrm{E}}$ of type $(\mathrm{1},K=\mathrm{A},j=2)$ to $c$ such that
$(v,v^{\mathrm{E}})$ is the EO pair (other types of EO vertices can be treated
similarly). Let the $N_{v}$-valent segments beginning at $v^{\mathrm{E}}$ and
terminating at the $N_{v}$ tri-valent vertices $\{v_{e_{1}},\dots,v_{e_{N_{v}%
}}\}$ be denoted by $\{s_{e_{1}},...,s_{e_{N_{v}}}\}$. As the vertex is of
type 1, all of these segments are charged in $\mathrm{U}(1)_{1}$. Let the
vertex $v^{\mathrm{E}}$ be along the \textquotedblleft straight
line\textquotedblright\ $\delta_{0}\langle\hat{E}_{2}^{a}(v)\rangle_{c}$ (in
the coordinate system that we have fixed once and for all).

Now consider the Hamiltonian constraint operator $\hat{H}_{T(\delta_{0}%
)}^{(i)}(v)$ at a given vertex $v\in V(c)$ is constructed out of products of
holonomies around loops. Each loop is constructed out of a segment along an
edge of $\gamma(c)$, the straight-line arc given by the quantum shift, and an
arc which joins $v^{\mathrm{E}}$ with one of the vertices in $\mathcal{S}%
_{v^{\mathrm{E}}}$. We need $N_{v}$ such arcs and upon choosing them to be
$(s_{e_{1}},...,s_{e_{N_{v}}})$ respectively,\footnote{This can always be done
as there is enough freedom in choosing the loops underlying the holonomies out
of which the Hamiltonian constraint is built.} $\hat{H}_{T(\delta_{0})}%
^{(i)}(v)|c\rangle$ will result in a linear combination of states, one of
which will clearly be $\bar{c}$. This completes the proof.

\subsubsection{Weakly Extraordinary Vertices}

One highly unpleasant feature of EO vertices is their background dependence.
As we require such vertices to lie in the coordinate neighborhood
$U_{\epsilon}(v,\gamma)$ of $v$, the property that we termed extraordinariness
is not a diffeomorphism-invariant notion. That is, if $v^{\mathrm{E}}$ is EO
with respect to $v\in V(c),$ then it does not imply that $\phi\cdot
v^{\mathrm{E}}$ is EO with respect to $\phi\cdot v\in V(\phi\cdot c)$. With
this drawback in mind, we introduce a generalization of extraordinary vertices
in this section. As we will see later, this generalization will play an
important role when we construct a habitat.

Let $\bar{c}$ be a charge network with an EO pair $(v,v^{\mathrm{E}}),$ where
$v^{\mathrm{E}}$ is an EO vertex of type $M=1$, say. Let $\phi$ be a
semi-analytic diffeomorphism of $\Sigma$ such that $\phi\cdot c=c$ and
consider the state $\phi\cdot\bar{c}$ such that $(\phi\cdot v,\phi\cdot
v^{\mathrm{E}})$ is not an EO pair. In this case we refer to the image of
$v^{\mathrm{E}}$ in this state as a \emph{weakly extraordinary vertex}. Notice
that as $\phi$ keeps $c$ invariant, $\phi\cdot\bar{c}$ has the same
\textquotedblleft topological structure\textquotedblright\ as $\bar{c}$. In
particular, such diffeomorphisms cannot change $N_{v}$ (defined in the
previous section). This implies the following:

\emph{Vertices which have all the properties stated above except property
{(7)} in \textbf{Set A} or \textbf{Set B} are weakly extraordinary
vertices.}\newline

We would like to emphasize that the real motivation behind introducing weakly
EO vertices will become clear in \cite{hat2} where we will analyze the issue
of diffeomorphism covariance of the Hamiltonian constraint.


\subsection{Action of a \textquotedblleft Second\textquotedblright%
\ Hamiltonian}

\subsubsection{Hints from the Classical Theory}

As observed in \cite{lm2} and explained in the introduction, one of the
reasons Thiemann's quantum Hamiltonian constraint can never produce a
non-trivial commutator (even if one worked with higher density constraints) is
due to the fact that it has trivial action on the vertices that it creates. At
first sight, it seems like we have run into the same problem. As we saw above,
the EO vertices created by $\hat{H}_{T(\delta)}^{(i)}[N]$ are degenerate,
whence the action of the Hamiltonian constraint on a state containing an EO
pair will act trivially at the EO vertex. It then seems plausible that an
analysis similar to the one done in \cite{lm2} would lead to a trivial
continuum commutator. However, following a simple observation in the classical
theory tells us how this triviality could be circumvented. The computation
done in Section \ref{classical} (which motivated our quantization choices in
the construction of $\hat{H}_{T(\delta)}^{(i)}[N]$) demonstrated how the
(classical) action of Hamiltonian constraint could be understood in terms of
spatial diffeomorphisms generated by triad fields. Thus the Poisson action of
two successive Hamiltonian constraints involves terms which would in turn act
on these triad fields. More precisely,
the triad field $E_{i}$ has a non-vanishing Poisson bracket with $H^{(i)}[N]$
and is given by (as usual all classical computation are done with density two
Hamiltonian constraint),
\begin{align}
\{H^{(i)}[N],E_{i}^{a}(x)\}  &  =-\epsilon^{ijk}E_{j}^{a}\partial_{a}\left(
NE_{k}^{b}\right)  (x)\label{eq:hamactontriad}\\
&  \approx-\epsilon^{ijk}\left(  E_{j}^{a}E_{k}^{b}\partial_{a}N\right)
(x)+\left(  NE_{j}^{a}E_{k}^{b}\partial_{a}E_{k}^{b}\right)  (x)\nonumber
\end{align}
where we have used the Gau\ss \ constraint $\partial_{a}E_{i}^{a}=0$. As the
Hamiltonian vector field action of $H[N]$ is approximated by a transformation
involving a triad-dependent diffeomorphism, as in (\ref{HAMVECK4}), we expect
the second Hamiltonian constraint to act non-trivially on an EO vertex via its
action on the generator of this diffeomorphism. In essence, this is the action
that is captured by the extra term in the Hamiltonian constraint when it acts
on EO vertices. More precisely, the extra term in $\hat{H}_{T(\delta)}%
^{(i)}[N]$ will induce an action on the EO vertex which will mirror the first
term in (\ref{eq:hamactontriad}) only.\footnote{This information suffices to
obtain an anomaly free commutator of the Hamiltonian constraints as will be
shown in Section \ref{comma}.}\footnote{It is possible to find a discrete
approximant to $X_{H[M]}X_{H[N]}f_{c}(A)$ which illustrates this point rather
clearly. Such a computation will produce terms involving $f_{\phi_{\vec{V}%
}^{\delta}\circ c}(A)$ where $\vec{V}$ is a triad dependent vector field of
the type given in (\ref{eq:hamactontriad}). However, this computation is
rather involved and as our primary motivation for considering such classical
computations is merely as guiding tools to make quantization choices, we do
not reproduce it here.}

\subsubsection{The Action of $\hat{H}_{T(\delta)}[N]$ on EO Pairs}

\label{comma}

Based on the classical insight of the previous section, we modify the
definition of $\hat{H}_{T(\delta)}^{(i)}[N]$ such that on any $|c\rangle$ not
containing an EO pair, it is still given by (\ref{dinky}). However, if
$|c\rangle$ contains an EO pair, then $\hat{H}_{T(\delta)}^{(i)}[N]$ contains
an additional term constructed to mimic (\ref{eq:hamactontriad}) on the
quantum shift. This term utilizes a dichotomy present between the classical
theory and loop quantized quantum field theories, which arises due to the
underlying representation of the holonomy-flux algebra.

Consider an edge $e$ and a transversal (co-dimension one) surface
$L_{e}^{\delta}$ which intersects $e$ in some interior point and whose
coordinate area scales with $\delta$. Classically, a quadratic functions of
fluxes like $E_{i}(L_{e}^{\delta})E_{j}(L_{e}^{\delta})$ is higher order in
$\delta$ than $E_{i}(L_{e}^{\delta})$, but in the quantum theory,
\begin{equation}
\frac{1}{\hbar}\hat{E}^{i}(L_{e}^{\delta})h_{e}^{\vec{n}_{e}}(A)=n_{e}%
^{i}h_{e}^{\vec{n}_{e}}(A),\qquad\frac{1}{\hbar^{2}}\hat{E}_{i}(L_{e}^{\delta
})\hat{E}_{j}(L_{e}^{\delta})h_{e}^{\vec{n}_{e}}(A)=n_{e}^{i}n_{e}^{j}%
h_{e}^{\vec{n}_{e}}(A).
\end{equation}
Thus owing to the peculiarity of the holonomy-flux representation, spectra of
flux operators do not carry the memory of coordinate area of the underlying
surfaces. We interpret this dichotomy as a quantization ambiguity, and it is
this ambiguity which we will use to modify $\hat{H}_{T(\delta)}^{(i)}[N]$.

In order to explain the most important non-triviality of the modification, we
will first work with density two constraints first.\footnote{Density two
constraints, when quantized, should have at finite triangulation an overall
factor of ($\delta\epsilon)^{-1}$, where $\epsilon$ comes from the
regularization of quantum shift. This $\epsilon$ is removed when one switches
to density $\frac{5}{4}$ constraint as quantization of $q^{-1/4}$ involves an
overall factor of $\epsilon$. Whence we will suppress the factor of $\frac
{1}{\epsilon}$ in the density two case, as it is not relevant in the final
result.} Finally we will switch to the density $\frac{5}{4}$ constraint by
choosing a particular operator ordering when including $\hat{q}^{-1/4}$.
\emph{This choice of operator ordering will lead to an anomaly-free quantum
Dirac algebra as we will see in the later sections.}

Let us first compute $\hat{H}_{T(\delta^{\prime})}^{(1)}[M]\hat{H}_{T(\delta
)}^{(2)}[N]|c_{1},c_{2},c_{3}\rangle$, with the constraint operators given in
(\ref{dinky}). Let $v\in V(c_{1}\cup c_{2}\cup c_{3})$ be the only vertex
which lies inside the support of $N$ and $M$. Then, suppressing all the
factors of $\hbar,$
\begin{align}
\hat{H}_{T(\delta^{\prime})}^{(1)}[M]\hat{H}_{T(\delta)}^{(2)}[N]|c_{1}%
,c_{2},c_{3}\rangle &  =\frac{1}{\delta}N(v)\hat{H}_{T(\delta^{\prime})}%
^{(1)}[M]\left[  |c_{1},c_{2}\cup\alpha_{v}^{\delta}(\langle\hat{E}_{3}%
\rangle_{c_{3}},n_{c_{1}}),c_{3}\rangle-|c_{1},c_{2}\cup\alpha_{v}^{\delta
}(\langle\hat{E}_{1}\rangle_{c_{1}},n_{c_{3}}),c_{3}\rangle\right] \nonumber\\
&  =\frac{1}{\delta\delta^{\prime}}N(v)M(v)\left[  \left(  |c_{1}\cup
\alpha_{v}^{\delta^{\prime}}(\langle\hat{E}_{2}\rangle_{c_{2}\cup\alpha
_{v}^{\delta}(\langle\hat{E}_{3}\rangle)},n_{c_{3}}),c_{2}\cup\alpha
_{v}^{\delta}(\langle\hat{E}_{3}\rangle_{c_{3}},n_{c_{1}}),c_{3}%
\rangle\right.  \right. \nonumber\\
&  \qquad\qquad\qquad\qquad\qquad-\left.  |c_{1}\cup\alpha_{v}^{\delta
^{\prime}}(\langle\hat{E}_{3}\rangle_{c_{3}},n_{c_{2}\cup\alpha_{v}^{\delta
}(\langle\hat{E}_{3}\rangle)}),c_{2}\cup\alpha_{v}^{\delta}(\langle\hat{E}%
_{3}\rangle_{c_{3}},n_{c_{1}}),c_{3}\rangle\right) \nonumber\\
&  \qquad\qquad\qquad\qquad-\left(  |c_{1}\cup\alpha_{v}^{\delta^{\prime}%
}(\langle\hat{E}_{2}\rangle_{c_{2}\cup\alpha_{v}^{\delta}(\langle\hat{E}%
_{3}\rangle)},n_{c_{3}}),c_{2}\cup\alpha_{v}^{\delta}(\langle\hat{E}%
_{1}\rangle_{c_{1}},n_{c_{3}}),c_{3}\rangle\right. \nonumber\\
&  \qquad\qquad\qquad\qquad\qquad-\left.  \left.  |c_{1}\cup\alpha_{v}%
^{\delta^{\prime}}(\langle\hat{E}_{3}\rangle_{c_{3}},n_{c_{2}\cup\alpha
_{v}^{\delta}(\langle\hat{E}_{3}\rangle)}),c_{2}\cup\alpha_{v}^{\delta
}(\langle\hat{E}_{1}\rangle_{c_{1}},n_{c_{3}}),c_{3}\rangle\right)  \right]
\end{align}


As before given a $(c_{1},c_{2},c_{3})$ let $\gamma(c)\equiv\gamma(c_{1}\cup
c_{2}\cup c_{3})$ and $e\in\ E(\gamma(c))$. Given a point $v^{\prime}%
\in\mathrm{Int}(e)$, let $L_{e}^{v^{\prime}}(\delta)$ be a surface of
co-dimension one (so $L_{e}^{v^{\prime}}$ is just a segment which intersects
$e$ transversally) whose coordinate length is of the order $\delta^{\prime
}=O(\delta^{2})$.\footnote{As we have a length scale in the theory
$\kappa\hbar$, one could use it to define $\delta^{\prime}\ =\ \frac
{\delta^{2}}{\kappa\hbar}$.} Consider a state $|c_{1},c_{2}^{\prime}%
,c_{3}\rangle$ which has an EO pair $(v,v^{\mathrm{E}})$ with $v^{\mathrm{E}}$
an EO vertex of type $(M=2,j=1)$. This state is of the type
\begin{equation}
|c_{1},c_{2}^{\prime},c_{3}\rangle=|c_{1},c_{2}\cup\alpha_{v}^{\delta_{0}%
}(\langle\hat{E}_{1}\rangle_{c_{1}},n_{c_{3}}),c_{3}\rangle
\end{equation}
for some fixed $\delta_{0}$. Our proposal for the action of $\hat{H}%
_{T(\delta)}^{(1)}[M]$ on $|c_{1},c_{2}\cup\alpha_{v}(\langle\hat{E}%
_{1}\rangle_{c_{1}},n_{c_{3}}),c_{3}\rangle$ is as follows (As $\delta_{0}$ is
fixed, we suppress it for the clarity of presentation):
\begin{align}
&  \hat{H}_{T(\delta)}^{(1)}[M]|c_{1},c_{2}\cup\alpha_{v}(\langle\hat{E}%
_{1}\rangle_{c_{1}},n_{c_{3}}),c_{3}\rangle\nonumber\\
&  =\sum_{v\in V(c_{1}\cup c_{2}\cup c_{3})}M(v)\hat{H}_{T(\delta)}%
^{(1)}(v)|c_{1},c_{2}\cup\alpha_{v}(\langle\hat{E}_{1}\rangle_{c_{1}}%
,n_{c_{3}}),c_{3}\rangle+\nonumber\\
&  \delta_{Supp(M),v^{\textrm{E}}}\ \epsilon\left(  \frac{1}{\delta}\sum_{e\in E(\gamma)}\langle\hat{E}%
_{2}(L_{e}^{v^{\prime}}(\delta^{\prime}))\rangle_{c_{2}}\left(  M(v+\delta
\dot{e}(0))-M(v)\right)  |c_{1},c_{2}\cup\alpha_{v}(\langle\hat{E}_{3}%
\rangle_{c_{3}},n_{c_{3}}),c_{3}\rangle\right. \nonumber\\
&  \qquad-\left.  \frac{1}{\delta}\sum_{e\in E(\gamma)}\langle\hat{E}%
_{3}(L_{e}^{v^{\prime}}(\delta^{\prime}))\rangle_{c_{2}}\left(  M(v+\delta
\dot{e}(0))-M(v)\right)  |c_{1},c_{2}\cup\alpha_{v}(\langle\hat{E}_{2}%
\rangle_{c_{2}},n_{c_{3}}),c_{3}\rangle\right)  \label{proposal1}%
\end{align}
where,%
\begin{figure}
\begin{center}
\includegraphics[scale=0.8]{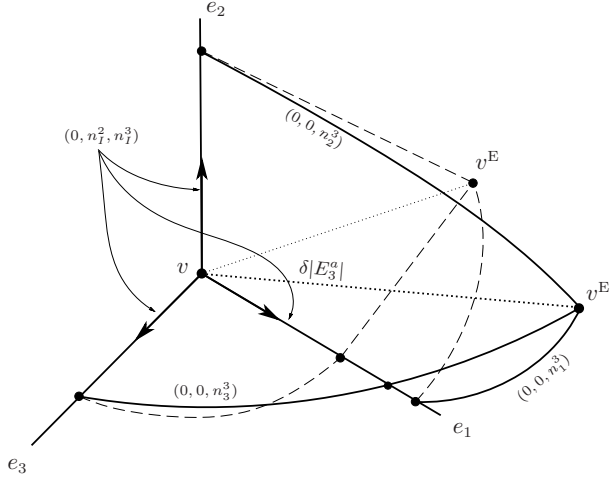}
\caption{One of the extra terms resulting from the modified action of $H^{(1)}%
_{T(\delta)}$ on $|c_1,c_2\cup\alpha_v(\langle\hat{E}_1\rangle,n^3),c_3\rangle
$. The dashed lines show the original position of $v^{\mathrm{E}%
}$. In this term, the $3^{\mathrm{rd}}$ quantum shift determines the $\beta
$ loops which are charged also in U(1)$_3$.}
\label{2Haction}
\end{center}
\end{figure}%

\begin{enumerate}
\item[(a)] The first term $\sum_{v\in V(c_{1}\cup c_{2}\cup c_{3})}M(v)\hat
{H}_{T(\delta)}^{(1)}(v)|c_{1},c_{2}\cup\alpha_{v}(\langle\hat{E}_{1}%
\rangle_{c_{1}},n_{c_{3}}),c_{3}\rangle$ is the unmodified action coming from
(\ref{dinky}).

\item[(b)] The second term is the proposed modification which is designed to
capture the displacement of EO vertex as motivated from
(\ref{eq:hamactontriad}).

\item[(c)] $\epsilon$ is a numerical coefficient which will choose to be $1$
and as we will see later, with this value, the off-shell closure condition is satisfied.

\item[(d)] $\delta_{Supp(M),v^{\textrm{E}}} = 1$ if $v^{\textrm{E}}$ lies inside the support of M, and is zero otherwise.

\end{enumerate}

But even as far as describing our proposal goes, we are not done yet. We need
to show that

\begin{enumerate}
\item[(1)] There exists an operator which when acting on $(c_{1},c_{2},c_{3})$
accomplishes (\ref{proposal1}), and

\item[(2)] The \textquotedblleft naive\textquotedblright\ classical limit of
this operator should give back the classical Hamiltonian.
\end{enumerate}

We proceed by defining an operator which yields (\ref{proposal1}) and then
argue that it differs from the unmodified operator by terms subleading in
$\delta,$ thus showing that it has the right \textquotedblleft
naive\textquotedblright\ classical limit.%

Once again we assume that $v^{\textrm{E}}$ is inside the support of the lapse $M$. If it lies outside the support then second and third terms are absent by definition.
\begin{align}
&  \hat{H}_{T(\delta)}^{(1)}[M]|c_{1},c_{2}\cup\alpha_{v}(\langle\hat{E}%
_{1}\rangle_{c_{1}},n_{c_{3}}),c_{3}\rangle\label{eq:proposal2}\\
&  =\frac{1}{\delta}M(v)\left[  (\hat{h}_{\alpha_{v}^{\delta}(\langle\hat
{E}_{2}\rangle_{c_{2}\cup\alpha_{v}(\langle\hat{E}_{1}\rangle_{c_{1}}%
,n_{c_{3}})},n_{c_{3}})}^{(1)})-(\hat{h}_{\alpha_{v}^{\delta}(\langle\hat
{E}_{3}\rangle_{c_{3}},n_{c_{2}}-n_{c_{3}})}^{(1)})\right]  |c_{1},c_{2}%
\cup\alpha_{v}(\langle\hat{E}_{1}\rangle_{c_{1}},n_{c_{3}}),c_{3}%
\rangle\nonumber\\
&  +\frac{1}{\kappa\hbar\delta}\left(  \sum_{e\in E(\gamma)}\langle\hat{E}%
_{2}(L_{e}^{v^{\prime}}(\delta^{\prime}))\rangle_{c_{2}}(M(v+\delta\dot
{e}(0))-M(v))(\hat{h}_{\alpha_{v}(\langle\hat{E}_{3}\rangle_{c_{3}},n_{c_{3}%
})}^{(2)}(\hat{h}_{\alpha_{v}(\langle\hat{E}_{1}\rangle_{c_{1}},n_{c_{3}}%
)}^{(2)})^{-1})\right. \nonumber\\
&  -\left.  \sum_{e\in E(\gamma)}\langle\hat{E}_{3}(L_{e}^{v^{\prime}}%
(\delta^{\prime}))\rangle_{c_{3}}(M(v+\delta\dot{e}(0))-M(v))(\hat{h}%
_{\alpha_{v}(\langle\hat{E}_{2}\rangle_{c_{2}},n_{c_{3}})}^{(2)}(\hat
{h}_{\alpha_{v}(\langle\hat{E}_{1}\rangle_{c_{1}},n_{c_{3}})}^{(2)}%
)^{-1})\right)  |c_{1},c_{2}\cup\alpha_{v}(\langle\hat{E}_{1}\rangle_{c_{1}%
},n_{c_{3}}),c_{3}\rangle\nonumber
\end{align}

The extra terms in the second and third line in (\ref{eq:proposal2}) are
subleading in $\delta$ as compared to the first (unmodified) term. This can be
seen as follows. The unmodified operator $\hat{H}_{T(\delta)}^{(1)}(v)$ is an
operator of the form $\frac{1}{\delta}\left[  \hat{h}_{\alpha(\delta)}-\hat
{h}_{\alpha(\delta)}^{-1}\right]  $ and hence it is $O(\delta)$. The second
and third terms are of the form $\frac{1}{\delta}\left(  \hat{E}^{i}%
(L_{e}(\delta^{\prime})(M(v+\delta)-M(v))\right)  $ and hence to leading order
in $\delta$ they are $O(\delta^{\prime})=O(\delta^{2})$.

These finite-triangulation operators, due to the structure of the extra terms,
are non-local in the sense that they can never be perceived as (quantum
counterparts of) the discretization of a classical local functional. A similar
feature was observed in the correction to the fundamental LQG curvature
operator, that was defined in \cite{amdiff} and led to an anomaly-free
quantization of the diffeomorphism constraint. Nonetheless, as we will see
later, the continuum limit of the Hamiltonian constraint operator will be
local in the sense that it will be expressed in terms of local differential operators.

This then is our proposal for the density two $\hat{H}_{T(\delta)}^{(1)}[N]$
when it acts on a state containing an EO pair $(v,v^{\mathrm{E}})$ with
$v^{\mathrm{E}}$ being an EO vertex of type $(K\in\{\mathrm{A,B}\},M=2,j=1)$
(i.e., it is either a type A or type B vertex with all incident edges charged
in U(1)$_{2}$ with the charge coming from the $\mathrm{U}(1)_{1}$ labels on
edges incident at vertices in $\mathcal{\bar{S}}_{v^{\mathrm{E}}}$. Other
cases can be considered similarly. We now modify our results appropriately for
the realistic case of density $\frac{5}{4}$ constraint. As we remarked
earlier, this amounts to choosing a particular operator ordering for $\hat
{q}^{-1/4},$ which is a scalar multiple of the identity operator on any charge
network state. The ordering we choose is given by%

\begin{align}
&  \hat{H}_{T(\delta)}^{(1)}[M]|c_{1},c_{2}\cup\alpha_{v}(\langle\hat{E}%
_{1}\rangle_{c_{1}},n_{c_{3}}),c_{3}\rangle\\
&  =\frac{1}{\delta}M(v)\langle\hat{q}(v)^{-1/4}\rangle_{c}\left[  (\hat
{h}_{\alpha_{v}^{\delta}(\langle\hat{E}_{2}\rangle_{c_{2}\cup\alpha
_{v}(\langle\hat{E}_{1}\rangle_{c_{1}},n_{c_{3}})},n_{c_{3}})}^{(1)})-(\hat
{h}_{\alpha_{v}^{\delta}(\langle\hat{E}_{3}\rangle_{c_{3}},(n_{c_{2}}%
-n_{c_{3}})}^{(1)})\right]  |c_{1},c_{2}\cup\alpha_{v}(\langle\hat{E}%
_{1}\rangle_{c_{1}},n_{c_{3}}),c_{3}\rangle\\
&  +\frac{1}{\kappa\hbar\delta}\left(  \sum_{e\in E(\gamma)}\langle\hat{E}%
_{2}(L_{e}^{v^{\prime}}(\delta^{\prime}))\rangle_{c_{2}}(M(v+\delta\dot
{e}(0))-M(v))(\hat{h}_{\alpha_{v}(\langle\hat{E}_{3}\rangle_{c_{3}},n_{c_{3}%
})}^{(2)})\hat{q}(v)^{-1/4}(\hat{h}_{\alpha_{v}(\langle\hat{E}_{1}%
\rangle_{c_{1}},n_{c_{3}})}^{(2)})^{-1})\right. \\
&  -\left.  \sum_{e\in E(\gamma)}\langle\hat{E}_{3}(L_{e}^{v^{\prime}}%
(\delta^{\prime}))\rangle_{c_{3}}(M(v+\delta\dot{e}(0))-M(v))(\hat{h}%
_{\alpha_{v}(\langle\hat{E}_{2}\rangle_{c_{2}},n_{c_{3}})}^{(2)}\hat
{q}(v)^{-1/4}(\hat{h}_{\alpha_{v}(\langle\hat{E}_{1}\rangle_{c_{1}},n_{c_{3}%
})}^{(2)})^{-1})\right)  |c_{1},c_{2}\cup\alpha_{v}(\langle\hat{E}_{1}%
\rangle_{c_{1}},n_{c_{3}}),c_{3}\rangle
\end{align}

Thus we finally have a definition of the Hamiltonian constraint operator on an
arbitrary charge network state. If the charge-network contains an EO pair then
the constraint operator has an additional piece which is non-local and can be
thought of as having a non-trivial action on the EO pair rather then acting on
a single isolated vertex. The complete implications of having an operator
which at finite triangulation not only changes charge network in the
neighborhood of a single vertex, but changes it in the neighborhood of a
sub-graph are not clear to us. However as we show in the end, these
quantization choices buy us a lot in the end, as the off-shell closure
conditions are satisfied.

\section{LMI Habitat}

In the previous section we completed the construction of the Hamiltonian
constraint operator at finite triangulation, which is densely defined on
$\mathcal{H}_{\mathrm{kin}}$. As is well known, due to the higher density
weight of the operator, it will not have a continuum limit (in any operator
topology) which is well-defined on $\mathcal{H}_{\mathrm{kin}}$. In this
section we construct an arena which we call the Lewandowski-Marolf-Inspired
(LMI) habitat, on which the net of finite-triangulation operators admits a
continuum limit. We will come back to the issue of operator topology later in
the section. First we engineer a habitat taking a cue from Lewandowski and
Marolf's seminal construction \cite{lm1}.

We want to build our habitat in such a way that not only does it admit some
sort of continuum limit of the Hamiltonian constraint, but that it admits a
representation of the entire Dirac algebra. We build our habitat keeping this
requirement in mind.\newline Starting with a charge network $c$ which has no
mono-colored vertex, construct a set $[c_{1},c_{2},c_{3}]_{(i)}$ as
follows.\newline%
\begin{equation}
[c_{1}, c_{2},c_{3}]_{(i=1)}=\bigcup_{c_{1}^{\prime}} \{\left( c_{1}^{\prime
},c_{2},c_{3}\right) \}
\end{equation}
where $c_{1}^{\prime}$ has at least one additional weakly extraordinary (WEO)
vertex as compared to $c_{1}$. $[c_{1},c_{2},c_{3}]_{(i=2)}$ and $[c_{1}%
,c_{2},c_{3}]_{(i=3)}$ are defined similarly. Now we consider\ the following
type of elements of $\mathrm{Cyl}^{\ast}$:%
\begin{equation}
\Psi_{\lbrack c_{1},c_{2},c_{3}]_{(i)}}^{f^{(i)}}=\sum_{(c_{1}^{\prime}%
,c_{2},c_{3})\in\lbrack c_{1},c_{2},c_{3}]_{(i)}}f^{(i)}(\bar{V}(c_{1}%
^{\prime}\cup c_{2}\cup c_{3}))\langle c_{1}^{\prime},c_{2},c_{3}|
\end{equation}

where

\begin{enumerate}
\item[(i)] $f^{(i)},i=1,2,3$ are smooth functions on $\Sigma^{|\cup
_{i}V(\mathbf{c}_{i})|}$

\item[(ii)] $\bar{V}(c_{i}^{\prime})$ is defined as follows: Let
$V(c_{i}^{\prime})=\{v_{0},v_{0}^{\mathrm{WE}},\dots,v_{K},v_{K}^{\mathrm{WE}%
},v_{K+1},\dots,v_{N}\}$ where $\{v_{0}^{\mathrm{WE}},...,v_{K}^{\mathrm{WE}%
}\}$ are WEO vertices for $c_{i}^{\prime}$ associated to $\{v_{0},...,v_{K}\}$ respectively.
\end{enumerate}

Then%
\begin{equation}
\bar{V}(c_{i}^{\prime}):=\{v_{0}^{\mathrm{WE}},...,v_{K}^{\mathrm{WE}}%
,v_{K+1},...,v_{M}\}
\end{equation}
Note that by construction $|\bar{V}(c_{i}^{\prime})|~=|\cup_{i}V(c_{i})|$ so
that $f^{(i)}$ are functions on $\Sigma^{|\cup_{i}V(c_{i})|}$. We define
$\mathcal{V}_{\mathrm{LMI}}$ as a subspace of $\mathrm{Cyl}^{\ast}$ spanned by
distributions of the type $\Psi_{\lbrack c_{1},c_{2},c_{3}]_{(i)}}^{f^{(i)}}$.

We will now show that $\hat{H}_{T(\delta)}^{(i)}[N]\forall i$ admits a
continuum limit as a linear operator from $\mathcal{V}_{\mathrm{LMI}%
}\rightarrow\mathrm{Cyl}^{\ast}$. The topology on the space of operators in
which we consider the continuum limit is defined via the following family of
seminorms: Given any pair $(\Psi,|c\rangle)\in\mathcal{V}_{\mathrm{LMI}}%
\times\mathcal{H}_{\mathrm{kin}}$, we say that $\hat{H}^{(i)}[N]^{\prime}$ is
a continuum limit of $\hat{H}_{T(\delta)}^{(i)}[N]$ if for $\epsilon>0$,
$\exists\ \delta_{0}=\delta_{0}(\Psi,c,N)$ such that
\begin{equation}
\left\vert (\hat{H}^{(i)}[N]^{\prime}\Psi)|c\rangle-\Psi(\hat{H}_{T(\delta
)}^{(i)}[N]|c\rangle)\right\vert <\epsilon
\end{equation}
$\forall\ \delta<\delta_{0}$ (we will generally decorate operators acting on
elements of $\mathrm{Cyl}^{\ast}$ with a prime). It turns out that the
continuum Hamiltonian constraint does not preserve the LMI habitat; rather
\begin{equation}
\hat{H}^{(i)}[N]^{\prime}:\mathcal{V}_{\mathrm{LMI}}\rightarrow\mathrm{Cyl}%
^{\ast}%
\end{equation}
This happens because when acting on a state, say $\Psi_{\lbrack c_{1}%
,c_{2},c_{3}]_{(1)}}^{f^{(1)}}\in\mathcal{V}_{\mathrm{LMI}}$, the resulting
states are still infinite linear combinations of (duals of) charge network
states, with amplitudes being functions of vertices. However, in contrast to
$f^{(1)}$ which is smooth on $\Sigma^{|V(c_{1}\cup c_{2}\cup c_{3})|}$,
coefficients of the charge networks in these linear combinations will be
discontinuous functions on $\Sigma^{|V(c_{1}\cup c_{2}\cup c_{3})|}$.

\subsection{The Continuum Limit}

Consider $\Psi_{\lbrack c_{1},c_{2},c_{3}]_{(1)}}^{f^{1}}$, where
$f^{(1)}:\Sigma^{|V(c_{1}\cup c_{2}\cup c_{3})|}\rightarrow%
\mathbb{R}
$. The action of the continuum Hamiltonian constraint $\hat{H}^{(1)}%
[N]^{\prime}+\hat{H}^{(2)}[N]^{\prime}+\hat{H}^{(3)}[N]^{\prime}$ on such
states can be deduced from Eqs. (\ref{eq:feb6-1}), (\ref{shitt}) and
(\ref{FUCK}) that are given below. Derivations of these results can be found
in Appendix \ref{A3}.

We first consider the action of $\hat{H}^{(1)}[N]^{\prime}$ on $\Psi_{\lbrack
c_{1},c_{2},c_{3}]_{(1)}}^{f^{1}}$:%
\begin{equation}
\hat{H}^{(1)}[N]^{\prime}\Psi_{\lbrack c_{1},c_{2},c_{3}]_{(1)}}^{f^{(1)}%
}=\sum_{v\in V(c_{1}\cup c_{2}\cup c_{3})}\left[  \Psi_{\lbrack c_{1}%
,c_{2},c_{3}]_{(1)}}^{\bar{f}_{v}^{(1)(1)}}-\Psi_{\lbrack c_{1},c_{2}%
,c_{3}]_{(1)}}^{\overline{\overline{f}}{}_{v}^{(1)(1)}}\right]
\label{eq:feb6-2}%
\end{equation}
where\footnote{To avoid notational clutter, we do not explicitly indicate the
dependence of $\overline{\overline{f}}{}_{v}^{(1)(1)}$ etc., on $N$%
,$(c_{1},c_{2},c_{3})$.} $\bar{f}_{v}^{(1)(1)}$ is given by (see below for
$\overline{\overline{f}}{}_{v}^{(1)(1)}$)
\begin{equation}
\bar{f}_{v}^{(1)(1)}(v_{1},\dots,v_{|V(c_{1}\cup c_{2}\cup c_{3})|}%
)=f^{(1)}(v_{1},\dots,v_{|V(c_{1}\cup c_{2}\cup c_{3})|})
\end{equation}
if the following hold:

\begin{enumerate}
\item[(1)] $\{v_{1},...,v_{|V(c_{1}\cup c_{2}\cup c_{3})|}\}\neq
V(c_{1}^{\prime}\cup c_{2}\cup c_{3})$ for any $(c_{1}^{\prime},c_{2}%
,c_{3})\in\lbrack c_{1},c_{2},c_{3}]_{(1)}$, or

\item[(2)] $\{v_{1},...,v_{|V(c_{1}\cup c_{2}\cup c_{3})|}\}=V(c_{1}^{\prime
}\cup c_{2}\cup c_{3})$ for some $(c_{1}^{\prime},c_{2},c_{3})\in\lbrack
c_{1},c_{2},c_{3}]_{(1)}$ but $v\notin\{v_{1},...,v_{|V(c_{1}\cup c_{2}\cup
c_{3})|}\}$.
\end{enumerate}

In the case that the complements of (1) and (2) hold, we have
\begin{equation}
\bar{f}_{v}^{(1)(1)}(v_{1},...,v_{|V(c_{1}\cup c_{2}\cup c_{3})|}%
)=N(v)\lambda(\vec{n}_{v}^{c})\langle\hat{E}_{2}^{a}(v)\rangle\frac{\partial
}{\partial v^{a}}f^{(1)}(v_{1},...,v_{|V(c_{1}\cup c_{2}\cup c_{3})|}).
\label{eq:feb6-1}%
\end{equation}
$\overline{\overline{f}}{}_{v}^{(1)(1)}$ is defined analogously, except that
$\langle\hat{E}_{2}^{a}(v)\rangle_{c_{2}}$ in (\ref{eq:feb6-1}) is replaced by
$\langle\hat{E}_{3}^{a}(v)\rangle_{c_{3}}$.

We now consider the action of $\hat{H}^{(2)}[N]^{\prime}$ on $\Psi_{\lbrack
c_{1},c_{2},c_{3}]_{(1)}}^{f^{(1)}}$:%
\begin{equation}
\hat{H}^{(2)}[N]^{\prime}\Psi_{\lbrack c_{1},c_{2},c_{3}]_{(1)}}^{f^{(1)}%
}=\sum_{v\in V(c_{1}\cup c_{2}\cup c_{3})}\left[  \Psi_{\lbrack c_{1}%
,c_{2},c_{3}]_{(1)}}^{\bar{f}_{v}^{(1)(2)}}-\Psi_{\lbrack c_{1},c_{2}%
,c_{3}]_{(1)}}^{\overline{\overline{f}}{}_{v}^{(1)(2)}}\right]
\label{eq:feb6-3}%
\end{equation}
where $\bar{f}_{v}^{(1)(2)}$ and $\overline{\overline{f}}{}_{v}^{(1)(2)}$ are
defined as follows: Let $\{v_{1},...,v_{|V(c_{1}\cup c_{2}\cup c_{3}%
)|}\}=V(c_{1}^{\prime}\cup c_{2}\cup c_{3})$ such that

\begin{enumerate}
\item[(1)] $(c_{1}^{\prime}\cup c_{2}\cup c_{3})\in\lbrack c_{1},c_{2}%
,c_{3}]_{(1)}$

\item[(2)] $v\in\{v_{1},...,v_{|V(c_{1}\cup c_{2}\cup c_{3})|}\}$ such that
$v\in V(c_{1}\cup c_{2}\cup c_{3})$ and there is an EO vertex $v^{\mathrm{E}}$
associated to $v$ of type $(M=1,j=2)$ which lies inside the support of $N$. In this case
\begin{subequations}
\label{shitt}%
\begin{align}
\bar{f}_{v}^{(1)(2)}(\bar{V}(c_{1}^{\prime}\cup c_{2}\cup c_{3}))  &  =\left[
\sum_{e\in E(c_{1}\cup c_{2}\cup c_{3})|b(e)=v}\langle\hat{E}_{1}%
(L_{e})\rangle_{c_{1}}\dot{e}^{a}(0)\partial_{a}N(v)\right]  f^{(1)}(\bar
{V}(c_{1}^{\prime}\cup c_{2}\cup c_{3}))\\
\overline{\overline{f}}{}_{v}^{(1)(2)}(\bar{V}(c_{1}^{\prime}\cup c_{2}\cup
c_{3}))  &  =\left[  \sum_{e\in E(c_{1}\cup c_{2}\cup c_{3})|b(e)=v}%
\langle\hat{E}_{3}(L_{e})\rangle_{c_{1}}\dot{e}^{a}(0)\partial_{a}N(v)\right]
f^{(1)}(\bar{V}(c_{1}^{\prime}\cup c_{2}\cup c_{3}))
\end{align}
where $L_{e}$ is as defined in Eq. (\ref{BORG}).
\end{subequations}
\end{enumerate}

In the case the complement of the two conditions {(1)} and {(2)} hold, we
have
\begin{align}
\bar{f}_{v}^{(1)(2)}(v_{1},\dots,v_{|V(c_{1}\cup c_{2}\cup c_{3})|})  &
=f^{(1)}(v_{1},\dots,v_{|V(c_{1}\cup c_{2}\cup c_{3})|}),\\
\overline{\overline{f}}{}_{v}^{(1)(2)}(v_{1},\dots,v_{|V(c_{1}\cup c_{2}\cup
c_{3})|})  &  =f^{(1)}(v_{1},\dots,v_{|V(c_{1}\cup c_{2}\cup c_{3})|})
\end{align}

The action of $\hat{H}^{(3)}[N]^{\prime}$ on $\Psi_{\lbrack c_{1},c_{2}%
,c_{3}]_{(1)}}^{f^{(1)}}$ can be written in analogy with (\ref{eq:feb6-3}).
\begin{equation}
\hat{H}^{(3)}[N]^{\prime}\Psi_{\lbrack c_{1},c_{2},c_{3}]_{(1)}}^{f^{(1)}%
}=\sum_{v\in V(c_{1}\cup c_{2}\cup c_{3})}\left[  \Psi_{\lbrack c_{1}%
,c_{2},c_{3}]_{(1)}}^{\bar{f}_{v}^{(1)(3)}}-\Psi_{\lbrack c_{1},c_{2}%
,c_{3}]_{(1)}}^{\overline{\overline{f}}{}_{v}^{(1)(3)}}\right]
\label{eq:feb6-4}%
\end{equation}
where $\bar{f}_{v}^{(1)(3)}$ and $\overline{\overline{f}}{}_{v}^{(1)(3)}$ are
defined as follows: Let $\{v_{1},...,v_{|V(c_{1}\cup c_{2}\cup c_{3}%
)|}\}=V(c_{1}^{\prime}\cup c_{2}\cup c_{3})$ such that

\begin{enumerate}
\item[(1)] $(c_{1}^{\prime}\cup c_{2}\cup c_{3})\in\lbrack c_{1},c_{2}%
,c_{3}]_{(1)}$

\item[(2)] $v\in\{v_{1},...,v_{|V(c_{1}\cup c_{2}\cup c_{3})|}\}$ such that
$v\in V(c_{1}\cup c_{2}\cup c_{3})$ and there is an EO vertex $v^{\mathrm{E}}$
associated to $v$ of type $(M=1,j=3)$. In this case
\begin{subequations}
\label{FUCK}%
\begin{align}
\bar{f}_{v}^{(1)(3)}(\bar{V}(c_{1}^{\prime}\cup c_{2}\cup c_{3}))  &  =\left[
\sum_{e\in E(c_{1}\cup c_{2}\cup c_{3})|b(e)=v}\langle\hat{E}_{2}%
(L_{e})\rangle_{c_{1}}\dot{e}^{a}(0)\partial_{a}N(v)\right]  f^{(1)}(\bar
{V}(c_{1}^{\prime}\cup c_{2}\cup c_{3}))\\
\overline{\overline{f}}{}_{v}^{(1)(3)}(\bar{V}(c_{1}^{\prime}\cup c_{2}\cup
c_{3}))  &  =\left[  \sum_{e\in E(c_{1}\cup c_{2}\cup c_{3})|b(e)=v}%
\langle\hat{E}_{1}(L_{e})\rangle_{c_{1}}\dot{e}^{a}(0)\partial_{a}N(v)\right]
f^{(1)}(\bar{V}(c_{1}^{\prime}\cup c_{2}\cup c_{3}))
\end{align}
As before, if the set $\{v_{1},...,v_{|V(c_{1}\cup c_{2}\cup c_{3})|}\}$ does
not satisfy conditions {(1)} or {(2)}, then the two functions $\bar{f}%
_{v}^{(1)(3)}$ and $\overline{\overline{f}}{}_{v}^{(1)(3)}$ take the same
value as $f^{(1)}$.
\end{subequations}
\end{enumerate}

The definitions of $\bar{f}_{v}^{(1)(i)}|_{i=1,2,3}$ make it rather clear that
the continuum Hamiltonian constraint does not preserve the LMI habitat. These
functions have a discontinuity as soon as one of their arguments is the vertex
$v$. This discontinuity is due to the discontinuous nature of the quantum
shift vector, which is in turn tied to the choice of representation we are
forced upon in LQG.


\subsection{The Action of the Hamiltonian Constraint on Irrelevant Vertices}

Before we compute the continuum limit of the commutator of two (regularized)
Hamiltonian constraints, we make two observations which vastly simplify the
structure of the computation (and indeed, without which, $\hat{H}[N]^{\prime}$
will not satisfy the off-shell closure condition). These observations are
related to the action of a Hamiltonian constraint on a charge network state
which lies in the image of $\hat{H}_{T}^{(i)}[N]$ for some $i\in
\{1,2,3\},N,\ \text{and}\ T$.

The action of such a finite triangulation Hamiltonian constraint on a charge
network which has no EO vertex, creates a set of vertices which we termed
irrelevant vertices (the name finds its justification in this section). When a
finite-triangulation Hamiltonian acts on a tri-valent irrelevant vertex, it
vanishes (as all such tri-valent vertices are in the kernel of $\hat{q}%
^{-1/4}$ operator)\textbf{.} Whence these vertices are irrelevant as far as
the action of a second Hamiltonian on such a charge network is concerned. This
is not quite true for the four-valent irrelevant vertices\footnote{It is
important to note that our entire construction, when generalized to three
dimensions would generically be free of such four-valent vertices!} However,
we now argue that the continuum limit of the action of a finite-triangulation
Hamiltonian constraint on a four-valent irrelevant vertex is trivial. This
feature is tied to the choice of our habitat (or more precisely to the
definition of $[c_{1},c_{2},c_{3}]_{(i)}$).

Let $(\tilde{c}_{1},c_{2},c_{3})$ be a charge-network with an EO vertex
$v_{0}^{\mathrm{E}}$, which for the sake of concreteness we consider to be of
type $(M=1,j=3)$. That is,
\begin{equation}
(\tilde{c}_{1},c_{2},c_{3})=(c_{1}\ \cup\alpha_{v_{0}}^{\delta_{0}}%
(\langle\hat{E}_{2}(v_{0})\rangle_{c_{2}},n_{c_{3}}),c_{2},c_{3})
\end{equation}
where $(c_{1},c_{2},c_{3})$ does not have any EO vertices and where
$v_{0}^{\mathrm{E}}$ is associated with $v_{0}$. There is a set of irrelevant
vertices in $V(\tilde{c}_{1}\cup c_{2}\cup c_{3})$, and let us denote this set
by $\{v_{1}^{v_{0}},...,v_{k}^{v_{0}}\}$. Let us consider one of them, say
$v_{1}^{v_{0}}$ and let the four edges incident on $v_{1}$ be $e_{v_{1}%
^{v_{0}}}^{1},\dots,e_{v_{1}^{v_{0}}}^{4}$ such that ($e_{v_{1}^{v_{0}}}^{1}%
$,$e_{v_{1}^{v_{0}}}^{3}$) and ($e_{v_{1}^{v_{0}}}^{2}$,$e_{v_{1}^{v_{0}}}%
^{4}$) are analytic pairs. Let us assume that ($e_{v_{1}^{v_{0}}}^{1}%
$,$e_{v_{1}^{v_{0}}}^{3}$) are charged only under $\mathrm{U}(1)_{1}$, whence
a simple computation shows that $\hat{H}_{T(\delta^{\prime})}^{(1)}%
(v_{1}^{v_{0}})$ acting on $|\tilde{c}_{1},c_{2},c_{3}\rangle$ vanishes.
However this is not true for $\hat{H}_{T(\delta^{\prime})}^{(2)}(v_{1}^{v_{0}%
})$ or $\hat{H}_{T(\delta^{\prime})}^{(3)}(v_{1}^{v_{0}})$. Their action will
produce EO pairs $(v_{1}^{v_{0}},(v_{1}^{v_{0}})^{\mathrm{E}})$ which are of
type $(M=2)$ or type $(M=3)$. Thus the action of the Hamiltonian constraint on
$|\tilde{c}_{1},c_{2},c_{3}\rangle$ produces a state which has a vertex
$v_{0}^{\mathrm{E}}$ charged in $\mathrm{U}(1)_{1}$ and a vertex
$(v_{1}^{v_{0}})^{\mathrm{E}}$ charged in $\mathrm{U}(1)_{2}$. As there exists
no set $[c_{1}^{\prime},c_{2}^{\prime},c_{3}^{\prime}]_{(i)}$ in which there
is ever a charge network with two mono-coloured vertices both charged in
different $\mathrm{U}(1)_{i}$, we have
\begin{equation}
\Psi_{\lbrack c_{1}^{\prime},c_{2}^{\prime},c_{3}^{\prime}]}^{f^{(i)}}\left(
\hat{H}_{T(\delta)}^{(i)}[N]|\tilde{c}_{1},c_{2},c_{3}\rangle\right)  =0
\end{equation}
$\forall\ \delta>0$ and $\forall\ i$. Hence we will ignore the action of the
Hamiltonian constraint on irrelevant vertices in what follows.

\section{Commutator of Two Hamiltonian Constraints}

In this section we embark upon the key computation performed in this paper. We
argue that the quantum Hamiltonian constraint that we have obtained above has
the right basic ingredients to achieve a anomaly-free representation of the
Dirac algebra. We will show that the commutator between two Hamiltonian
constraint is, in a precise sense \emph{a} quantization of the right hand side
of the corresponding classical Poisson bracket.

Let us first describe briefly, precisely what it is that we want to show.
Recall that%
\begin{equation}
\{H[M],H[N]\}=V[\vec{\omega}],\qquad\omega^{a}:=q^{-1/2}E_{i}^{a}E_{i}%
^{b}\left(  N\partial_{b}M-M\partial_{b}N\right)  .
\end{equation}
Our aim is to show that the above equality holds at the quantum level. That
is, schematically we want to prove that
\begin{equation}
\lbrack\hat{H}[M],\hat{H}[N]]=\hat{V}[\vec{\omega}] \label{mar6-1}%
\end{equation}

Our strategy will be the following. As the continuum Hamiltonian constraint
does not preserve the habitat $\mathcal{V}_{\mathrm{LMI}},$ but maps it
elements into elements of $\mathrm{Cyl}^{\ast}$, the commutator of two
continuum Hamiltonians does not make sense on $\mathcal{V}_{\mathrm{LMI}}$.
However, things are not as bad as they look. Let us assume for a moment an
ideal scenario where we had a habitat $\mathcal{V}_{\mathrm{grand}}$ on which
any product of a finite number of continuum Hamiltonian constraints is a well
defined operator. Then $\forall\Psi\in\mathcal{V}_{\mathrm{grand}}$, we would
have
\begin{align}
&  \sum_{i,j}\left(  [\hat{H}^{(i)}[M]^{\prime},\hat{H}^{(j)}[N]^{\prime}%
]\Psi\right)  |c_{1},c_{2},c_{3}\rangle\label{oct4-2}\\
&  =\sum_{i,j}\left(  \left(  \hat{H}^{(i)}[M]\hat{H}^{(j)}[N]-\left(
M\leftrightarrow N\right)  \right)  \Psi\right)  |c_{1},c_{2},c_{3}%
\rangle\nonumber\\
&  =\sum_{i,j}\lim_{\delta\rightarrow0}\lim_{\delta^{\prime}\rightarrow0}%
\Psi\left(  \hat{H}_{T(\delta^{\prime})}^{(j)}[N]\hat{H}_{T(\delta)}%
^{(i)}[M]-\left(  M\leftrightarrow N)\right)  \right)  |c_{1},c_{2}%
,c_{3}\rangle\nonumber
\end{align}
As we show below, the right hand side of this equation is well-defined and
constitutes our definition of the \emph{continuum commutator}:
\begin{equation}
\left(  \lbrack\hat{H}[M],\hat{H}[N]]^{\prime}\Psi\right)  |c\rangle
:=\sum_{i,j}\lim_{\delta\rightarrow0}\lim_{\delta^{\prime}\rightarrow0}%
\Psi\left(  \left(  \hat{H}_{T(\delta^{\prime})}^{(j)}[N]\hat{H}_{T(\delta
)}^{(i)}[M]-\left(  M\leftrightarrow N\right)  \right)  |c\rangle\right)
\end{equation}

In light of (\ref{oct4-2}), the equality in (\ref{mar6-1}) amounts to proving that%

\begin{equation}
\sum_{i,j}\lim_{\delta\rightarrow0}\lim_{\delta^{\prime}\rightarrow0}%
\Psi\left(  \hat{H}_{T(\delta^{\prime})}^{(j)}[N]\hat{H}_{T(\delta)}%
^{(i)}[M]-\left(  M\leftrightarrow N\right)  \right)  |c_{1},c_{2}%
,c_{3}\rangle=\lim_{\delta^{\prime\prime}\rightarrow0}\left(  -\mathrm{i}%
\hbar\right)  \Psi\left(  \hat{V}_{T(\delta^{\prime\prime})}[\vec{\omega
}]|c_{1},c_{2},c_{3}\rangle\right)  \label{eq:mar6-2}%
\end{equation}
$\forall\Psi\in\mathcal{V}_{\mathrm{LMI}},\ \forall\ (c_{1},c_{2},c_{3})$ and,
where $\vec{\omega}\ =\ N\vec{\nabla}M\ -\ M\vec{\nabla}N$

The minus sign in $-i\hbar$ may seem surprising but it arises due to the
argument given in appendix \ref{A2point5}.\newline

Our strategy towards proving (\ref{eq:mar6-2}) will be as follows.

\begin{enumerate}
\item[(1)] The first step in obtaining the continuum commutator on
$\mathcal{V}_{\mathrm{LMI}}$ is computing $\sum_{i,j}[\hat{H}_{T(\delta
^{\prime})}^{(i)}[N]\hat{H}_{T(\delta)}^{(j)}[M]-\left(  M\leftrightarrow
N\right)  ]|c_{1}^{\prime},c_{2}^{\prime},c_{3}^{\prime}\rangle$ for any
charge-network state $|c^{\prime}\rangle$. As we show in Section
\ref{nonoverlaplapse}, the regularized commutator vanishes $\forall
\ \delta,\delta^{\prime}$ if $\mathrm{supp}(N)\cap\mathrm{supp}(M)=\emptyset$.
For the case when $\mathrm{supp}(N)\cap\mathrm{supp}(M)\neq\emptyset$, the
computation is slightly more involved and details are provided in Appendix
\ref{A4}.

\item[(2)] In Section \ref{overlaplapse} we use the results of Appendix
\ref{A4} to derive the continuum limit of the right hand side of Equation
(\ref{eq:mar6-2}).

\item[(3)] In Section \ref{diffside}, we define $\hat{V}_{T(\delta
,\delta^{\prime})}[\vec{\omega}]$ such that $\hat{V}_{T(\delta,\delta^{\prime
})}[\vec{\omega}]|c^{\prime}\rangle$ precisely equals the relevant
terms\footnote{By relevant terms we mean those states which do not vanish once
they are \textquotedblleft dotted\textquotedblright\ with a state in the
habitat.} in
\begin{equation}
\sum_{i,j}[\hat{H}_{T(\delta^{\prime})}^{(i)}[N]\hat{H}_{T(\delta)}%
^{(j)}[M]-\left(  M\leftrightarrow N\right)  ]|c_{1}^{\prime},c_{2}^{\prime
},c_{3}^{\prime}\rangle.
\end{equation}
This will finally lead us to our main result.
\end{enumerate}

\subsubsection{Analyzing the Case When $\mathrm{s{}upp}(N)\cap\mathrm{s{}%
upp}(M)=\emptyset$}

\label{nonoverlaplapse}

Consider a state $|c_{1}^{\prime},c_{2}^{\prime},c_{3}^{\prime}\rangle$ such
that the only vertices in $V(c_{1}^{\prime}\cup c_{2}^{\prime}\cup
c_{3}^{\prime})$ which lie in the support of $N$ and $M$ respectively are,
$v_{N}$ and $v_{M}$. Then for any $i,j,$
\begin{align}
&  \left[  \hat{H}_{T(\delta^{\prime})}^{(i)}[N]\hat{H}_{T(\delta)}%
^{(j)}[M]-\left(  M\leftrightarrow N\right)  \right]  |c_{1}^{\prime}%
,c_{2}^{\prime},c_{3}^{\prime}\rangle\nonumber\\
&  =\frac{1}{\delta\delta^{\prime}}\left[  \hat{O}_{T(\delta^{\prime})}%
^{(i)}(v_{N})\hat{O}_{T(\delta)}^{j}(v_{M})\left(  \mathcal{F}[N;v_{N}%
)\mathcal{G}[M;v_{M})-\left(  M\leftrightarrow N\right)  \right)  \right]
|c_{1}^{\prime},c_{2}^{\prime},c_{3}^{\prime}\rangle
\end{align}
where $\mathcal{F}[N;v_{N})$ and $\mathcal{G}[M;v_{M})$ are in general a local
functional of $N$ and $M$ evaluated at $v_{N}$ and $v_{M}$
respectively.\footnote{The precise form of these functionals, depend on the
nature of $|c^{\prime}\rangle$, but what is important for our purposes is that
it can at most involve the first derivative of the lapse evaluated at $v_{N}$
or $v_{M}$.} Locality of these functionals implies that%

\begin{equation}
\left[  \hat{H}_{T(\delta^{\prime})}^{(i)}[N]\hat{H}_{T(\delta)}%
^{(j)}[M]-\left(  M\leftrightarrow N\right)  \right]  |c_{1}^{\prime}%
,c_{2}^{\prime},c_{3}^{\prime}\rangle=0
\end{equation}
$\forall\ \delta,\delta^{\prime}$ and $\forall\ |c^{\prime}\rangle
\ \in\mathcal{H}_{\mathrm{kin}}$. Whence for any state in $\mathcal{V}%
_{\mathrm{LMI}}$ we have
\begin{equation}
\lim_{\delta\rightarrow0}\lim_{\delta^{\prime}\rightarrow0}\Psi_{\lbrack
c_{1},c_{2},c_{3}]_{(k)}}^{f^{(k)}}\left(  \left[  \hat{H}_{T(\delta^{\prime
})}^{(i)}[N]\hat{H}_{T(\delta)}^{(j)}[M]-\left(  M\leftrightarrow N\right)
\right]  |c^{\prime}\rangle\right)  =0
\end{equation}
This implies that as long as $N$,$M$ have non-intersecting supports,%
\begin{equation}
\lbrack\hat{H}[M],\hat{H}[N]]\Psi_{\lbrack c_{1},c_{2},c_{3}]_{(k)}}^{f^{(k)}%
}=0 \label{eq:mar10-2}%
\end{equation}

\subsubsection{Analyzing the Case When $\mathrm{s{}upp}(N)\cap\mathrm{s{}%
upp}(M)\neq\emptyset$}

\label{overlaplapse}

Let $\Psi_{\lbrack c_{1},c_{2},c_{3}]_{(i)}}^{f^{(i)}}$ be such that there
exists a single vertex in $V(c_{1}\cup c_{2}\cup c_{3})$ which falls inside
s$\text{upp}(N)\cap\text{supp}(M)$. The case where more then one element of
the vertex set $V(c_{1}\cup c_{2}\cup c_{3})$ falls in the overlap region is a
straightforward generalization of the analysis given here. In this case, in
order to evaluate
\begin{equation}
\Psi_{\lbrack c_{1},c_{2},c_{3}]_{(i)}}^{f^{(i)}}\left(  \sum_{i,j}[\hat
{H}_{T(\delta^{\prime})}^{(i)}[N]\hat{H}_{T(\delta)}^{(j)}[M]-\left(
M\leftrightarrow N\right)  ]|c_{1}^{\prime},c_{2}^{\prime},c_{3}^{\prime
}\rangle\right)  ,
\end{equation}
it suffices to consider only those states $|c^{\prime}\rangle$ which have only
one vertex in the support of $V(c_{1}\cup c_{2}\cup c_{3})$, so we first
consider a state $|c^{\prime}\rangle$ with precisely one vertex $v_{0}$ in
$V(c_{1}^{\prime}\cup c_{2}^{\prime}\cup c_{3}^{\prime})$ which lies inside
the support of $N$ and $M$. As shown in Appendix \ref{A4}, we have the
following:
\begin{equation}
\sum_{i\neq j}[\hat{H}_{T(\delta^{\prime})}^{(i)}[N]\hat{H}_{T(\delta)}%
^{(j)}[M]-\left(  M\leftrightarrow N\right)  ]|c_{1}^{\prime},c_{2}^{\prime
},c_{3}^{\prime}\rangle=|\psi_{1}^{\delta,\delta^{\prime}}\left(
v_{0},c^{\prime},[M,N]\right)  \rangle+|\psi_{2}^{\delta,\delta^{\prime}%
}\left(  v_{0},c^{\prime},[M,N]\right)  \rangle+|\psi_{3}^{\delta
,\delta^{\prime}}\left(  v_{0},c^{\prime},[M,N]\right)  \rangle
\label{eq:comm-1}%
\end{equation}
where the $|\psi_{i}^{\delta,\delta^{\prime}}\left(  v_{0},c^{\prime
},[M,N]\right)  \rangle$ are given in Appendix \ref{A4} in (\ref{eq:termA}),
(\ref{eq:termB}), and (\ref{eq:termC}), respectively.

\textbf{Claim}: $\lim_{\delta,\delta^{\prime}\rightarrow0}\Psi_{\lbrack
c_{1},c_{2},c_{3}]_{(i)}}^{f^{i}}\left(  |\psi_{j}^{\delta,\delta^{\prime}%
}\left(  v_{0},c^{\prime},[N,M]\right)  \rangle\right)  =0$ $\forall i\neq
j,(c,c^{\prime})$.

\textbf{Proof}: The proof is straightforward, as $|\psi_{j}^{\delta
,\delta^{\prime}}\left(  v_{0},c^{\prime},[N,M]\right)  \rangle$ is a linear
combination of four states with each of this state contains (with respect to
$c^{\prime}$) one EO state of type $j$. Thus clearly these states are
orthogonal to all states in $[c_{1},c_{2},c_{3}]_{(i)}$ for any $c$ as long as
$i\neq j$. This completes the proof.

Then
\begin{align}
&  \lim_{\delta,\delta^{\prime}\rightarrow0}\Psi_{\lbrack c_{1},c_{2}%
,c_{3}]_{(i)}}^{f^{(i)}}\left(  \sum_{j\neq k}[\hat{H}_{T(\delta^{\prime}%
)}^{(j)}[N]\hat{H}^{(k)}[M]-\left(  M\leftrightarrow N\right)  ]|c_{1}%
^{\prime},c_{2}^{\prime},c_{3}^{\prime}\rangle\right) \label{eq:mar20-1}\\
&  =\lim_{\delta,\delta^{\prime}\rightarrow0}\Psi_{\lbrack c_{1},c_{2}%
,c_{3}]_{(i)}}^{f^{(i)}}\left(  |\psi_{i}^{\delta,\delta^{\prime}}\left(
v_{0},c^{\prime},[M,N]\right)  \rangle\right) \nonumber
\end{align}
Without loss of generality, we consider the $i=1$ case.

\textbf{Lemma}: $\forall\ c^{\prime},c,N,M,$
\begin{equation}
\lim_{\delta,\delta^{\prime}\rightarrow0}\Psi_{\lbrack c_{1},c_{2}%
,c_{3}]_{(1)}}^{f^{(1)}}\left(  |\psi_{1}^{\delta,\delta^{\prime}}\left(
v_{0},c^{\prime},[M,N]\right)  \rangle\right)  =:\left(  \Psi_{\lbrack
c_{1},c_{2},c_{3}]_{(1)}}^{f_{v_{0}}^{(1)(3,1)}[M,N]}-\Psi_{\lbrack
c_{1},c_{2},c_{3}]_{(1)}}^{f_{v_{0}}^{(1)(1,3)}[M,N]}\right.  \left.
-\Psi_{\lbrack c_{1},c_{2},c_{3}]_{(1)}}^{f_{v_{0}}^{(1)(1,2)}[M,N]}%
+\Psi_{\lbrack c_{1},c_{2},c_{3}]_{(1)}}^{f_{v_{0}}^{(1)(2,1)}[M,N]}\right)
|c^{\prime}\rangle
\end{equation}
where $\Psi_{\lbrack c_{1},c_{2},c_{3}]_{(1)}}^{f_{v_{0}}^{(1)(i,j)}%
[M,N]}\notin\mathcal{V}_{\mathrm{LMI}}$ are distributions with the vertex
functions $f_{v_{0}}^{(1)(i,j)}[M,N]:\Sigma^{|V(c_{1}\cup c_{2}\cup c_{3}%
)|}\rightarrow%
\mathbb{R}
$ defined as
\begin{equation}
f_{v_{0}}^{(1)(i,j)}[M,N](v_{1},\dots,v_{|V(c_{1}\cup c_{2}\cup c_{3}%
)|})=f^{(1)}(v_{1},\dots,v_{|V(c_{1}\cup c_{2}\cup c_{3})|})
\end{equation}
if the one of the following hold:

\begin{enumerate}
\item[(1)] $\{v_{1},...,v_{|V(c_{1}\cup c_{2}\cup c_{3})|}\}\neq
V(c_{1}^{\prime}\cup c_{2}\cup c_{3})$ for any $(c_{1}^{\prime},c_{2}%
,c_{3})\in\lbrack c_{1},c_{2},c_{3}]_{(1)}$, or

\item[(2)] $\{v_{1},...,v_{|V(c_{1}\cup c_{2}\cup c_{3})|}\}=V(c_{1}^{\prime
}\cup c_{2}\cup c_{3})$ for some $(c_{1}^{\prime},c_{2},c_{3})\in\lbrack
c_{1},c_{2},c_{3}]_{(1)}$ but $v_{0}\notin\{v_{1},...,v_{|V(c_{1}\cup
c_{2}\cup c_{3})|}\}$.
\end{enumerate}

In the case that the complements of (1) and (2) hold, we have
\begin{align}
f_{v_{0}}^{(1)(i,j)}[M,N](v_{1},...,v_{|V(c_{1}\cup c_{2}\cup c_{3})|})  &
=\tfrac{1}{4}\left(  \frac{\hbar}{\mathrm{i}}\right)  ^{2}\lambda^{2}(\vec
{n}_{v}^{c})\label{eq:mar2-1}\\
&  \times\sigma(1;i,j)\left(  \sum_{e\in E(c_{1}\cup c_{2}\cup c_{3})}%
\langle\hat{E}_{i}(L(e))\rangle\dot{e}^{a}(0)\left(  M(v_{0})\partial
_{a}N(v_{0})-N(v_{0})\partial_{a}M(v_{0})\right)  \right) \nonumber\\
&  \times\langle\hat{E}_{j}^{b}(v_{0})\rangle\frac{\partial}{\partial v^{a}%
}f^{(1)}(v_{1},...,v_{|V(c_{1}\cup c_{2}\cup c_{3})|})\nonumber
\end{align}

where
\begin{equation}
\sigma(1;i,j)=\left\{
\begin{array}
[c]{ll}%
0,\qquad\qquad & i=j\\
0, & \text{at least one of }i\text{ or }j\neq1\\
+1, & i\neq j=1\\
-1, & i=1\neq j
\end{array}
\right.
\end{equation}
The proof is exactly analogous to the proof for the continuum limit of
$\hat{H}_{T(\delta)}^{(1)}[N]$ on $\mathcal{V}_{\mathrm{LMI}}$, hence we do
not give further details here.

Thus finally, in the topology that we have put on the space of
(finite-triangulation) operators, the continuum limit of the commutator is as
follows:
\begin{equation}
\lbrack\hat{H}[M],\hat{H}[N]]^{\prime}\left(  \Psi_{\lbrack c_{1},c_{2}%
,c_{3}]_{(1)}}^{f^{\left(  1\right)  }}+\Psi_{\lbrack c_{1},c_{2},c_{3}%
]_{(1)}}^{f^{(2)}}+\Psi_{\lbrack c_{1},c_{2},c_{3}]_{(1)}}^{f^{(3)}}\right)
=\sum_{i\in(1,2,3)}\sum_{j\neq i}\left(  \Psi_{\lbrack c_{1},c_{2}%
,c_{3}]_{(i)}}^{f_{v_{0}}^{(i)(j,i)}[M,N]}-\Psi_{\lbrack c_{1},c_{2}%
,c_{3}]_{(i)}}^{f_{v_{0}}^{(i)(i,j)}[M,N]}\right)  \label{eq:mar2-2}%
\end{equation}
$f_{v_{0}}^{(i)(j,i)}[M,N]$ and $f_{v_{0}}^{(i)(i,j)}[M,N]$ are defined above
in (\ref{eq:mar2-1}) for $i=1$. For $i=2,3$ they can be defined similarly. We
reminder the reader that our analysis has been restricted to $\mathcal{K}%
:=V(c_{1}\cup c_{2}\cup c_{3})\cap\ $supp$(N)\cap\ $supp$(M)=\{v_{0}\}$. The
most general case is when this set contains more then one element and in this
case Equation (\ref{eq:mar2-2}) generalizes to
\begin{equation}
\widehat{\lbrack H[M],H[N]]}{}^{\prime}\left(  \Psi_{\lbrack c_{1},c_{2}%
,c_{3}]_{(1)}}^{f^{1}}+\Psi_{\lbrack c_{1},c_{2},c_{3}]_{(1)}}^{f^{(2)}}%
+\Psi_{\lbrack c_{1},c_{2},c_{3}]_{(1)}}^{f^{(3)}}\right)  =\sum
_{v\in\mathcal{K}}\sum_{i\in(1,2,3)}\sum_{j\neq i}\left(  \Psi_{\lbrack
c_{1},c_{2},c_{3}]_{(i)}}^{f_{v}^{(i)(j,i)}[M,N]}-\Psi_{\lbrack c_{1}%
,c_{2},c_{3}]_{(i)}}^{f_{v}^{(i)(i,j)}[M,N]}\right)
\label{commutator-generalcase}%
\end{equation}


Next we quantize $V[\vec{\omega}]$ on $\mathcal{V}_{\mathrm{LMI}}$ and show
that (\ref{eq:mar6-2}) is satisfied.

\subsection{Quantization of $V[\vec{\omega}]$}

\label{diffside}

Recall that%
\begin{equation}
\{H[M],H[N]\}=V[\vec{\omega}]=\int_{\Sigma}\mathrm{d}^{2}x~q^{-1/2}E_{i}%
^{a}E_{i}^{b}\left(  N\partial_{b}M-M\partial_{b}N\right)  F_{ac}^{j}E_{j}%
^{c}\equiv V[q^{-1}\left[  N,M\right]  ]
\end{equation}
Before quantizing this classical functional, we rewrite it in a particular
way, such that the resulting quantum operator on LMI habitat will equal
$[\hat{H}[M],\hat{H}[N]]^{\prime}$:
\begin{align}
V[q^{-1}\left[  N,M\right]  ]  &  =\int_{\Sigma}\mathrm{d}^{2}x~\left(
N\partial_{a}M-M\partial_{a}N\right) \nonumber\\
&  \times\left(  \sum_{i\neq1}\left(  E_{i}^{a}E_{i}^{b}\right)  F_{bc}%
^{1}E_{1}^{c}+\sum_{i\neq2}(E_{i}^{a}E_{i}^{b})F_{bc}^{2}E_{2}^{c}+\sum
_{i\neq3}(E_{i}^{a}E_{i}^{b})F_{bc}^{3}E_{3}^{c}\right)  q^{-1/2}\nonumber\\
&  \equiv V^{1}([N,M])+V^{2}([N,M])+V^{3}([N,M])
\end{align}
where%
\begin{align}
V^{1}([N,M])  &  :=\int_{\Sigma}\mathrm{d}^{2}x~\left(  N\partial
_{a}M-M\partial_{a}N\right)  \left(  \sum_{i\neq1}(E_{i}^{a}E_{i}^{b}%
)F_{bc}^{1}E_{1}^{c}\right)  q^{-1/2}\label{eq:mar2-11}\\
&  =\int_{\Sigma}\mathrm{d}^{2}x~\left(  N\partial_{a}M-M\partial_{a}N\right)
\left(  E_{2}^{a}E_{1}^{c}F_{bc}^{1}E_{2}^{b}+E_{3}^{a}E_{1}^{c}F_{bc}%
^{1}E_{3}^{b}\right)  q^{-1/2}\nonumber\\
&  =\int_{\Sigma}\mathrm{d}^{2}x~\left(  N\partial_{a}M-M\partial_{a}N\right)
\left(  \left(  E_{2}^{a}E_{1}^{c}-E_{1}^{a}E_{2}^{c}\right)  F_{bc}^{1}%
E_{2}^{b}+\left(  E_{3}^{a}E_{1}^{c}-E_{1}^{a}E_{3}^{c}\right)  F_{bc}%
^{1}E_{3}^{b}\right)  q^{-1/2}\nonumber
\end{align}
where we have subtracted classically trivial terms like $E_{1}^{a}E_{2}%
^{c}F_{bc}^{1}E_{2}^{b}$ which will give a non-trivial contribution in quantum
theory (these terms upon quantization are higher order in $\hbar$ whence there
is no contradiction). $V^{2}([N,M])$ and $V^{3}([N,M])$ are defined similarly,
and involve terms containing $F^{2}$ and $F^{3}$ respectively. We will
quantize $V[q^{-1}\left[  N,M\right]  ]$ as a sum of quantum operators
corresponding to $V^{i}|_{i=\{1,2,3\}}$ respectively.

\subsubsection{Quantization of $V^{1}([N,M])$}

Before presenting the quantization of $V^{1}([N,M])$ in detail, we try to
explain the underlying ideas. Spiritually the quantization is similar to the
quantization of $\hat{H}_{T(\delta)}^{(i)}[N]$, but there are some
differences. Consider a graph $\gamma$ and a triangulation $T(\gamma,\delta)$
adapted to $\gamma$ such that every vertex $v\in V(\gamma),$ whose valence is
$N$, is contained in $N$ \textquotedblleft rectangles\textquotedblright\ whose
area is $\delta^{2}$ with the area of overlap region being $O(\delta^{3})$.

From (\ref{eq:mar2-11}) we see that $V^{1}([N,M])$ is the sum of two terms.
Let us focus on one of them which involves $F_{bc}^{1}E_{2}^{b}$:%
\begin{align}
V^{1}([N,M]) &  =V^{1,2}([N,M])+V^{1,3}([N,M])\nonumber\\
&  :=\int_{\Sigma}\mathrm{d}^{2}x~(N\partial_{a}M-M\partial_{a}N)\left(
\left(  E_{2}^{a}E_{1}^{c}-E_{1}^{a}E_{2}^{c}\right)  F_{bc}^{1}E_{2}%
^{b}\right)  q^{-1/2}+\text{term containing $F_{bc}^{1}E_{3}^{b}$}%
\end{align}
We will quantize $V^{1,2}([N,M])$ as a product of elementary operators with
the following quantization scheme:
\begin{align*}
E_{i=1,2}^{a}\left(  N\partial_{a}M-M\partial_{a}N\right)  \hspace*{0.3in} &
\rightarrow\hspace*{0.3in}\text{flux}\\
E_{j=1,2}^{c}q^{-1/4}|_{\epsilon}\hspace*{0.3in} &  \rightarrow\hspace
*{0.3in}\text{quantum shift}\ =:V_{j=1,2}^{c}(x)\\
E_{j=1,2}^{c}q^{-1/4}(x)|_{\epsilon}F_{bc}^{1}E_{2}^{b}\hspace*{0.3in} &
\rightarrow\hspace*{0.3in}\left.
\begin{array}
[c]{c}%
\text{holonomy around a loop generated by $V_{j=1,2}^{c}$ and }\\
\text{charged with eigenvalue of flux associated to $E_{2}^{b}$}%
\end{array}
\right.  \\
q_{\epsilon}^{-1/4}\hspace*{0.3in} &  \rightarrow\hspace*{0.3in}\text{quantize
separately}%
\end{align*}
The regularized ``flux" quantization of $E_{i=1,2}^{a}\left(  N\partial_{a}M-M\partial_{a}N\right)$ is defined as follows.
\begin{equation}
\begin{array}{lll}
\widehat{E_{i=1,2}^{a}\left(  N\partial_{a}M-M\partial_{a}N\right)}(v)\vert_{\delta,\delta^{\prime}}\vert c\rangle\ :=\\
\vspace*{0.1in}
\hspace*{1.0in}\sum_{e\in E(\gamma)|b(e)=v}\frac{\hat{E}_{2}(L_{e}^{\delta
})\left(  N(v)\left(  M(v+\delta^{\prime}\dot{e}(0))-M(v)\right)  -\left(
N\leftrightarrow M\right)  \right)  }{|L_{e}^{\delta}\vert\delta^{\prime}}
\end{array}
\end{equation}
Now we follow essentially the same strategy used in quantizing $\hat
{H}_{T(\delta)}^{(i)}[N]$. However, there is one key technical difference in
the construction of the quantum shift as compared to the quantum shift used in
$\hat{H}_{T(\delta)}^{(i)}[N]$. Recall that the definition of the quantum
shift $\hat{V}_{j}^{c}(v)|_{\epsilon}$ at a given point $v$ depended on the
regularization of $\hat{E}_{j}^{a}(v)$ and $q_{\epsilon}^{-1/4}(v)$
separately. The regularization required $B(v,\epsilon)$ which was used as a
smearing object for regularizing the distribution $\hat{E}_{j}^{a}(v)$ and
which also went in the construction of $q_{\epsilon}^{-1/4}(v)$ (as shown in
Appendix \ref{A1}). These regularizations gave rise to
\begin{equation}
\hat{E}_{j}^{a}(v)=\frac{1}{\epsilon}\hat{O}_{1j}^{a}(v),\qquad q_{\epsilon
}^{-1/4}(v)=\epsilon\hat{O}_{2}(v,\epsilon)
\end{equation}
where $\hat{O}_{1},\ \hat{O}_{2}$ are densely-defined operators on
$\mathcal{H}_{\mathrm{kin}},$ in contrast to being operator-valued
distributions, and $\hat{O}_{2}$ implicitly depends on $\epsilon$. This
construction implied that the quantum shift $\hat{V}_{j}^{a}(v)=\hat{O}%
_{1j}^{a}(v)\hat{O}_{2}(v,\epsilon)$ was (explicitly) independent of
$\epsilon$. However, for defining the quantum shift in $\hat{V}([N,M])$ we use
a different regularization, where the ball $B(v,\epsilon)$ used for smearing
$\hat{E}_{j}^{a}$ is four times as large as the ball used in constructing
$q_{\epsilon}^{-1/4}$. This would in turn imply that
\begin{equation}
\hat{V}_{j}^{a}(v)=\tfrac{1}{4}\hat{O}_{1j}^{a}(v)\hat{O}_{2}(v,\epsilon)
\end{equation}
The $\frac{1}{4}$ factor will account for the overall $\frac{1}{4}$ factor
that we obtained on the LHS.

We are now ready to put all the pieces together. Given a graph $\gamma$ and a
triangulation $T(\gamma,\delta)$ adapted to $\gamma,$ a quantization of
$V^{1,2}_{T(\delta),\delta^{\prime}}([N,M])$ is given by
\begin{align}
\hat{V}^{1,2}_{T(\delta),\delta^{\prime}}([N,M])|c^{\prime}\rangle &  =-\frac{\hbar^{2}}{4\mathrm{i}%
}\left[  \left(  \sum_{v\in V(\gamma)}\sum_{\triangle_{v}|v\in V(\gamma
)}|\triangle_{v}|\sum_{e\in E(\gamma)|b(e)=v}\frac{\hat{E}_{2}(L_{e}^{\delta
})\left(  N(v)\left(  M(v+\delta^{\prime}\dot{e}(0))-M(v)\right)  -\left(
N\leftrightarrow M\right)  \right)  }{|L_{e}^{\delta}|\delta^{\prime}}\right.
\right. \nonumber\\
&  \qquad\qquad\qquad\left.  \sum_{e^{\prime}\in E(\gamma)|b(e^{\prime}%
)=v}\frac{1}{|L_{e^{\prime}}^{\epsilon}|\delta^{2}}\left( (\hat{h}^{1}
_{\alpha(e^{\prime},\langle\hat{E}_{1}(v)\rangle)})^{n_{e^{\prime}}^{2}%
}-1\right)  \hat{q}_{\epsilon}^{-1/4}(v)\hat{O}_{2}(v,\epsilon)\right) \nonumber\\
&  \qquad\qquad-\left(  \sum_{v\in V(\gamma)}\sum_{\triangle_{v}|v\in
V(\gamma)}|\triangle_{v}|\sum_{e\in E(\gamma)|b(e)=v}\frac{\hat{E}_{1}%
(L_{e}^{\delta})\left(  N(v)\left(  M(v+\delta^{\prime}\dot{e}(0))-M(v)\right)
-\left(  N\leftrightarrow M\right)  \right)  }{|L_{e}^{\delta}|\delta^{\prime
}}\right. \nonumber\\
&  \qquad\qquad\qquad\left.  \left.  \sum_{e^{\prime}\in E(\gamma
)|b(e^{\prime})=v}\frac{1}{|L_{e^{\prime}}^{\epsilon}|\delta^{2}}\left(
(\hat{h}^{1}_{\alpha(e^{\prime},\langle\hat{E}_{2}(v)\rangle)})^{n_{e^{\prime}}^{2}%
}-1\right)  \hat{q}_{\epsilon}^{-1/4}(v)\hat{O}_{2}(v,\epsilon)\right)  \right]
|c^{\prime}\rangle
\end{align}
where
\begin{enumerate}
 \item[(1)] $L_{e}^{\delta}$ is a (co-dimension one) surface transversal to
$e$, intersecting in a point which is in the coordinate neighborhood of $v$
and the length of $L_{e}^{\delta}$ is $\delta$.

\item[(2)] $\alpha(e^{\prime},\langle\hat{E}_{1}\rangle(v))$ is the loop starting at $v$, spanned by a straight-line arc along the
$\hat{E}_{1}^{a}(v)$,\textbf{ }and a segment along $e^{\prime}$ such that the
area of the loop is $\delta^{2}$.

\item[(3)] The factor of $|L_{e^{\prime}}^{\epsilon}|$ in the denominator
comes from the fact the quantization of $F_{ab}^{i}E_{j}^{b}$ (see the Section
\ref{HatFiniteTriangulation}) along an edge $e^{\prime}$ by \textquotedblleft
charging\textquotedblright\ the loop along with holonomy of $A^{i}$ is defined
by the eigenvalue of flux $E_{j}(L_{e^{\prime}}^{\epsilon}),$ which requires
dividing the resulting operator by $|L_{e^{\prime}}^{\epsilon}|$. Note that on
choosing $|L_{e^{\prime}}^{\epsilon}|~=\epsilon$, this factor cancels with the
factor of $\epsilon$ present in the quantization of $\hat{q}^{-1/4}$.

\item[(4)] The factor of $\frac{1}{4}$ in front is due to the quantum shift
being $\frac{1}{4}\hat{V}^{a}$ and the factor of ($-1$) is due to the
classical object being $E_{i}^{c}F_{bc}^{1}E_{2}^{b}=-E_{i}^{b}F_{bc}^{1}%
E_{2}^{c},$ the latter of which we actually quantize.

\item[(5)] The two factors of $\hbar$ come from the quantum shift $\langle
\hat{E}_{i}^{a}\rangle$, and the flux eigenvalue under which the holonomy
around the loop is charged (this is the same convention we used when
quantizing the Hamiltonian constraint at finite triangulation).\footnote{Note:
Our convention is always that $\langle\hat{E}_{i}^{a}\rangle$ is without a
factor of $\hbar$.}

\item[(6)] The factor of i$^{-1}$ comes from expressing the curvature in terms
of a loop holonomy.
\end{enumerate}

We now follow the same steps that we followed in Section
\ref{HatFiniteTriangulation}, and replace the sum over holonomies by a
product. The resulting final operator at finite triangulation is
\begin{align}
\hat{V}_{T(\delta),\delta^{\prime}}^{1,2}([N,M])|c^{\prime}\rangle &
=\frac{\mathrm{i}\hbar^{2}}{4\delta\delta^{\prime}}\left[  \left(  \sum_{v\in
V(\gamma)}\sum_{e\in E(\gamma)|b(e)=v}\hat{E}_{2}(L_{e}^{\delta})\left(
N(v)\left(  M(v+\delta^{\prime}\dot{e}(0))-M(v)\right)  -\left(
N\leftrightarrow M\right)  \right)  \right.  \right.  \nonumber\\
&  \qquad\qquad\qquad\times\left.  \left[  \prod_{e^{\prime}\in E(\gamma
)|b(e^{\prime})=v}\hat{h}_{\alpha(e^{\prime},\langle\hat{E}_{1}^{a}%
(v)\rangle)}^{n_{e^{\prime}}^{2}}-1\right]  \hat{q}_{\epsilon}^{-1/4}%
(v)\hat{O}_{2}(v)\right)  \nonumber\\
&  \qquad\qquad-\left(  \sum_{v\in V(\gamma)}\sum_{e\in E(\gamma)|b(e)=v}%
\hat{E}_{1}(L_{e}^{\delta})\left(  N(v)\left(  M(v+\delta^{\prime}\dot
{e}(0))-M(v)\right)  -\left(  N\leftrightarrow M\right)  \right)  \right.
\nonumber\\
&  \qquad\qquad\qquad\times\left.  \left.  \left[  \prod_{e^{\prime}\in
E(\gamma)|b(e^{\prime})=v}\hat{h}_{\alpha(e^{\prime},\langle\hat{E}_{2}%
^{a}(v)\rangle)}^{n_{e^{\prime}}^{2}}-1\right]  \hat{q}_{\epsilon}%
^{-1/4}(v)\hat{O}_{2}(v)\right)  \right]  |c^{\prime}\rangle
\end{align}
which finally yields the following linear combination of charge network
states
\begin{align}
\hat{V}_{T(\delta),\delta^{\prime}}^{1,2}([N,M])|c^{\prime}\rangle &
=\frac{\mathrm{i}\hbar^{3}}{4\delta\delta^{\prime}}\frac{1}{\hbar^{2}}\left[
\left(  \sum_{v\in V(\gamma)}\lambda^{2}(\vec{n}_{v}^{c})\sum_{e\in
E(\gamma)|b(e)=v}\langle{E_{2}(L_{e}^{\delta})\rangle\left(  N(v)\left(
M(v+\delta^{\prime}\dot{e}(0))-M(v)\right)  -\left(  N\leftrightarrow M\right)
\right)  }\right.  \right.  \nonumber\\
&  \hspace*{1in}\times\left.  \left[  |c_{1}^{\prime}\cup\alpha_{v}^{\delta
}(\langle\hat{E}_{1}(v)\rangle,n^{2}),c_{2}^{\prime},c_{3}^{\prime}%
\rangle-|c^{\prime}\rangle\right]  \right)  \nonumber\\
&  \qquad\qquad\qquad-\left(  \sum_{v\in V(\gamma)}\lambda^{2}(\vec{n}_{v}%
^{c})\sum_{e\in E(\gamma)|b(e)=v}\langle{\hat{E}_{1}(L_{e}^{\delta}%
)\rangle\left(  N(v)\left(  M(v+\delta^{\prime}\dot{e}(0))-M(v)\right)
-\left(  N\leftrightarrow M\right)  \right)  }\right.  \nonumber\\
&  \hspace*{1in}\qquad\times\left.  \left.  \left[  |c_{1}^{\prime}\cup
\alpha_{v}^{\delta}(\langle\hat{E}_{2}(v)\rangle,n^{2}),c_{2}^{\prime}%
,c_{3}^{\prime}\rangle-|c^{\prime}\rangle\right]  \right)  \right]
\end{align}
$\hat{V}_{T(\delta),\delta^{\prime}}^{1,3}([N,M])$ can be defined analogously.
Whence finally it is rather easy to see that,
\begin{align}
\hat{V}_{T(\delta),\delta^{\prime}}^{1}([N,M])|c^{\prime}\rangle &
=-\frac{\hbar}{\mathrm{i}}\frac{1}{4\delta\delta^{\prime}}\sum_{v\in
V(\gamma)}N(v)\lambda^{2}(\vec{n}_{v}^{c})\sum_{e\in E(c_{1}^{\prime}\cup
c_{2}^{\prime}\cup c_{3}^{\prime})|b(e)=v}\left(  M(v+\delta^{\prime}\dot
{e}(0))-M(v)\right)  \nonumber\\
&  \left[  \left(  \langle\hat{E}_{3}(L_{e})\rangle\left(  |c_{1}^{\prime}%
\cup\alpha_{v}^{\delta}(\langle\hat{E}_{1}\rangle(v),n_{c_{3}^{\prime}}%
),c_{2}^{\prime},c_{3}^{\prime}\rangle-|c^{\prime}\rangle\right)  -\langle
\hat{E}_{1}(L_{e})\rangle\left(  |c_{1}^{\prime}\cup\alpha_{v}^{\delta
}(\langle\hat{E}_{3}\rangle(v),n_{c_{3}^{\prime}}),c_{2}^{\prime}%
,c_{3}^{\prime}\rangle-|c^{\prime}\rangle\right)  \right)  \right.
\nonumber\\
&  -\left.  \left(  \langle\hat{E}_{1}(L_{e})\rangle\left(  |c_{1}^{\prime
}\cup\alpha_{v}^{\delta}(\langle\hat{E}_{2}\rangle(v),n_{c_{2}^{\prime}%
}),c_{2}^{\prime},c_{3}^{\prime}\rangle-|c^{\prime}\rangle\right)
-\langle\hat{E}_{2}(L_{e})\rangle\left(  |c_{1}^{\prime}\cup\alpha_{v}%
^{\delta}(\langle\hat{E}_{2}\rangle(v),n_{c_{2}^{\prime}}),c_{2}^{\prime
},c_{3}^{\prime}\rangle-|c^{\prime}\rangle\right)  \right)  \right]
\nonumber\\
&  -\left(  N\leftrightarrow M\right)  \label{mar9-1}%
\end{align}
Without loss of generality let us assume that the only vertex in
$V(c_{1}^{\prime}\cup c_{2}^{\prime}\cup c_{3}^{\prime})$ which is contained
in the supports of both $N,M$ is $v_{0}$, so that then\footnote{If $N$ and $M$
have support containing different vertices of the underlying state then, it is
easy to see that the operator vanishes at finite triangulation $\forall
\ \delta^{\prime}$ whence its continuum limit will vanish on $\mathcal{V}%
_{\mathrm{LMI}}$. Whence in this case, the equality of RHS and LHS of
(\ref{eq:mar6-2}) is obvious.}
\begin{align}
&  \hat{V}_{T(\delta),\delta^{\prime}}^{1}([N,M])|c^{\prime}\rangle
\label{eq:mar9-2}\\
&  =-\frac{\hbar}{\mathrm{i}}\frac{1}{4\delta\delta^{\prime}}N(v_{0}%
)\lambda^{2}(\vec{n}_{v}^{c})\sum_{e\in E(c_{1}^{\prime}\cup c_{2}^{\prime
}\cup c_{3}^{\prime})|b(e)=v_{0}}\left(  M(v_{0}+\delta^{\prime}\dot
{e}(0))-M(v_{0})\right)  \nonumber\\
&  \times\left[  \left(  \langle\hat{E}_{3}(L_{e})\rangle\left(
|c_{1}^{\prime}\cup\alpha_{v_{0}}^{\delta}(\langle\hat{E}_{1}\rangle
(v_{0}),n_{c_{3}^{\prime}}),c_{2}^{\prime},c_{3}^{\prime}\rangle-|c^{\prime
}\rangle\right)  -\langle\hat{E}_{1}(L_{e})\rangle\left(  |c_{1}^{\prime}%
\cup\alpha_{v_{0}}^{\delta}(\langle\hat{E}_{3}\rangle(v_{0}),n_{c_{3}^{\prime
}}),c_{2}^{\prime},c_{3}^{\prime}\rangle-|c^{\prime}\rangle\right)  \right)
\right.  \nonumber\\
&  -\left.  \left(  \langle\hat{E}_{1}(L_{e})\rangle\left(  |c_{1}^{\prime
}\cup\alpha_{v_{0}}^{\delta}(\langle\hat{E}_{2}\rangle(v_{0}),n_{c_{2}%
^{\prime}}),c_{2}^{\prime},c_{3}^{\prime}\rangle-|c^{\prime}\rangle\right)
-\langle\hat{E}_{2}(L_{e})\rangle\left(  |c_{1}^{\prime}\cup\alpha_{v_{0}%
}^{\delta}(\langle\hat{E}_{2}\rangle(v_{0}),n_{c_{2}^{\prime}}),c_{2}^{\prime
},c_{3}^{\prime}\rangle-|c^{\prime}\rangle\right)  \right)  \right]
\nonumber\\
&  -\left(  N\leftrightarrow M\right)  \nonumber
\end{align}
Now notice that (using (\ref{eq:termA}) and (\ref{eq:mar9-2})),
\begin{equation}
\left(  -\mathrm{i}\hbar\right)  \hat{V}_{T(\delta),\delta^{\prime}}%
^{1}([N,M])|c^{\prime}\rangle=|\psi_{1}^{\delta,\delta^{\prime}}\left(
v_{0},c^{\prime},[M,N]\right)  \rangle\label{eq:mar9-3}%
\end{equation}

The remaining $\hat{V}_{T(\delta),\delta^{\prime}}^{i,j}([N,M])$ operators are
defined analogously. The sum of all these operators constitutes a quantization
of the RHS $V\left[  q^{-1}[N,M]\right]  $ on $\mathcal{H}_{\mathrm{kin}}$:%

\begin{equation}
\left(  -\mathrm{i}\hbar\right)  \hat{V}_{T(\delta),\delta^{\prime}%
}([N,M])|c^{\prime}\rangle=\left(  -\mathrm{i}\hbar\right)  \sum_{i}\hat
{V}_{T(\delta),\delta^{\prime}}^{i}([N,M])|c^{\prime}\rangle=\sum_{i}|\psi
_{i}^{\delta,\delta^{\prime}}\left(  v_{0},c^{\prime},[M,N]\right)
\rangle\label{eq:mar9-3'}%
\end{equation}

We are now ready to state our main result:

\textbf{Theorem}: $\widehat{[H[M],H[N]]}^{\prime}\Psi=\left(  -\mathrm{i}%
\hbar\right)  \hat{V}([N,M])^{\prime}\Psi$, $\forall\ \Psi\in\mathcal{V}%
_{\mathrm{LMI}}$

\textbf{Proof}: For any $\Psi_{\lbrack c_{1},c_{2},c_{3}]_{(i)}}^{f^{(i)}}$,
the LHS is given by (\ref{eq:mar2-2}) and it is a result of the continuum
limit of the net given in (\ref{eq:mar20-1}). As the RHS is the continuum
limit of the net given in (\ref{eq:mar9-3'}), which is precisely the same net
as in (\ref{eq:mar20-1}), the result follows.

Thus there exists a quantization of $\{H[M],H[N]\}$ as an operator from
$\mathcal{V}_{\mathrm{LMI}}$ to $\mathrm{Cyl}^{\ast}$ (as a limit point of a
net of finite-triangulation operators in a particular topology which is
similar to the weak *-topology) which equals a quantization of $V\left[
q^{-1}[N,M]\right]  $ as an operator from $\mathcal{V}_{\mathrm{LMI}}$ to
$\mathrm{Cyl}^{\ast}$. This, in our opinion, demonstrates that the
quantization of Hamiltonian constraint we have proposed in this paper has the
right structural properties such that it can give rise to a faithful
representation of Dirac algebra.

\section{Conclusions}

A satisfactory definition of quantum dynamics in canonical LQG is still
missing. Even in the Euclidean sector of the theory, the progress is rather
fragmented and is mainly achieved in a variety of mini-superspace models. To
the best of our knowledge the only midi-superspace models where a completely
satisfactory definition of the quantum constraints in generally covariant loop
quantized field theories is known are, in fact, non-gravitational theories,
namely two-dimensional PFT and the HK model. However, these models miss
perhaps the most interesting aspect of the constraint algebra of canonical
gravity: That it is a Lie algebroid instead of being a Lie algebra
\cite{weinstein}; that is, the fact that the Poisson bracket of two
Hamiltonian constraints involves phase space-dependent structure functions. In
this paper, we proposed a toy model which has the same Dirac algebra as
Euclidean three-dimensional canonical gravity but here, in a certain sense,
non-linear aspects of gravity are absent. We focused on a part of the
constraint algebra and finally derived a quantization of continuum Hamiltonian
constraint, which has the potential to give rise to a faithful representation
of the Dirac algebra. To the best of our knowledge, our work along with
\cite{madtom} is a first attempt towards a quantum realization of off-shell
closure of the Dirac algebra within the LQG framework.

As the theory is topological Abelian gauge theory, one might perceive this
model as being too simplistic. However this is not quite true. As our main
focus has been on understanding the off-shell closure condition in the quantum
theory (and as the theory is topological only on shell), we work in a genuine
field-theoretic context. It is quite straightforward to generalize our results
to $3+1$ dimensional $\mathrm{U}(1)^{3}$ theory (which is precisely the model
studied in \cite{madtom}) and, in fact, some of the technical annoyance that
we face in two dimensions (e.g., the presence of irrelevant vertices) can be
evaded in three spatial dimensions. On the other hand, in $2+1$ dimensions the
physical spectrum of this topological gauge theory is well understood (as
states supported on the moduli space of flat $\mathrm{U}(1)^{3}$ connections),
and a complete set of Dirac observables is known. Hence one could investigate
the consistency of the quantum theory defined here by investigating issues
associated to the kernel of constraints, and the representation of quantum
Dirac observables.

We now recap the most salient aspects of our constructions before highlighting
the key open issues and some of the unsatisfactory aspects of our work.

The usual construction of composite operators in LQG is via some classical
polynomial function of holonomies and fluxes. However, as we were motivated to
look for a quantization of Hamiltonian constraint which mimicked a certain
discrete approximant of the classical geometric action involving phase
space-dependent diffeomorphisms, our quantization choices involved quantizing
$F_{ab}^{i}E_{j}^{b}$ as a holonomy operator (with the holonomy being in a
state-dependent representation) and quantizing the remaining triad $E_{k}^{a}$
(or more precisely $Nq^{-1/4}E_{k}^{a}$) as a quantum shift which generated
the loop underlying the holonomy associated to $F_{ab}^{i}E_{j}^{b}$. Thus
these choices have a \textquotedblleft spiritual\textquotedblright\ similarity
to the `$\bar{\mu}$-scheme' which led to a physically viable quantization of
the Hamiltonian constraint in LQC \cite{aplimproved}.\footnote{However notice
that the analogy is only superficial. In the $\bar{\mu}$-scheme, the triad
dependence underlying the holonomy operator does not come from the
$q^{-1/2}E\wedge E$ term. We thank Martin Bojowald for pointing this out to
us.}

The Hamiltonian constraint at finite triangulation $\hat{H}_{T(\delta)}[N]$
created very specific types of vertices that we called extraordinary (EO)
vertices. As the underlying gauge group is $\mathrm{U}(1)^{3}$, these vertices
are \emph{always} degenerate (i.e., they are in the kernel of the inverse
volume operator). This aspect of the construction bares some similarity to
Thiemann's Hamiltonian, in which the newly created vertices are also
degenerate.\footnote{We should note here that the similarity is only
restricted to $\mathrm{U}(1)^{3}$. The superficial extension of our analysis
to $\mathrm{SU}(2)$ suggests that the EO vertices created in that case will
not be degenerate.} Whence at first sight it seemed as if one faces the same
problem as Thiemann's Hamiltonian does, in that the action of a second
successive Hamiltonian will have no non-trivial action at EO vertices and the
objections raised in \cite{lm2} will remain true in our case. However, this
expected triviality overlooked a key fact about the EO vertices: Their
\textquotedblleft location\textquotedblright\ with respect to the original
charge network was state-dependent, which was in turn due to the fact that
these EO vertices were created along straight-line arcs of the quantum shift.
This fact along with a classical Poisson bracket computation suggested a
plausible modification of Hamiltonian constraint when it acted on EO vertices.
The modification was such that its precise interpretation was a
\textquotedblleft non-local\textquotedblright\ action (i.e., these actions
were generated by operators which involved holonomies around finite loops) not
only on the EO vertex but on a subgraph containing the EO pair. However, the
detailed understanding of such terms in the context of the underlying
continuum classical theory is not clear to us and should be investigated
further.\cite{martineft}

The action of the finite-triangulation Hamiltonian constraint $\hat
{H}_{T(\delta)}[N]$ on a charge network $|c\rangle$ is remarkably different
than the action considered so far in LQG, and it is worth summarizing its
three main features:

\begin{enumerate}
\item[(1)] The deformations of the graph underlying $c$ are state-dependent
(as dictated by the quantum shift).

\item[(2)] The $\mathrm{U}(1)^{3}$ edge labels in $c$ change, with the change
itself being state-dependent.

\item[(3)] A certain special class of degenerate vertices move under the
action of $\hat{H}_{T(\delta)}[N]$.
\end{enumerate}

All three of these conditions hint at a rather rich structure of quantum
dynamics in the model that could be of interest to discrete approaches
inspired by canonical loop quantum gravity.

We then constructed a habitat which was designed in such a way that the
continuum limit of the finite-triangulation Hamiltonian constraint could be
taken in the topology induced by a family of seminorms. The continuum limit
Hamiltonian constraint $\hat{H}[N]^{\prime}$ does not preserve this habitat
and can only be interpreted as a linear operator from $\mathcal{V}%
_{\mathrm{LMI}}$ to $\mathrm{Cyl}^{\ast}$.
Although this implies that the commutator of $\hat{H}[N]^{\prime}$ with itself
is ill-defined, it turns out that the limit of finite-triangulation
commutators is still well-defined on $\mathcal{V}_{\mathrm{LMI}}$ and is
non-vanishing. We finally showed that there exists a quantization of the RHS,
\emph{which is not quantized as an ordinary diffeomorphism with
triad-dependent shift, but requires a specific operator ordering.} This
quantization matches with the continuum quantum commutator which is the LHS of
the off-shell closure relation.

We now come to the open issues and certain related unsatisfactory aspects of
our work. Our entire construction is based upon decomposing the Hamiltonian
constraint into three pieces involving $F^{1}$, $F^{2}$, and $F^{3}$
respectively. Although each of these pieces is gauge-invariant in the present
model, this is not true in the case of $\mathrm{SU}(2)$. Thus more careful
analysis is needed to extend our proposal to a quantization of the Hamiltonian
constraint in the $\mathrm{SU}(2)$ case.

The second issue lies in the choice of the habitat. Our experience of how to
construct habitats on which higher density operators in LQG admit a continuum
limit is rather limited. After the seminal work done in \cite{lm1}, where a
habitat was constructed in which Thiemann's (regularized) Hamiltonian
constraint admitted a continuum limit (once again in a seminorm topology
induced by the habitat states and states in $\mathcal{H}_{\mathrm{kin}}$), the
only places where habitats have been utilized have been in
\cite{amhamcons,amdiff}. In these two examples habitats even turned out to be
a physically-appropriate home for the quantum constraints, as the kernel
(which was known via other methods) was a subspace of the habitat. However, in
our case, the nature of the regularized constraints makes it rather difficult
to construct a suitable habitat on which not only do the regularized
constraints admit a continuum limit but also that all the details of the
regularized constraint operators remain intact when we take continuum limit
(for example, the change in edge-labels induced by the action of $\hat
{H}_{T(\delta)}[N]$ go amiss when we consider the dual action on the habitat).
It is important to note that we have constructed a habitat only with two goals
in mind:

\begin{enumerate}
\item[(1)] $\hat{H}_{T(\delta)}^{i}[N]$ admits a continuum limit; and

\item[(2)] $[\hat{H}_{T}[N],\hat{H}_{T^{\prime}}[M]]$ admits a continuum limit.
\end{enumerate}

$\mathcal{V}_{\mathrm{LMI}}$ need not be a physically relevant habitat as

\begin{enumerate}
\item[(a)] $H[N]^{\prime}$ does not preserve $\mathcal{V}_{\mathrm{LMI}}$.

\item[(b)] We do not know if the states in the moduli space of flat
$\mathrm{U}(1)^{3}$ connections are included in $\mathcal{V}_{\mathrm{LMI}}%
$\footnote{Naively speaking, the habitat states with constant vertex functions
can indeed be thought of as states with support only on flat connections.
However this issue has not been investigated in detail.}

\item[(c)] The classical theory is a completely integrable system, but we do
not know if the habitat admits a representation of quantum observables and is
there a precise sense in which these observables commute with the quantum constraints.
\end{enumerate}

Detailed investigations of all these three issues could be key in constructing
physically interesting habitats.

There is an alternate viewpoint one could adhere to. Let us assume that we can
extend our constructions to three spatial dimensions, and appropriately
density-weighted constraints that satisfy the off-shell closure condition on
some habitat. As far as the $\mathrm{U}(1)^{3}$ theory is concerned, since the
inverse volume operator is just a multiple of the identity operator on charge
network states, the higher density-weighted operator induces an unambiguous
definition of a density one operator. It is quite plausible that in the
Uniform-Rovelli-Smolin topology (or some suitable generalization thereof),
this density one operator converges to a densely defined operator on
$\mathcal{H}_{\mathrm{kin}}$. Then the requirement of off-shell closure would
only be used to select, out of an infinitude of possible choices, a density
one quantum Hamiltonian constraint, and one could just choose to work on
$\mathcal{H}_{\mathrm{kin}}$.\footnote{We do not believe these ideas can work
in two spatial dimensions, due to the presence of irrelevant vertices which
would be absent in three dimensions.}

A faithful representation of the Dirac algebra entails not only the second
equation in Eq.(\ref{eq:qdirac}), but all three of them. In particular, the
construction of a finite-triangulation Hamiltonian constraint operator
involved certain background structure, which survived in the continuum limit.
The definition of the quantum shift involved a certain regularization scheme
and it is far from clear if this scheme is diffeomorphism-covariant. The
notion of extraordinary vertices, which are required to lie inside certain
balls around non-degenerate vertices is certainly a non-covariant notion, as
one can map an EO vertex with respect to some charge network $c$ into a WEO
vertex with respect to the same charge network via some diffeomorphism. Hence
in light of the non-covariant structures which have gone into the construction
of the (continuum) Hamilonian, it is far from clear if the third equation in
(\ref{eq:qdirac}) is satisfied. We will come back to these issues in
\cite{hat2}.

\section*{Acknowledgements}

We are grateful to the Penn State gravity group, especially Abhay Ashtekar,
Ivan Agullo, Martin Bojowald, Will Nelson and Artur Tsobanjan for useful
discussions and constant encouragement. We are indebted to Miguel Campliglia
for collaboration during the initial stages, countless discussions on the
subject, for explaining to us the weak coupling interpretation of
$\mathrm{U}(1)^{3}$ theory, and for making several prescient remarks during
the construction of the habitat which clarified many of our misunderstandings.
We are especially grateful to Madhavan Varadarajan for many enlightening
discussions and for sharing the much more sophisticated results obtained by
him in collaboration with CT prior to publication. We thank Martin Bojowald,
Miguel Campiglia and Will Nelson for their comments on the manuscript. The
work of AH and AL is supported by NSF grant PHY-0854743 and by the Eberly
Endowment fund. The work of CT is supported by NSF grant PHY-0748336 and the
Mebus Fellowship, as well as the generosity of the Raman Research Institute,
where a portion of this work was conducted.

\section*{Appendices}

\appendix

\section{Volume and Inverse Volume Operators}

\label{A1}

In this appendix we discuss the construction of a U(1)$^{3}$ volume operator,
as well as an operator corresponding to $q^{-1/4},$ which are used in the main
body of the paper. We closely follow Thiemann \cite{qsd4}.

\subsection{Volume Operator}

In \cite{qsd4}, an operator-valued distribution corresponding to the
degeneracy vector $E^{i}=\frac{1}{2}\epsilon^{ijk}\eta_{ab}E_{j}^{a}E_{k}^{b}$
is constructed by Thiemann when the gauge group is SU(2). In the case of
U(1)$^{3}$, the construction proceeds analogously and leads to the operator
action
\begin{equation}
\hat{E}^{i}(x)|c\rangle=\tfrac{1}{8}(\kappa\hbar)^{2}\sum_{v\in V(c)}%
\delta^{(2)}(x,v)\sum_{e_{I}\cap e_{I^{\prime}}=\{v\}}\epsilon(e_{I}%
,e_{I^{\prime}})\epsilon^{ijk}n_{I}^{j}n_{I^{\prime}}^{k}|c\rangle
\end{equation}
where $V(c)$ is the vertex set of $c$, and
\begin{equation}
\epsilon(e,e^{\prime}):=\frac{\eta_{ab}\dot{e}^{a}(0)\dot{e}^{\prime b}%
(0)}{\left\vert \eta_{ab}\dot{e}^{a}(0)\dot{e}^{\prime b}(0)\right\vert }%
=\pm1,0.
\end{equation}
The additional factor of $\frac{1}{4}$ comes from evaluating the $\delta
$-functions at endpoints of integration over $t,t^{\prime}$ (we have arranged
all edges as outgoing at vertices). Classically $q=E^{i}E^{i},$ but the
presence of the $\delta$-function requires an additional regularization as an
intermediate step. One can show that the regularized $\hat{q}$ is the square
of an essentially self-adjoint operator, and it is positive semi-definite, so
its square root is well-defined as an operator-valued distribution:
\begin{equation}
\sqrt{\hat{q}(x)}|c\rangle=\tfrac{1}{8}(\kappa\hbar)^{2}\sum_{v\in V(c)}%
\delta^{(2)}(x,v)\sqrt{\left(  \sum_{e_{I}\cap e_{I^{\prime}}=\{v\}}%
\epsilon(e_{I},e_{I^{\prime}})\epsilon^{ijk}n_{I}^{j}n_{I^{\prime}}%
^{k}\right)  ^{2}}|c\rangle
\end{equation}
We can use this to define a regular operator corresponding to the volume of a
region $R\subset\Sigma$:%
\begin{equation}
\hat{V}(R)|c\rangle:=\int_{R}\mathrm{d}^{2}x~\sqrt{\hat{q}(x)}|c\rangle
=\tfrac{1}{8}(\kappa\hbar)^{2}\sum_{v\in V(c)\cap R}\sqrt{\left(  \sum
_{e_{I}\cap e_{I^{\prime}}=\{v\}}\epsilon(e_{I},e_{I^{\prime}})\epsilon
^{ijk}n_{I}^{j}n_{I^{\prime}}^{k}\right)  ^{2}}|c\rangle
\end{equation}
We observe that since $\hat{E}^{i}$ vanishes on states charged in only one
copy of U(1), the volume also vanishes on these states. Moreover, due to the
orientation factor $\epsilon(e_{I},e_{I^{\prime}}),$ $\hat{V}$ vanishes at
vertices at which there are not at least two edges with linearly-independent tangents.

\subsection{A $q^{-1/4}$ Operator}

In this subsection we derive a Thiemann-like classical identity for $q^{-1/4}$
which is then promoted to a regularized operator on $\mathcal{H}%
_{\mathrm{kin}}.$ One can imagine several variants on the following
construction, but here we settle on one that satisfies two properties that we
use in the main text:

\begin{enumerate}
\item[(i)] $\hat{q}^{-1/4}$ vanishes everywhere except at charge network
vertices whose outgoing edges have linearly-independent tangents.

\item[(ii)] $\hat{q}^{-1/4}$ vanishes at vertices , all whose incident edges are charged in a single
U(1)$_{i}$.
\end{enumerate}

We will refer to those charge network vertices at which $\hat{q}^{-1/4}$ is
non-vanishing as \textit{non-trivial}, and those for which $\hat{q}^{-1/4}$
vanishes will be called \textit{degenerate}.

We now proceed with the construction. We begin by noticing that classically,
for $x\in R\subset\Sigma,$%
\begin{equation}
\eta^{ab}\epsilon_{ijk}\{A_{a}^{j}(x),V(R)^{3/8}\}\{A_{b}^{k}(x),V(R)^{3/8}%
\}=2\left(  \tfrac{3}{8}\right)  ^{2}V(R)^{-5/4}E^{i}(x).
\end{equation}
Now if $B(x,\epsilon)$ is a (coordinate) circular ball of coordinate radius
$\epsilon$ centered at $x,$ then
\begin{equation}
\lim_{\epsilon\rightarrow0}\frac{V(x,\epsilon)}{\pi\epsilon^{2}}=\sqrt
{q(x)},\qquad\text{where}\qquad V(x,\epsilon):=\int_{B(x,\epsilon)}%
\mathrm{d}^{2}x~\sqrt{q(x)}.
\end{equation}
Therefore the Poisson bracket identity above allows us to write%
\begin{equation}
q^{-5/8}E^{i}(x)=\lim_{\epsilon\rightarrow0}\epsilon^{5/2}\frac{\pi^{5/4}%
\eta^{ab}\epsilon_{ijk}}{2(3/8)^{2}}\{A_{a}^{j}(x),V(x,\epsilon)^{3/8}%
\}\{A_{b}^{k}(x),V(x,\epsilon)^{3/8}\}
\end{equation}
We can form pure inverse powers of $q$ by taking even powers of this identity.
For instance, $q^{-1/4}=(q^{-5/8}E^{i})^{2},$ which is the quantity we are
interested in. Before such an identity can be quantized on $\mathcal{H}%
_{\mathrm{kin}}$, we must replace $A$ by holonomies. Letting $h_{a}^{i}(x)$ be
a coordinate length $\epsilon$ holonomy along an edge in the $a$-direction
which crosses or terminates at $x,$ we have
\begin{equation}
q^{-5/8}E^{i}(x)=\lim_{\epsilon\rightarrow0}\epsilon^{5/2}\frac{\pi^{5/4}%
\eta^{ab}\epsilon_{ijk}}{2\kappa^{2}(3/8)^{2}}\frac{(h_{a}^{j}(x))^{-1}%
}{\mathrm{i}\epsilon}\{h_{a}^{j}(x),V(x,\epsilon)^{3/8}\}\frac{(h_{b}%
^{k}(x))^{-1}}{\mathrm{i}\epsilon}\{h_{b}^{k}(x),V(x,\epsilon)^{3/8}\}.
\end{equation}
Removing the $\epsilon\rightarrow0$ limit, and making the replacements
$V\rightarrow\hat{V}$ and $\{~,~\}\rightarrow(\mathrm{i}\hbar)^{-1}[~,~],$ we
obtain a well-defined ($\epsilon$-regularized) operator on $\mathcal{H}%
_{\mathrm{kin}}.$ Squaring the resulting operator, we arrive at%
\begin{equation}
\hat{q}_{\epsilon}^{-1/4}=\epsilon\frac{81\pi^{5/2}\eta^{ab}\eta^{cd}%
}{32(\kappa\hbar)^{4}}\sum_{i,j}(h_{a}^{i})^{-1}[h_{a}^{i},\hat{V}^{\frac
{3}{8}}](h_{b}^{j})^{-1}[h_{b}^{j},\hat{V}^{\frac{3}{8}}](h_{c}^{i}%
)^{-1}[h_{c}^{i},\hat{V}^{\frac{3}{8}}](h_{d}^{j})^{-1}[h_{d}^{j},\hat
{V}^{\frac{3}{8}}]. \label{q14}%
\end{equation}
where we have dropped the various arguments for notational clarity. We will
choose the holonomies in (\ref{q14}) based on the state $|c\rangle$ on which
the operator acts. Specifically, given\ a vertex $v\in\gamma\left(  c\right)
$ which is at least bi-valent with linearly-independent tangents, we single
out a pair of linearly-independent edges to define the $x$ and $y$ coordinate
axes of a coordinate system with origin at $v$. We let the holonomies of
(\ref{q14}) lie along these coordinate axes (so that they partially overlap
the edges of $\gamma(c)$) and have endpoints (or beginning points) at $v.$
Then in this coordinate system,%
\begin{equation}
\hat{q}_{\epsilon}^{-1/4}=\epsilon\frac{81\pi^{5/2}\epsilon^{IJ}\epsilon^{KL}%
}{32(\kappa\hbar)^{4}}\sum_{i,j}(h_{I}^{i})^{-1}[h_{I}^{i},\hat{V}^{\frac
{3}{8}}](h_{J}^{j})^{-1}[h_{J}^{j},\hat{V}^{\frac{3}{8}}](h_{K}^{i}%
)^{-1}[h_{K}^{i},\hat{V}^{\frac{3}{8}}](h_{L}^{j})^{-1}[h_{L}^{j},\hat
{V}^{\frac{3}{8}}].
\end{equation}
If there are not two linearly-independent directions defined by tangents of
edges at $v,$ then we pick some orthogonal direction by hand along which to
lay holonomies; we will see shortly that in this case of `linear vertices,'
the operator has trivial action.

Since each action of the volume operator gives an eigenvalue proportional to
$(\kappa\hbar)^{2},$ and $4\times\frac{3}{8}=\frac{3}{2},$ the eigenvalues of
$\hat{q}_{\epsilon}^{-1/4}$ are proportional to $(\kappa\hbar)^{-4}%
((\kappa\hbar)^{2})^{3/2}=(\kappa\hbar)^{-1},$ and we separate this
dimensionful dependence, as well as the $\epsilon$-dependence, from the
dimensionless part of the eigenvalue, writing%
\begin{equation}
\langle c|\hat{q}_{\epsilon}^{-1/4}(v)|c\rangle=:\frac{\epsilon}{\kappa\hbar
}\lambda\left(  \vec{n}_{v}^{c}\right)
\end{equation}
for the particular case where $v$ is a non-trivial vertex of $|c\rangle$, and
$\vec{n}_{v}^{c}$ denotes the collection of charge labels on the edges there.
This operator is completely regular in its action on $\mathcal{H}%
_{\mathrm{kin}},$ so taking $\epsilon\rightarrow0$ one obtains the zero
operator. The strategy in the main text is to combine $\hat{q}_{\epsilon
}^{-1/4}$ with a regularized $\hat{E}_{i}^{a}$ which behaves like
$\epsilon^{-1}$ in the regulating parameter, and hence the combination remains
regular as $\epsilon\rightarrow0.$

To see that the two properties (i) and (ii) are satisfied, observe that
$\hat{q}_{\epsilon}^{-1/4}$ acting on a state $|c\rangle$ at some vertex $v$
is a sum of products, where each factor is of the form%
\begin{equation}
\hat{V}^{\frac{3}{8}}-(h_{I}^{i})^{-1}\hat{V}^{\frac{3}{8}}h_{I}^{i}.
\end{equation}
If there are not two linearly-independent tangents at $v,$ then $\hat
{V}^{\frac{3}{8}}$ vanishes, and since $\hat{V}^{\frac{3}{8}}$ in the term
$(h_{I}^{i})^{-1}\hat{V}^{\frac{3}{8}}h_{I}^{i}$ also sees a linear vertex,
this term vanishes also, whence $\hat{q}_{\epsilon}^{-1/4}$ vanishes.

If $|c_{i}\rangle$ is charged in a single U(1)$_{i},$ again we notices that
each term in $\hat{q}_{\epsilon}^{-1/4}$ includes some factor of the form%
\begin{equation}
\left(  \hat{V}^{\frac{3}{8}}-(h_{I}^{i})^{-1}\hat{V}^{\frac{3}{8}}h_{I}%
^{i}\right)  |c_{i}\rangle,
\end{equation}
and $h_{I}^{i}|c_{i}\rangle$ is also charged in U(1)$_{i}$ only, so $\hat
{V}^{\frac{3}{8}}h_{I}^{i}|c_{i}\rangle=0.$

Before concluding this section, we mention that the overall factor of
$\epsilon$ is not the only possibility (nor is the overall constant); as noted
in \cite{lm1}, one may choose the regions associated with each instance of
$\hat{V}$ independently so as to obtain an arbitrary power of the regulating
parameter (or some combination of several regulating parameters). Here and in
the main text, we have merely followed an economical prescription, where all
regulating parameters scale in the same way.

\section{Characterization of Type B Extraordinary Vertices}

\label{AtypeB}

In this section we classify the type B EO vertices. Recall that if, given a
charge-network state $|c\rangle,$ a quantum shift is such that straight-line
arc associated to it lies along one of the edges, then the corresponding
vertex is called type B (depicted in Figure \ref{HactionB}). The conditions
are a minor modification of the conditions characterizing type A vertices.

We will call an $N_{v}$-valent vertex $v^{\mathrm{E}}$ an extraordinary vertex
of type $(M,j,B)$ \emph{iff} following conditions are satisfied.

\textbf{Set B}:

\begin{enumerate}
\item[(1)] Exactly two edges incident at $v^{\mathrm{E}}$ are analytic
continuations of each other and these two edges are necessarily charged in
more than one copy of $\mathrm{U}(1)^{3}$. The remaining $N_{v}-2$ edges are
colored in only $\mathrm{U}(1)_{M}|_{M\in\{1,2,3\}}$. Let us refer to the two
multi-colored edges as $e_{v^{\mathrm{E}}}^{(1)},e_{v^{\mathrm{E}}}^{(2)}$.

\begin{enumerate}
\item[(1a)] Tangents to all the edges incident at $v^{\mathrm{E}}$ are
parallel or anti-parallel.
\end{enumerate}

\item[(2)] Let us denote the $N_{v}-2$ vertices which are the end points of
the $N_{v}-2$ edges beginning at $v^{\mathrm{E}}$ and are distinct from
$e_{v^{\mathrm{E}}}^{(1)},e_{v^{\mathrm{E}}}^{(2)}$ by the set $\mathcal{S}%
_{v^{\mathrm{E}}}:=\{v_{(1)}^{\mathrm{E}},...v_{(N_{v}-2)}^{\mathrm{E}}%
\}$.\footnote{This set satisfies exactly same conditions that $\mathcal{S}%
_{v^{\mathrm{E}}}$ satisfies in the case of type A vertices.} The valence of
all these vertices is bounded between three and four.

\begin{enumerate}
\item[(2a)] At most two vertices in $\mathcal{S}_{v^{\mathrm{E}}}$ are tri-valent.
\end{enumerate}

\item[(3)] The tri-valent vertices are such that the edges which are not
incident at $v^{\mathrm{E}}$ are analytic extensions of each other and the
four-valent vertices are such that two of the edges which are not incident at
$v^{\mathrm{E}}$ are analytic extensions of each other and fourth edge is the
analytic extension of the edge which is incident at $v^{\mathrm{E}}$.

\begin{enumerate}
\item[(3a)] Any four-valent vertex defined in {(3}) is such that if the four
edges incident on it $(e_{1},e_{2},e_{3},e_{4})$ are such that $e_{1}\circ
e_{2}$ is entire analytic and $e_{3}\circ e_{4}$ is entire analytic, then
$\vec{n}_{e_{1}}=\vec{n}_{e_{2}}$, $\vec{n}_{e_{3}}=\vec{n}_{e_{4}}$.
\end{enumerate}

\item[(4)] Let $e_{v^{\mathrm{E}}}$ be an edge beginning at $v^{\mathrm{E}}$
which ends in a four-valent vertex $f(e_{v^{\mathrm{E}}})$. By {(3)}, there
exists an analytic extension $\tilde{e}_{v^{\mathrm{E}}}$ of $e_{v^{\mathrm{E}%
}}$ in $E(\bar{c})$ beginning at $v^{\mathrm{E}}$. The final vertex
$f(\tilde{e}_{v^{\mathrm{E}}})$ of $\tilde{e}_{v^{\mathrm{E}}}$ is always
tri-valent.\newline Thus restricting attention to analytic extensions of each
of the edges beginning at $v^{\mathrm{E}}$, all such edges end in tri-valent
vertices, and all of these tri-valent vertices are such that the remaining two
edges incident on them are analytic extensions of each other. The set of these
$N_{v}-2$ tri-valent vertices \textquotedblleft associated\textquotedblright%
\ to $v^{\mathrm{E}}$ is $\mathcal{\bar{S}}_{v^{\mathrm{E}}}:=\ \{\bar
{v}_{\left(  1\right)  }^{\mathrm{E}},\dots,\bar{v}_{\left(  N_{v}-2\right)
}^{\mathrm{E}}\}$.\footnote{Note that $\mathcal{S}_{v^{\mathrm{E}}}%
\cap\mathcal{\bar{S}}_{v^{\mathrm{E}}}\ =\ \text{tri-valent vertices in
$\mathcal{S}_{v^{\mathrm{E}}}$}$.\newline}

\item[(5)] Let us denote these (maximally analytic inside $E(\gamma)$) edges
beginning at $v^{\mathrm{E}}$ by $\{\tilde{e}_{v^{\mathrm{E}}}^{(1)}%
,...\tilde{e}_{v^{\mathrm{E}}}^{N_{v}-2}\}$.
Without loss of generality, consider the case when all the edges incident on
$v^{\mathrm{E}}$ except $e_{v^{\mathrm{E}}}^{(1)},e_{v^{\mathrm{E}}}^{(2)}$
are charged in first copy\footnote{In this case we will say that
$v^{\mathrm{E}}$ is of type $(M=1,j\ \in\ \{2,3\},\mathrm{B})$.} Let the
charges on these edges are $\{(n_{\tilde{e}_{v^{\mathrm{E}}}^{(1)}%
},0,0),\text{\'{E}},(n_{\tilde{e}_{v^{\mathrm{E}}}^{(N_{v}-2)}},0,0)\}$.
\newline If $\tilde{e}_{v^{\mathrm{E}}}^{(k)}$ ($k\ \in\ \{1,\acute{E}%
,N_{v}-2\}$) ends in a three valent vertex $f(\tilde{(}e)_{v^{\mathrm{E}}%
}^{(k)})$ and if the charges on the remaining two (analytically related )
edges $e_{v^{\mathrm{E}}}^{(k)\prime},e_{v^{\mathrm{E}}}^{(k)\prime\prime}$
incident on $f(\tilde{(}e)_{v^{\mathrm{E}}}^{(k)})$ are $(n_{e_{v^{\mathrm{E}%
}}^{(k)\prime}}^{(1)},n_{e^{(k)\prime}{v^{\mathrm{E}}}}^{(2)}%
,n_{e_{v^{\mathrm{E}}}^{(k)\prime}}^{(3)})$ and $(n_{e_{v^{\mathrm{E}}%
}^{(k)\prime\prime}}^{(1)},n_{e_{v^{\mathrm{E}}}^{(k)\prime\prime}}%
^{(2)}=n_{e_{v^{\mathrm{E}}}^{(k)\prime}}^{(2)},n_{e_{v^{\mathrm{E}}%
}^{(k)\prime\prime}}^{(3)}=n_{e_{v^{\mathrm{E}}}^{(k)\prime}}^{(3)}$, then either

\begin{enumerate}
\item[(a)] $n_{\tilde{e}_{v^{\mathrm{E}}}^{(k)}}^{(1)}=n_{e_{v^{\mathrm{E}}%
}^{(k)\prime}}^{(2)}$ or

\item[(b)] $n_{\tilde{e}_{v^{\mathrm{E}}}^{(k)}}^{(1)}\ =n_{e_{v^{\mathrm{E}}%
}^{(k)\prime}}^{(3)}$
\end{enumerate}

\item[(6)] Now consider the set $\mathcal{\bar{S}}_{v^{\mathrm{E}}}$. Recall
that each element in this set is a tri-valent vertex. Consider the vertex
$f(\tilde{e}_{v^{\mathrm{E}}})$ whose three incident edges are $\tilde
{e}_{v^{\mathrm{E}}},\tilde{e}_{v^{\mathrm{E}}}^{\prime},\ \text{and}%
\ \tilde{e}_{v^{\mathrm{E}}}^{\prime\prime}$ respectively. Recall that
$\tilde{e}_{v^{\mathrm{E}}}^{\prime},\ \tilde{e}_{v^{\mathrm{E}}}%
^{\prime\prime}$ are analytic continuations of each other. Depending on
whether $n_{\tilde{e}_{v^{\mathrm{E}}}}^{(1)}\lessgtr0$, choose one out of the
two edges, $\tilde{e}_{v^{\mathrm{E}}}^{\prime},\ \tilde{e}_{v^{\mathrm{E}}%
}^{\prime\prime}$ which has lesser or greater charge (depending on whether
$n_{\tilde{e}_{v^{\mathrm{E}}}}^{(1)}$ is greater or less then zero) in
U(1)$_{1}$ than the other edge. Consider the set of all such chosen edges for
each vertex in $\mathcal{\bar{S}}_{v^{\mathrm{E}}}$. We refer to this set as
$\overline{\mathcal{T}}_{v^{\mathrm{E}}}$.\newline Now consider $\sum
_{\tilde{e}\in\mathcal{\bar{S}}_{v^{\mathrm{E}}}}n_{\tilde{e}_{v^{\mathrm{E}}%
}}^{(1)}$. Suppose this quantity is positive (negative). Then among the two
edges $e_{v^{\mathrm{E}}}^{(1)}$ and $e_{v^{\mathrm{E}}}^{(2)}$ pick the edge
whose charge in U$(1)_{1}$ is lesser or greater than the charge in U$(1)_{1}$
of the other edge. Let us assume it is $e_{v^{\mathrm{E}}}^{(1)}$. Then
\begin{equation}
\mathcal{T}_{v^{\mathrm{E}}}:=\overline{\mathcal{T}}_{v^{\mathrm{E}}}%
\cup\{e_{v^{\mathrm{E}}}^{(1)}\}
\end{equation}

\begin{enumerate}
\item[(6a)] All edges in $\mathcal{T}_{v^{\mathrm{E}}}$ meet at a vertex $v$
which is such that if the number of edges incident at $v$ is greater then
$N_{v}$ and \emph{if the charges $\{\tilde{e}_{k}^{v^{\mathrm{E}}%
}\}_{k=1,\dots,N_{v}}$ are the }U\emph{$(1)_{j}$ charges on the edges in
$\mathcal{T}_{v^{\mathrm{E}}}$, then the }U\emph{$(1)_{j}$ charge on the edges
incident at $v$ which are not in $\mathcal{T}_{v^{\mathrm{E}}}$ is zero.} As
shown in Appendix \ref{A2}, $v,$ if it exists, is unique.
\end{enumerate}


\item[(7)] Finally, consider the graph $\gamma:=\gamma(\bar{c})-\{\tilde
{e}_{(1)}^{v^{\mathrm{E}}},\dots,\tilde{e}_{(N_{v}-2)}^{v^{\mathrm{E}}}\}$ and
a charge-network $c$ based on $\gamma$ obtained by deleting $\{\tilde{e}%
_{(1)}^{v^{\mathrm{E}}},\dots,\tilde{e}_{(N_{v}-2)}^{v^{\mathrm{E}}}\}$ along
with the charges on them, and also deleting exactly same amount of charges
from the edges in $\mathcal{T}_{v^{\mathrm{E}}}$. Note that by construction
$v$ belongs to $\gamma$. Now consider $U_{\epsilon}(\gamma,v)$. The final and
key feature of the EO vertex $v^{\mathrm{E}}$ is\newline$v^{\mathrm{E}}\in
U_{\epsilon}(\gamma,v)$ and $v^{\mathrm{E}}$ is the endpoint of the
\textquotedblleft straight-line curve\textquotedblright\ $\delta\cdot\langle
E_{j}^{a}\rangle$ for some $\delta$, where $j=2$ if in ({6}) condition (a) is
satisfied, and $j=3$ if in ({5}), (b) is satisfied.
\end{enumerate}

We call the pair $(v,v^{\mathrm{E}})$ \emph{extraordinary} with $v^{\mathrm{E}%
}$ a type B EO vertex.

\pagebreak

\section{Uniqueness of $v$ Associated to $v^{\mathrm{E}}$}

\label{A2}

\textbf{Lemma}: Consider a charge-network $\bar{c}$ containing a vertex
$v^{\mathrm{E}}$ of type $(M=1,j=2,K\in\{\mathrm{A,B}\})$ satisfying
conditions {(1)}-{(7)} as listed in \textbf{Set A} or \textbf{Set B}. Then the
vertex $v$ described in condition(s) {(7)} is unique.

\textbf{Proof}: We will only prove the lemma for a type A EO vertex. The proof
for type B case is exactly analogous.

Let us assume the contrary. So there exists $v$ and $v^{\prime}$ in $V(\bar
{c})$ with respect to which $v^{\mathrm{E}}$ is EO. This clearly implies that
for all tri-valent vertices in $\mathcal{\bar{S}}_{v^{\mathrm{E}}},$ two edges
(which are analytic extensions of each other) begin, one such set of edges end
in $v$ and the other set ends in $v^{\prime}$. Moreover all the edges incident
at $v$ and $v^{\prime}$ are such that all the edges apart from the ones which
begin at a vertex in $\mathcal{\bar{S}}_{v^{\mathrm{E}}}$ have zero charge in
U$(1)_{2}$. Now consider one such vertex $v_{1}\in\mathcal{\bar{S}%
}_{v^{\mathrm{E}}}$. If the charge (in U$(1)_{1}$) on edge $e_{v^{\mathrm{E}%
},v_{1}}$ with bounded by $v^{\mathrm{E}}$ and $v_{1}$ is positive, then
depending on whether the charge in U$(1)_{2}$ on edge $e_{v_{1},v}$ (which
equals the U$(1)_{2}$ charge on $e_{v_{1},v^{\prime}}$) is positive or
negative, check on which of these two edges the U$(1)_{1}$ charge is greater.
This singles out one of $v$ or $v^{\prime}$ with respect to which
$v^{\mathrm{E}}$ is EO.

\section{Continuum Limit of the Hamiltonian Constraint Operator}

\label{A3}

In this Appendix we derive (\ref{eq:feb6-2}), (\ref{eq:feb6-3}) and
(\ref{eq:feb6-4}). We do this by showing that for any $N$, $|\tilde{c}\rangle$
and any $\Psi_{\lbrack c_{1},c_{2},c_{3}]_{(1)}}^{f^{(1)}}$,%
\begin{equation}
\left(  \hat{H}^{(i)}[N]^{\prime}\Psi_{\lbrack c_{1},c_{2},c_{3}]_{(1)}%
}^{f^{(1)}}\right)  |\tilde{c}\rangle=\lim_{\delta\rightarrow0}\Psi_{\lbrack
c_{1},c_{2},c_{3}]_{(1)}}^{f^{(1)}}\left(  \hat{H}_{T(\delta)}^{(i)}%
[N]|\tilde{c}\rangle\right)
\end{equation}
for $i=1,2,3$. The left hand side is given in (\ref{eq:feb6-2}),
(\ref{eq:feb6-3}), (\ref{eq:feb6-4}) respectively.

\textbf{Proofs}: We want to show that for any $c,\tilde{c},N,$ and
$\Psi_{\lbrack c_{1},c_{2},c_{3}]_{(i)}}^{f^{(i)}},$ the following holds:
\begin{equation}
\lim_{\delta\rightarrow0}\Psi_{\lbrack c_{1},c_{2},c_{3}]_{(i)}}^{f^{(i)}%
}\left(  \hat{H}_{T(\delta)}^{(j)}[N]|\tilde{c}\rangle\right)  =\sum_{v\in
V(c_{1}\cup c_{2}\cup c_{3})}\left(  \Psi_{\lbrack c_{1},c_{2},c_{3}]_{(i)}%
}^{\bar{f}_{v}^{(i,j)}}-\Psi_{\lbrack c_{1},c_{2},c_{3}]_{(i)}}^{\overline
{\overline{f}}{}_{v}^{(i,j)}}\right)  |\tilde{c}\rangle\label{eq:contham1}%
\end{equation}
for any $i,j\in\{1,2,3\}$. where $\Psi_{\lbrack c_{1},c_{2},c_{3}]_{(i)}%
}^{\overline{f}_{v}^{(i,j)}},\ \Psi_{\lbrack c_{1},c_{2},c_{3}]_{(i)}%
}^{\overline{\overline{f}}{}_{v}^{(i,j)}}$ are in $\mathrm{Cyl}^{\ast}$. The
computation will be divided into following cases:

\textbf{Type A}: The $c$ on which $\Psi_{\lbrack c_{1},c_{2},c_{3}]_{(i)}%
}^{f^{(i)}}$ is based is such that all the EO vertices which can be created
from this state necessary lie off $\gamma(c)$; i.e., all the EO vertices are
type A.

\textbf{Type B}: Compliment of type A case.

We will analyze only the type A case here as the complimentary case can be
analyzed in a similar manner but requires more bookkeeping. The results proven
here hold for both cases. We further divide the analysis of the type A case
into following sub-classes:

Case (A,1): $i=j$. Without loss of generality, we take $i=j=1$.

Case (A,1,a): $\tilde{c}=c$.

Then%

\begin{align}
\mathrm{LHS} &  =\lim_{\delta\rightarrow0}\Psi_{\lbrack c_{1},c_{2}%
,c_{3}]_{(1)}}^{f^{(1)}}\left(  \hat{H}_{T(\delta)}^{(1)}[N]|c\rangle\right)
\\
&  =\lim_{\delta\rightarrow0}\frac{1}{\delta}\Psi_{\lbrack c_{1},c_{2}%
,c_{3}]_{(1)}}^{f^{(1)}}\sum_{v\in V(c)}N(v)\lambda(\vec{n}_{c}^{v})\left(
|c_{1}\cup\alpha_{v}^{\delta}(\langle\hat{E}_{2}\rangle,n^{3}),c_{2}%
,c_{3}\rangle-|c_{1}\cup\alpha_{v}^{\delta}(\langle\hat{E}_{3}\rangle
,n^{2}),c_{2},c_{3}\rangle\right)  \nonumber\\
&  =\lim_{\delta\rightarrow0}\sum_{v\in V(c)}N(v)\lambda(\vec{n}_{c}^{v}%
)\frac{1}{\delta}\left(  f_{[c_{1},c_{2},c_{3}]_{(1)}}^{(1)}(\bar{V}(c_{1}%
\cup\alpha_{v}^{\delta}(\langle\hat{E}_{2}\rangle,n^{3}))-f_{[c_{1}%
,c_{2},c_{3}]_{(1)}}^{(1)}(\bar{V}(c_{1}\cup\alpha_{v}^{\delta}(\langle\hat
{E}_{3}\rangle,n^{2})))\right)  \nonumber\\
&  =\lim_{\delta\rightarrow0}\sum_{v\in V(c)}N(v)\lambda(\vec{n}_{c}^{v}%
)\frac{1}{\delta}\left(  f_{[c_{1},c_{2},c_{3}]_{(1)}}^{(1)}(\bar{V}(c_{1}\cup
c_{2}\cup c_{3}\cup v_{\mathrm{E},(1,2)}^{\delta}))-\ f_{[c_{1},c_{2}%
,c_{3}]_{(1)}}^{(1)}(\bar{V}(c_{1}\cup c_{2}\cup c_{3}\cup v_{\mathrm{E}%
,(1,3)}^{\delta}))\right)  \nonumber
\end{align}

where $v_{\mathrm{E},(1,2)}^{\delta}$ is the EO vertex of type $(K\in
\{\mathrm{A,B}\},M=1,j=2)$. It is associated to $v$ and which is at the
\textquotedblleft apex\textquotedblright\ of $\alpha_{v}^{\delta}(\langle
\hat{E}_{2}\rangle,n^{3})$. Similarly $v_{\mathrm{E},(1,3)}^{\delta}$ is the
EO vertex that is associated to $v$ and which is at the \textquotedblleft
apex\textquotedblright\ of $\alpha_{v}^{\delta}(\langle\hat{E}_{3}%
\rangle,n^{2})$. Note that
\begin{equation}
\bar{V}(c_{1}\cup\alpha_{v}^{\delta}(\langle\hat{E}_{2}\rangle,n^{3})\cup
c_{2}\cup c_{3})=\bar{V}(c_{1}\cup c_{2}\cup c_{3}\cup v_{\mathrm{E}%
,(I,2)}^{\delta}))
\end{equation}
as the \emph{fiducial vertices} do not contribute to the set of vertices which
are in $\bar{V}(\tilde{c})$.\newline

we can now take the continuum limit of the above matrix elements, to get
\begin{equation}
\lim_{\delta\rightarrow0}\Psi_{\lbrack c_{1},c_{2},c_{3}]_{(1)}}^{f^{(1)}%
}\left(  \hat{H}_{T(\delta)}^{(1)}[N]|c\rangle\right)  =\sum_{v\in
V(c)}N(v)\lambda(\vec{n}_{c}^{v})\left[  \langle\hat{E}_{2}^{a}\rangle
(v)\frac{\partial}{\partial v^{a}}f^{1}(V(c))-\langle\hat{E}_{3}^{a}%
\rangle(v)\frac{\partial}{\partial v^{a}}f^{1}(V(c))\right]
\label{eq:feb6lhs}%
\end{equation}
Let us now look at the RHS of (\ref{eq:feb6-2}).
\begin{align}
\mathrm{RHS}  & =\sum_{v\in V(c)}N(v)\lambda(\vec{n}_{c}^{v})\left[
\Psi_{\lbrack c_{1},c_{2},c_{3}]_{(1)}}^{\bar{f}_{v}^{(1)(1)}}\left(
|c\rangle\right)  -\Psi_{\lbrack c_{1},c_{2},c_{3}]_{(1)}}^{\overline
{\overline{f}}{}_{v}^{(1)(1)}}\left(  |c\rangle\right)  \right]
\label{eq:feb6rhs}\\
& =\sum_{v\in V(c)}N(v)\lambda(\vec{n}_{c}^{v})\left[  \bar{f}_{v}%
^{(1)(1)}(V(c_{1}\cup c_{2}\cup c_{3}))-\overline{\overline{f}}{}_{v}%
^{(1)(1)}(V(c_{1}\cup c_{2}\cup c_{3}))\right]  \nonumber
\end{align}
which using the definition of $\bar{f}_{v}^{(1)(1)}$ and $\overline
{\overline{f}}{}_{v}^{(1)(1)}$ given in Eq. (\ref{eq:feb6-1}) clearly matches with
the LHS in (\ref{eq:feb6lhs}).

Case (A,1,b): $\tilde{c}\neq c,$ but $\tilde{c}\in\lbrack c_{1},c_{2}%
,c_{3}]_{(i)}$. There are three separate sub-cases in (A,1,b): Let $v_{0}\in
V(\tilde{c}_{1}\cup c_{2}\cup c_{3})$ and let s$\text{upp}(N)=B(v_{0}%
,\epsilon).$ Then we have Case (A,1,b,i): $v_{0}\in V(\tilde{c}_{1}\cup
c_{2}\cup c_{3})$ is mono-colored. As $v_{0}$ is mono-colored and as
$(\tilde{c}_{1},c_{2},c_{3})\in\lbrack c_{1},c_{2},c_{3}]_{(1)}$ there is a
$v_{0}^{\prime}\in V(c_{1}\cup c_{2}\cup c_{3})$ with respect to which
$v_{0}^{\prime}$ is WEO. Then%
\begin{equation}
\mathrm{RHS}=\lim_{\delta\rightarrow0}\Psi_{\lbrack c_{1},c_{2},c_{3}]_{(1)}%
}^{(1)}\left(  \hat{H}_{T(\delta)}^{(1)}[N]|\tilde{c}_{1},c_{2},c_{3}%
\rangle\right)  =0
\end{equation}
On the other hand,
\begin{align}
\mathrm{LHS}  & =\sum_{v\in V(c)}\Psi_{\lbrack c_{1},c_{2},c_{3}]_{(1)}}%
^{\bar{f}_{v}^{(1)}}\left(  |\tilde{c}_{1},c_{2},c_{3}\rangle\right)
\nonumber\\
& =\bar{f}_{v_{0}^{\prime}}^{(1)}(\bar{V}(\tilde{c}_{1}\cup c_{2}\cup
c_{3}))-\overline{\overline{f}}{}_{v_{0}^{\prime}}^{(1)}(\bar{V}(\tilde{c}%
_{1}\cup c_{2}\cup c_{3}))\nonumber\\
& =\overline{f}_{v_{0}^{\prime}}^{(1)}(V(c_{1}\cup c_{2}\cup c_{3}%
)/v_{0}^{\prime}\cup\{v_{0}\})-\overline{\overline{f}}{}_{v_{0}^{\prime}%
}^{(1)}(V(c_{1}\cup c_{2}\cup c_{3})/v_{0}^{\prime}\cup\{v_{0}\})\nonumber\\
& =f_{v_{0}^{\prime}}^{(1)}(V(c_{1}\cup c_{2}\cup c_{3})/v_{0}^{\prime}%
\cup\{v_{0}\})-f_{v_{0}^{\prime}}^{(1)}(V(c_{1}\cup c_{2}\cup c_{3}%
)/v_{0}^{\prime}\cup\{v_{0}\})\nonumber\\
& =0
\end{align}
Case (A,1,b,ii): $v_{0}\in V(\tilde{c}_{1}\cup c_{2}\cup c_{3})\cap
V(c_{1}\cup c_{2}\cup c_{3})$ such that $v_{0}$ has an associated WEO vertex
$v_{0}^{\prime}\in V(\tilde{c}_{1}\cup c_{2}\cup c_{3})$. Let $N$ be such that
the support of $N$ only includes $v_{0},$ and no other vertex of $V(c_{1}\cup
c_{2}\cup c_{3})$ lies inside the support of $N$. We prove a small lemma which
will be useful while analyzing this case.

\textbf{Lemma}\label{lemmamar16-1}: There exists no charge network in
$[c_{1},c_{2},c_{3}]_{(1)}$ which corresponds to $(\tilde{c}_{1}\cup
\alpha_{v_{0}}^{\delta}(\langle\hat{E}_{2}\rangle/\langle\hat{E}_{3}%
\rangle,n_{c_{3}}/n_{c_{2}}),c_{2},c_{3})$.

\textbf{Proof}: $V(\tilde{c}_{1}\cup\alpha_{v_{0}}^{\delta}(\langle\hat{E}%
_{2}\rangle/\langle\hat{E}_{3}\rangle,n_{c_{3}}/n_{c_{2}})\cup c_{2}\cup
c_{3})$ has a vertex which is not in $V(c_{1}\cup c_{2}\cup c_{3})$ and is not
WEO with respect to any vertex in $V(c_{1}\cup c_{2}\cup c_{3})$.

Then%
\begin{align}
\mathrm{LHS}  & =\lim_{\delta\rightarrow0}\Psi_{\lbrack c_{1},c_{2}%
,c_{3}]_{(1)}}^{f^{(1)}}\left(  \hat{H}_{T(\delta)}^{(1)}[N]|\tilde{c}%
_{1},c_{2},c_{3}\rangle\right)  \nonumber\\
& =\lim_{\delta\rightarrow0}\Psi_{\lbrack c_{1},c_{2},c_{3}]_{(1)}}^{f^{(1)}%
}N(v_{0})\lambda(v_{0},\tilde{c}_{1},c_{2},c_{3})\left(  |\tilde{c}_{1}%
\cup\alpha_{v}^{\delta}(\langle\hat{E}_{2}\rangle,n_{c_{3}}),c_{2}%
,c_{3}\rangle-|\tilde{c}_{1}\cup\alpha_{v}^{\delta}(\langle\hat{E}_{3}%
\rangle,n_{c_{2}}),c_{2},c_{3}\rangle\right)  \nonumber\\
& =\lim_{\delta\rightarrow0}\Psi_{\lbrack c_{1},c_{2},c_{3}]_{(1)}}^{f^{(1)}%
}N(v_{0})\lambda(\vec{n}_{c}^{v_{0}})\left(  |\tilde{c}_{1}\cup\alpha
_{v}^{\delta}(\langle\hat{E}_{2}\rangle,n_{c_{3}}),c_{2},c_{3}\rangle
-|\tilde{c}_{1}\cup\alpha_{v}^{\delta}(\langle\hat{E}_{3}\rangle,n_{c_{2}%
}),c_{2},c_{3}\rangle\right)
\end{align}
where in the last line we have used $\lambda(v_{0},\tilde{c}_{1},c_{2}%
,c_{3})=\lambda(\vec{n}_{c}^{v_{0}})$.

Using the above lemma, we have
\begin{equation}
\Psi_{\lbrack c_{1},c_{2},c_{3}]_{(1)}}^{f^{(1)}}\left(  |\tilde{c}_{1}%
\cup\alpha_{v}^{\delta}(\langle\hat{E}_{2}\rangle,n_{c_{3}}),c_{2}%
,c_{3}\rangle\right)  =\Psi_{\lbrack c_{1},c_{2},c_{3}]_{(1)}}^{f^{(1)}%
}\left(  |\tilde{c}_{1}\cup\alpha_{v}^{\delta}(\langle\hat{E}_{3}%
\rangle,n_{c_{2}}),c_{2},c_{3}\rangle\right)  =0
\end{equation}
$\forall\ \delta$. Thus $\mathrm{LHS}=0$.

Let us now look at the $\mathrm{RHS.}$ As $N$ has support only in the
neighborhood of $v_{0}\in V(c)$, the RHS equals
\begin{equation}
\left(  \Psi_{\lbrack c_{1},c_{2},c_{3}]_{(1)}}^{\bar{f}_{v_{0}}^{(1)}}%
-\Psi_{\lbrack c_{1},c_{2},c_{3}]_{(1)}}^{\overline{\overline{f}}{}_{v_{0}%
}^{(1)}}\right)  |\tilde{c}_{1},c_{2},c_{3}\rangle=\left(  \bar{f}_{v_{0}%
}^{(1)}(\bar{V}(\tilde{c}_{1}\cup c_{2}\cup c_{3}))\ -\overline{\overline{f}%
}{}_{v_{0}}^{(1)}(\bar{V}(\tilde{c}_{1}\cup c_{2}\cup c_{3})\right)
\end{equation}
But as $v_{0}$ has a WEO vertex $v_{0}^{\prime}$ associated with it, the
arguments of $\bar{f}_{v_{0}}^{(1)}$ and $\overline{\overline{f}}{}_{v_{0}%
}^{(1)}$ contain $v_{0}^{\prime}$ in place of $v_{0}$ and as both functions
agree with $f^{(1)}$ when no argument is $v_{0},$ $\mathrm{RHS}=0$.

Case (A,1,b,iii): $v_{0}\in V(\tilde{c}_{1}\cup c_{2}\cup c_{3})\cap
V(c_{1}\cup c_{2}\cup c_{3})$ such that $v_{0}$ has an associated WEO vertex
$v_{0}^{\prime}\in V(\tilde{c}_{1}\cup c_{2}\cup c_{3})$ which lies inside the
support of $N$. The above argument goes through and both LHS and RHS vanish.

Case (A,1,b,iv): $v_{0}\in V(\tilde{c}_{1}\cup c_{2}\cup c_{3})\cap
V(c_{1}\cup c_{2}\cup c_{3})$ such that $v_{0}$ has no associated WEO vertex.
Let $N$ be such that s$\text{upp}(N)\subset B(v_{0},\epsilon)$ and $v\in
V(c)\cap B(v_{0},\epsilon)\implies v=v_{0}$. Then%
\begin{align}
\mathrm{LHS}  & =\lim_{\delta\rightarrow0}\Psi_{\lbrack c_{1},c_{2}%
,c_{3}]_{(1)}}^{f^{(1)}}\left(  \hat{H}_{T(\delta)}^{(1)}[N]|\tilde{c}%
_{1},c_{2},c_{3}\rangle\right)  \label{eq:feb5-1}\\
& =\lim_{\delta\rightarrow0}\Psi_{\lbrack c_{1},c_{2},c_{3}]_{(1)}}^{f^{(1)}%
}\left(  |\tilde{c}_{1}\cup\alpha_{v_{0}}^{\delta}(\langle\hat{E}_{2}%
\rangle,n^{3}),c_{2},c_{3}\rangle-|\tilde{c}_{1}\cup\alpha_{v_{0}}^{\delta
}(\langle\hat{E}_{3}\rangle,n^{2}),c_{2},c_{3}\rangle\right)  \nonumber
\end{align}
Note that as $(\tilde{c}_{1},c_{2},c_{3})\in\lbrack c_{1},c_{2},c_{3}]_{(1)}$,
it is clear that $|\tilde{c}_{1}\cup\alpha_{v_{0}}^{\delta}(\langle\hat{E}%
_{2}\rangle,n^{3}),c_{2},c_{3}\rangle$ and $|\tilde{c}_{1}\cup\alpha_{v_{0}%
}^{\delta}(\langle\hat{E}_{3}\rangle,n^{2}),c_{2},c_{3}\rangle$ both belong to
$[c_{1},c_{2},c_{3}]_{(1)}$. Whence, we have
\begin{align}
\mathrm{LHS}  & =\lim_{\delta\rightarrow0}\frac{1}{\delta}N(v_{0})\lambda
(\vec{n}_{c}^{v_{0}})\left(  f^{(1)}(\bar{V}(\tilde{c}_{1}\cup\alpha_{v_{0}%
}^{\delta}(\langle\hat{E}_{2}\rangle,n^{3})\cup c_{2}\cup c_{3}))-f^{(1)}%
(\overline{V}(\tilde{c}_{1}\cup\alpha_{v_{0}}^{\delta}(\langle\hat{E}%
_{3}\rangle,n^{2})\cup c_{2}\cup c_{3}))\right)  \nonumber\\
& =N(v_{0})\lambda(\vec{n}_{c}^{v_{0}})\left[  \langle\hat{E}_{2}^{a}%
\rangle\frac{\partial}{\partial v^{a}}f^{(1)}(V(c))-\langle\hat{E}_{3}%
^{a}\rangle\frac{\partial}{\partial v^{a}}f^{(1)}(V(c))\right]  ,
\end{align}
where in the first line we have used $\lambda(v_{0};(\tilde{c}_{1},c_{2}%
,c_{3}))=\lambda(\vec{n}_{c}^{v_{0}})$ and in the second line we have used
$\bar{V}(c)=V(c)$.

We now evaluate the RHS: As $N$ has support only in the neighborhood of the
vertex $v_{0}$ in $V(c)$, only non-vanishing contributions are through
$\bar{f}_{v_{0}},\ \overline{\overline{f}}_{v_{0}}$. $\bar{f}_{v_{0}}%
:\Sigma^{|V(c)|}\rightarrow%
\mathbb{R}
\mathbf{\ }$such that
\begin{equation}
\bar{f}_{v_{0}}^{(1)}(v_{1},...,v_{|V(c)|})=f^{(1)}(v_{1},...,v_{|V(c)|}%
)\qquad\text{if}\qquad\{v_{1},...,v_{|V(c_{1}\cup c_{2}\cup c_{3})|}\}\neq
V(c)
\end{equation}
and%
\begin{equation}
\bar{f}_{v_{0}}^{(1)}(V(c))=N(v_{0})\lambda(\vec{n}_{c}^{v_{0}})\langle\hat
{E}_{2}\rangle(v_{0})\frac{\partial}{\partial v_{0}^{a}}f^{(1)}(V(c))
\end{equation}
Similarly, $\overline{\overline{f}}_{v_{0}}:\Sigma^{|V(c)|}\rightarrow%
\mathbb{R}
\ $such that%
\begin{equation}
\bar{f}_{v_{0}}^{(1)}(v_{1},...,v_{|V(c)|})=f^{(1)}(v_{1},...,v_{|V(c)|}%
)\qquad\text{if}\qquad(v_{1},...,v_{|V(c)|})\neq V(c)
\end{equation}
and%
\begin{equation}
\bar{f}_{v_{0}}^{(1)}(V(c))=N(v_{0})\lambda(\vec{n}_{c}^{v_{0}})\langle\hat
{E}_{2}\rangle(v_{0})\frac{\partial}{\partial v_{0}^{a}}f^{(1)}(V(c)).
\end{equation}

Case (A,1,c): $\tilde{c}\notin\lbrack c_{1},c_{2},c_{3}]_{(i)}$. It is rather
straightforward to show that both the $\text{RHS}$ and $\text{LHS}$ vanish in
this case.

Case (A,2): $i\neq j$. let $i=1,\ j=2$. Other cases can be analyzed
analogously. Thus our aim is to show that, given $\Psi_{\lbrack c_{1}%
,c_{2},c_{3}]_{(1)}}^{f^{(1)}}$,%
\begin{equation}
\lim_{\delta\rightarrow0}\Psi_{\lbrack c_{1},c_{2},c_{3}]_{(1)}}^{f^{(1)}%
}\left(  \hat{H}_{T(\delta)}^{(2)}[N]|\tilde{c}\rangle\right)  =\sum_{v\in
V(c)}\left(  \Psi_{\lbrack c_{1},c_{2},c_{3}]_{(1)}}^{\bar{f}_{v}^{(1)(2)}%
}-\Psi_{\lbrack c_{1},c_{2},c_{3}]_{(1)}}^{\overline{\overline{f}}{}_{v_{0}%
}^{(1)(2)}}\right)  |\tilde{c}\rangle
\end{equation}
$\forall\ N,|\tilde{c}\rangle$, and where $\bar{f}_{v}^{(1)(2)}$ and
$\overline{\overline{f}}{}_{v_{0}}^{(1)(2)}$ are defined in (\ref{eq:feb6-3}).

As before we consider different cases: 

Case (A,2,a): $\tilde{c}$ does not have an EO vertex of type $M=1$. In this
case, $\forall\ \delta>0,$ $\hat{H}_{T(\delta)}^{(2)}[N]$ will create states
with EO vertices of type $M=2,3$ and type $M=2,1$, when acting on $|\tilde
{c}\rangle$. However, as $[c_{1},c_{2},c_{3}]_{(1)}$ has no states with EO
vertices of type $M=2$, $\mathrm{LHS}=\mathrm{RHS}=0$.

Case (A,2,b): $\tilde{c}$ does have an EO vertex of type $M=1,j=2$. This
condition implies that $\tilde{c}=c_{1}^{\prime}\cup\alpha_{v_{0}}^{\delta
_{0}}(\langle\hat{E}_{2}\rangle_{\tilde{c}_{2}},n_{\tilde{c}_{3}}),\tilde
{c}_{2},\tilde{c}_{3})$ for some $c_{1}^{\prime}$.  Let us also assume that this vertex is inside the support of the lapse function $N$. Now both the $\text{LHS}$
and $\text{RHS}$ are non-zero iff $(c_{1}^{\prime},\tilde{c}_{2},\tilde{c}%
_{3})=(c_{1},c_{2},c_{3})$. In this case we have (using the equation for the
action of the Hamiltonian constraint on charge networks involving these
specific type of EO vertices as given in (\ref{proposal1}))
\begin{align}
\mathrm{LHS}  & =\lim_{\delta\rightarrow0}\Psi_{\lbrack c_{1},c_{2}%
,c_{3}]_{(1)}}^{f^{(1)}}\left(  \hat{H}_{T(\delta)}^{(2)}[N]|c_{1}\cup
\alpha_{v_{0}}^{\delta_{0}}(\langle\hat{E}_{2}\rangle_{c_{2}}(v_{0}),n_{c_{3}%
}),c_{2},c_{3}\rangle\right)  \label{eq:equationforsomelhs}\\
& =\lim_{\delta\rightarrow0}\left(  \frac{1}{\delta}\Psi_{\lbrack c_{1}%
,c_{2},c_{3}]_{(1)}}^{f^{(1)}}\left[  \sum_{e\in E(c)|b(e)=v_{0}}\left(
\langle\hat{E}_{3}(L_{e}(\delta^{\prime}))\rangle\dot{e}^{a}(0)(N(v_{0}%
+\delta\dot{e}(0))-N(v_{0})\right)  |c_{1}\cup\alpha_{v_{0}}^{\delta^{0}%
}(\langle\hat{E}_{1}\rangle(v_{0}),n^{3}),c_{2},c_{3}\rangle\right]  \right.
\nonumber\\
& -\left.  \frac{1}{\delta}\Psi_{\lbrack c_{1},c_{2},c_{3}]_{(1)}}^{f^{(1)}%
}\left[  \sum_{e\in E(c)|b(e)=v_{0}}\left(  \langle\hat{E}_{1}(L_{e}%
(\delta^{\prime}))\rangle\dot{e}^{a}(0)(N(v_{0}+\delta\dot{e}(0))-N(v_{0}%
)\right)  |c_{1}\cup\alpha_{v_{0}}^{\delta^{0}}(\langle\hat{E}_{3}%
\rangle(v_{0}),n^{3}),c_{2},c_{3}\rangle\right]  \right)  \nonumber
\end{align}
Where $\delta^{\prime}\rightarrow0$ faster then $\delta\rightarrow0$. However
note that the (net of) flux expectation values remains constant in the limit
$\delta^{\prime}\rightarrow0$ (as they are simply equal to $n^{1}$ or $n^{3}%
$), whence $\langle\hat{E}_{3}(L_{e}(\delta^{\prime}))\rangle$ is independent
of $\delta^{\prime}$ and we denote it simply as $\langle\hat{E}_{3}%
(L_{e})\rangle$ where $L_{e}$ could be any surface fixed once and for all. It
is easy to see that the $\text{LHS}$ simplifies to%
\begin{align}
\mathrm{LHS} &  =\sum_{e\in E(c)|b(e)=v_{0}}\langle\hat{E}_{3}(L_{e}%
)\rangle\dot{e}^{a}(0)\partial_{a}N(v_{0})f^{(1)}(\bar{V}(c_{1}\cup
\alpha_{v_{0}}^{\delta_{0}}(\langle\hat{E}_{1}\rangle(v_{0}),n^{3})\cup
c_{2}\cup c_{3})\nonumber\\
&  \qquad-\sum_{e\in E(c)|b(e)=v_{0}}\langle\hat{E}_{1}(L_{e})\rangle\dot
{e}^{a}(0)\partial_{a}N(v_{0})f^{(1)}(\bar{V}(c_{1}\cup\alpha_{v_{0}}%
^{\delta_{0}}(\langle\hat{E}_{3}(v_{0})\rangle,n^{3})\cup c_{2}\cup
c_{3})\label{BORG}%
\end{align}
But upon using (\ref{eq:feb6-1}) we see that, this precisely equals the
$\text{RHS}$.

\section{On the minus sign in Eq. (\ref{eq:mar6-2})}

\label{A2point5}

Consider the ideal scenario under which the continuum Hamiltonian constraint $\hat{H}[N]^{\prime}$ preserved ${\cal V}_{LMI}$. In that case we would like to prove that

\begin{equation}
[\hat{H}[N]^{\prime}, \hat{H}[M]^{\prime}]\Psi\ =\ \left(i\hbar\right)V_{diff}[M,N]\Psi
\end{equation}
$\forall\ \Psi\ \in {\cal V}_{LMI}$. This implies that $\forall\ \vert c\rangle$, we need

\begin{equation}
\lim_{(T, T^{\prime})\rightarrow\ 0}\Psi\left(\hat{H}_{T^{\prime}}[M]\hat{H}_{T}[N]\ -\ N\leftrightarrow M\right)\vert c\rangle\ =\ \lim_{T\rightarrow 0}\Psi\left((-i\hbar)\hat{V}_{T}[M,N]\right)\vert c\rangle
\end{equation}

However notice that the LHS  of above equation is 
\begin{equation}\nonumber
\left([\hat{H}[N],\hat{H}[M]]^{\prime}\Psi\right)\vert c\rangle
\end{equation} 
 
and the RHS is 
\begin{equation}\nonumber
-i\hbar\left(\hat{V}([M,N])^{\prime}\Psi\right)\vert c\rangle
\end{equation}

Whence we are led to prove that

\begin{equation}
[\hat{H}[N],\hat{H}[M]]^{\prime}\Psi\ =\ -i\hbar \hat{V}([M,N])\Psi
\end{equation}

Technically we are seeking an anti-representation of Hamiltonian constraint on ${\cal V}_{LMI}$.\\

\section{Details of the Commutator Computation}

\label{A4}

In this Appendix we derive equation (\ref{eq:comm-1}). A key ingredient in
this derivation is the fact that the action of the Hamiltonian constraint on
irrelevant vertices is trivial.

Given a charge-network state $|c^{\prime}\rangle$, our first objective is to
evaluate%
\begin{equation}
\sum_{i,j}[\hat{H}_{T(\delta^{\prime})}^{(i)}[N],\hat{H}_{T(\delta)}%
^{(j)}[M]]|c^{\prime}\rangle
\end{equation}
The nine terms in the commutator can be grouped in the following way:%

\begin{equation}
\left(  \sum_{i}\left(  \hat{H}_{T(\delta^{\prime})}^{(i)}[N]\hat{H}%
_{T(\delta)}^{(i)}[M]-\left(  N\leftrightarrow M\right)  \right)  +\sum_{i\neq
j}\left(  \hat{H}_{T(\delta^{\prime})}^{(i)}[N]\hat{H}_{T(\delta)}%
^{(j)}[M]-\left(  N\leftrightarrow M\right)  \right)  \right)  |c^{\prime
}\rangle
\end{equation}
It is easy to see that $[\hat{H}_{T(\delta^{\prime})}^{(i)}[N],\hat
{H}_{T(\delta)}^{(i)}[M]]$ do not contribute for any $i$. Let us consider the
action of $[\hat{H}_{T(\delta^{\prime})}^{(1)}[N],\hat{H}_{T(\delta)}%
^{(1)}[M]]$ on $|c^{\prime}\rangle$. Recall that
\begin{equation}
\hat{H}_{T(\delta)}^{(1)}[N]|c^{\prime}\rangle:=N(v)\hat{H}_{T(\delta)}%
^{(1)}(v)|c^{\prime}\rangle,
\end{equation}
and let
\begin{equation}
\hat{H}_{T(\delta)}^{(1)}(v)|c^{\prime}\rangle=\sum_{j=1}^{2}|c_{\delta
\ j}^{\prime v}\rangle.
\end{equation}
Then we have%
\begin{align}
\lbrack\hat{H}_{T(\delta^{\prime})}^{(1)}[N],\hat{H}_{T(\delta)}%
^{(1)}[M]]|c^{\prime}\rangle & =\sum_{v\in V(c^{\prime})}\left(  \hat
{H}_{T(\delta^{\prime})}^{(1)}[N]M(v)\hat{H}_{T(\delta)}^{(1)}(v)-\left(
N\leftrightarrow M\right)  \right)  |c^{\prime}\rangle\nonumber\\
& =\sum_{j=1}^{2}\sum_{v\in V(c^{\prime})}\sum_{v^{\prime}\in V(c_{\delta
\ j}^{\prime v})}\left(  N(v^{\prime})M(v)-M(v^{\prime})N(v)\right)  \hat
{H}_{T(\delta^{\prime})}^{(1)}(v^{\prime})\hat{H}_{T(\delta)}^{(1)}%
(v)|c^{\prime}\rangle\nonumber\\
& =\sum_{j=1}^{2}\sum_{v\in V(c^{\prime})}\sum_{v^{\prime}\in V(c^{\prime}%
)}\left(  N(v^{\prime})M(v)-M(v^{\prime})N(v)\right)  \hat{H}_{T(\delta
^{\prime})}^{(1)}(v^{\prime})\hat{H}_{T(\delta)}^{(1)}(v)|c^{\prime}%
\rangle\nonumber\\
& =0
\end{align}
In the third line we have used the fact that action of $\hat{H}_{T(\delta
)}^{(1)}(v)$ on EO vertices of type $M=1$ is zero and that the action of
$\hat{H}_{T(\delta^{\prime})}^{(1)}$ on four-valent irrelevant vertices
resulting from the action of $\hat{H}_{T(\delta)}^{(1)}$ is zero (due to the
specific charge-configuration on the incident edges). This shows that the
first set of terms $\sum_{i}\left(  \hat{H}_{T(\delta\prime)}^{(i)}[N]\hat
{H}_{T(\delta)}^{(i)}[M]-\left(  N\leftrightarrow M\right)  \right)  $ do not contribute.

For the second set of terms with $\sum_{i\neq j}$, we first group them as
follows:
\begin{align}
\sum_{i\neq j}\left(  \hat{H}_{T(\delta^{\prime})}^{(i)}[N]\hat{H}_{T(\delta
)}^{(j)}[M]-\left(  N\leftrightarrow M\right)  \right)  |c^{\prime}\rangle &
=\left[  \left(  \hat{H}_{T(\delta^{\prime})}^{(1)}[N]\left(  \hat
{H}_{T(\delta)}^{(2)}[M]+\hat{H}_{T(\delta)}^{(3)}[M]\right)  -\left(
N\leftrightarrow M\right)  \right)  \right.  \label{eq:oct91}\\
& +\left(  \hat{H}_{T(\delta^{\prime})}^{(2)}[N]\left(  \hat{H}_{T(\delta
)}^{(3)}[M]+\hat{H}_{T(\delta)}^{(1)}[M]\right)  -\left(  N\leftrightarrow
M\right)  \right)  \nonumber\\
& +\left.  \left(  \hat{H}_{T(\delta^{\prime})}^{(3)}[N]\left(  \hat
{H}_{T(\delta)}^{(1)}[M]+\hat{H}_{T(\delta)}^{(2)}[M]\right)  -\left(
N\leftrightarrow M\right)  \right)  \right]  |c^{\prime}\rangle\nonumber
\end{align}
Due to the anti-symmetrization in the lapse functions, all the terms that are
ultra-local in the lapses vanish. Thus, the above equation simplifies to%

\begin{align}
& \sum_{i\neq j}\left(  \hat{H}_{T(\delta^{\prime})}^{(i)}[N]\hat{H}%
_{T(\delta)}^{(j)}[M]-\left(  N\leftrightarrow M\right)  \right)  |c^{\prime
}\rangle\label{eq:feb28-1}\\
& \approx\tfrac{1}{4}\left(  \frac{\hbar}{\mathrm{i}}\right)  ^{2}\frac
{1}{\delta\delta^{\prime}}M(v_{0})\lambda(\vec{n}_{v_{0}}^{c^{\prime}}%
)^{2}\sum_{e\in E(c^{\prime})|b(e)=v_{0}}\left(  N(v_{0}+\delta^{\prime}%
\dot{e}(0))-N(v_{0})\right)  \nonumber\\
& \times\left[  \left(  \langle\hat{E}_{3}(L_{e})\rangle|c_{1}^{\prime}%
\cup\alpha_{v_{0}}^{\delta}(\langle\hat{E}_{1}\rangle(v_{0}),n_{c_{3}^{\prime
}}),c_{2}^{\prime},c_{3}^{\prime}\rangle-\langle\hat{E}_{1}(L_{e}%
)\rangle|c_{1}^{\prime}\cup\alpha_{v_{0}}^{\delta}(\langle\hat{E}_{3}%
\rangle(v_{0}),n_{c_{3}^{\prime}}),c_{2}^{\prime},c_{3}^{\prime}%
\rangle\right)  \right.  \nonumber\\
& -\left(  \langle\hat{E}_{1}(L_{e})\rangle|c_{1}^{\prime}\cup\alpha_{v_{0}%
}^{\delta}(\langle\hat{E}_{2}\rangle(v_{0}),n_{c_{2}^{\prime}}),c_{2}^{\prime
},c_{3}^{\prime}\rangle-\langle\hat{E}_{2}(L_{e})\rangle|c_{1}^{\prime}%
\cup\alpha_{v_{0}}^{\delta}(\langle\hat{E}_{2}\rangle(v_{0}),n_{c_{2}^{\prime
}}),c_{2}^{\prime},c_{3}^{\prime}\rangle\right)  \nonumber\\
& +\left(  \langle\hat{E}_{1}(L_{e})\rangle|c_{1}^{\prime},c_{2}^{\prime}%
\cup\alpha_{v_{0}}^{\delta}(\langle\hat{E}_{2}\rangle(v_{0}),n_{c_{1}^{\prime
}}),c_{3}^{\prime}\rangle-\langle\hat{E}_{2}(L_{e})\rangle|c_{1}^{\prime
},c_{2}^{\prime}\cup\alpha_{v_{0}}^{\delta}(\langle\hat{E}_{1}\rangle
(v_{0}),n_{c_{1}^{\prime}}),c_{3}^{\prime}\rangle\right)  \nonumber\\
& -\left(  \langle\hat{E}_{2}(L_{e})\rangle|c_{1}^{\prime},c_{2}^{\prime}%
\cup\alpha_{v_{0}}^{\delta}(\langle\hat{E}_{3}\rangle(v_{0}),n_{c_{3}^{\prime
}}),c_{3}^{\prime}\rangle-\langle\hat{E}_{3}(L_{e})\rangle|c_{1}^{\prime
},c_{2}^{\prime}\cup\alpha_{v_{0}}^{\delta}(\langle\hat{E}_{2}\rangle
(v_{0}),n_{c_{3}^{\prime}}),c_{3}^{\prime}\rangle\right)  \nonumber\\
& +\left(  \langle\hat{E}_{2}(L_{e})\rangle|c_{1}^{\prime},c_{2}^{\prime
},c_{3}^{\prime}\cup\alpha_{v_{0}}^{\delta}(\langle\hat{E}_{3}\rangle
(v_{0}),n_{c_{3}^{\prime}})\rangle-\langle\hat{E}_{3}(L_{e})\rangle
|c_{1}^{\prime},c_{2}^{\prime},c_{3}^{\prime}\cup\alpha_{v_{0}}^{\delta
}(\langle\hat{E}_{2}\rangle(v_{0}),n_{c_{3}^{\prime}})\rangle\right)
\nonumber\\
& -\left.  \left(  \langle\hat{E}_{3}(L_{e})\rangle|c_{1}^{\prime}%
,c_{2}^{\prime},c_{3}^{\prime}\cup\alpha_{v_{0}}^{\delta}(\langle\hat{E}%
_{1}\rangle(v_{0}),n_{c_{1}^{\prime}})\rangle-\langle\hat{E}_{1}(L_{e}%
)\rangle|c_{1}^{\prime},c_{2}^{\prime},c_{3}^{\prime}\cup\alpha_{v_{0}%
}^{\delta}(\langle\hat{E}_{3}\rangle(v_{0}),n_{c_{1}^{\prime}})\rangle\right)
\right]  \nonumber\\
& -\left(  N\leftrightarrow M\right)  \nonumber
\end{align}
Some remarks are in order:  Without loss of generality we are assuming that $\delta$ is small enough such that all EO vertices that are created in the neighborhood of $v_{0}$ are in the support of both $N$ and $M$.\\
The weak equality $\approx$ indicates that we have
thrown away all the terms resulting from the action of the second Hamiltonian
on irrelevant vertices, as these terms will not contribute once we
\textquotedblleft dot\textquotedblright\ them with a habitat state. Henceforth
we will understand that these additional terms have been thrown away and we
will replace $\approx$ with exact equality. Due to reasons explained below
(\ref{eq:equationforsomelhs}), we have omitted the $\delta^{\prime}$ label
from the surface $L_{e}$. The overall plus sign comes from the fact that we
chose $\epsilon=1$ in (\ref{proposal1}).

Due to the underlying symmetry among the terms on the right hand side, we can
rewrite the above equation as%

\begin{align}
& \sum_{i\neq j}\left(  \hat{H}_{T(\delta^{\prime})}^{(i)}[N]\hat{H}%
_{T(\delta)}^{(j)}[M]-\left(  N\leftrightarrow M\right)  \right)  |c^{\prime
}\rangle\label{eq:finitetricom}\\
& =\tfrac{1}{4}\left(  \frac{\hbar}{\mathrm{i}}\right)  ^{2}\frac{1}%
{\delta\delta^{\prime}}M(v_{0})\lambda(\vec{n}_{v_{0}}^{c^{\prime}})^{2}%
\sum_{e\in E(c^{\prime})|b(e)=v_{0}}\left(  N(v_{0}+\delta^{\prime}\dot
{e}(0))-N(v_{0})\right)  \nonumber\\
& \times\left[  \left(  \langle\hat{E}_{3}(L_{e})\rangle\left(  |c_{1}%
^{\prime}\cup\alpha_{v_{0}}^{\delta}(\langle\hat{E}_{1}\rangle(v_{0}%
),n_{c_{3}^{\prime}}),c_{2}^{\prime},c_{3}^{\prime}\rangle-|c^{\prime}%
\rangle\right)  -\langle\hat{E}_{1}(L_{e})\rangle\left(  |c_{1}^{\prime}%
\cup\alpha_{v_{0}}^{\delta}(\langle\hat{E}_{3}\rangle(v_{0}),n_{c_{3}^{\prime
}}),c_{2}^{\prime},c_{3}^{\prime}\rangle-|c^{\prime}\rangle\right)  \right)
\right.  \nonumber\\
& -\left(  \langle\hat{E}_{1}(L_{e})\rangle\left(  |c_{1}^{\prime}\cup
\alpha_{v_{0}}^{\delta}(\langle\hat{E}_{2}\rangle(v_{0}),n_{c_{2}^{\prime}%
}),c_{2}^{\prime},c_{3}^{\prime}\rangle-|c^{\prime}\rangle\right)
-\langle\hat{E}_{2}(L_{e})\rangle\left(  |c_{1}^{\prime}\cup\alpha_{v_{0}%
}^{\delta}(\langle\hat{E}_{2}\rangle(v_{0}),n_{c_{2}^{\prime}}),c_{2}^{\prime
},c_{3}^{\prime}\rangle-|c^{\prime}\rangle\right)  \right)  \nonumber\\
& +\left(  \langle\hat{E}_{1}(L_{e})\rangle\left(  |c_{1}^{\prime}%
,c_{2}^{\prime}\cup\alpha_{v_{0}}^{\delta}(\langle\hat{E}_{2}\rangle
(v_{0}),n_{c_{1}^{\prime}}),c_{3}^{\prime}\rangle-|c^{\prime}\rangle\right)
-\langle\hat{E}_{2}(L_{e})\rangle\left(  |c_{1}^{\prime},c_{2}^{\prime}%
\cup\alpha_{v_{0}}^{\delta}(\langle\hat{E}_{1}\rangle(v_{0}),n_{c_{1}^{\prime
}}),c_{3}^{\prime}\rangle-|c^{\prime}\rangle\right)  \right)  \nonumber\\
& -\left(  \langle\hat{E}_{2}(L_{e})\rangle\left(  |c_{1}^{\prime}%
,c_{2}^{\prime}\cup\alpha_{v_{0}}^{\delta}(\langle\hat{E}_{3}\rangle
(v_{0}),n_{c_{3}^{\prime}}),c_{3}^{\prime}\rangle-|c^{\prime}\rangle\right)
-\langle\hat{E}_{3}(L_{e})\rangle\left(  |c_{1}^{\prime},c_{2}^{\prime}%
\cup\alpha_{v_{0}}^{\delta}(\langle\hat{E}_{2}\rangle(v_{0}),n_{c_{3}^{\prime
}}),c_{3}^{\prime}\rangle-|c^{\prime}\rangle\right)  \right)  \nonumber\\
& +\left(  \langle\hat{E}_{2}(L_{e})\rangle\left(  |c_{1}^{\prime}%
,c_{2}^{\prime},c_{3}^{\prime}\cup\alpha_{v_{0}}^{\delta}(\langle\hat{E}%
_{3}\rangle(v_{0}),n_{c_{3}^{\prime}})\rangle-|c^{\prime}\rangle\right)
-\langle\hat{E}_{3}(L_{e})\rangle\left(  |c_{1}^{\prime},c_{2}^{\prime}%
,c_{3}^{\prime}\cup\alpha_{v_{0}}^{\delta}(\langle\hat{E}_{2}\rangle
(v_{0}),n_{c_{3}^{\prime}})\rangle-|c^{\prime}\rangle\right)  \right)
\nonumber\\
& -\left.  \left(  \langle\hat{E}_{3}(L_{e})\rangle\left(  |c_{1}^{\prime
},c_{2}^{\prime},c_{3}^{\prime}\cup\alpha_{v_{0}}^{\delta}(\langle\hat{E}%
_{1}\rangle(v_{0}),n_{c_{1}^{\prime}})\rangle-|c^{\prime}\rangle\right)
-\langle\hat{E}_{1}(L_{e})\rangle\left(  |c_{1}^{\prime},c_{2}^{\prime}%
,c_{3}^{\prime}\cup\alpha_{v_{0}}^{\delta}(\langle\hat{E}_{3}\rangle
(v_{0}),n_{c_{1}^{\prime}})\rangle-|c^{\prime}\rangle\right)  \right)
\right]  -\left(  N\leftrightarrow M\right)  \nonumber
\end{align}
We have added and subtracted $|c^{\prime}\rangle$ to ensure that the
commutator has a well-defined continuum limit on the LMI habitat.\newline

We will divide the RHS of (\ref{eq:finitetricom}) into three pieces. This will
aide us in analyzing the continuum limit in a rather straightforward manner.%
\begin{align}
& |\psi_{1}^{\delta,\delta^{\prime}}(c^{\prime},[M,N])\rangle\label{eq:termA}%
\\
& :=\tfrac{1}{4}\left(  \frac{\hbar}{\mathrm{i}}\right)  ^{2}\frac{1}%
{\delta\delta^{\prime}}M(v_{0})\lambda(\vec{n}_{v_{0}}^{c^{\prime}})^{2}%
\sum_{e\in E(c^{\prime})|b(e)=v_{0}}\left(  N(v_{0}+\delta^{\prime}\dot
{e}(0))-N(v_{0})\right)  \nonumber\\
& \times\left[  \left(  \langle\hat{E}_{3}(L_{e})\rangle\left(  |c_{1}%
^{\prime}\cup\alpha_{v_{0}}^{\delta}(\langle\hat{E}_{1}\rangle(v_{0}%
),n_{c_{3}^{\prime}}),c_{2}^{\prime},c_{3}^{\prime}\rangle-|c^{\prime}%
\rangle\right)  -\langle\hat{E}_{1}(L_{e})\rangle\left(  |c_{1}^{\prime}%
\cup\alpha_{v_{0}}^{\delta}(\langle\hat{E}_{3}\rangle(v_{0}),n_{c_{3}^{\prime
}}),c_{2}^{\prime},c_{3}^{\prime}\rangle-|c^{\prime}\rangle\right)  \right)
\right.  \nonumber\\
& -\left.  \left(  \langle\hat{E}_{1}(L_{e})\rangle\left(  |c_{1}^{\prime}%
\cup\alpha_{v_{0}}^{\delta}(\langle\hat{E}_{2}\rangle(v_{0}),n_{c_{2}^{\prime
}}),c_{2}^{\prime},c_{3}^{\prime}\rangle-|c^{\prime}\rangle\right)
-\langle\hat{E}_{2}(L_{e})\rangle\left(  |c_{1}^{\prime}\cup\alpha_{v_{0}%
}^{\delta}(\langle\hat{E}_{2}\rangle(v_{0}),n_{c_{2}^{\prime}}),c_{2}^{\prime
},c_{3}^{\prime}\rangle-|c^{\prime}\rangle\right)  \right)  \right]  -\left(
N\leftrightarrow M\right)  \nonumber
\end{align}%
\begin{align}
& |\psi_{2}^{\delta,\delta^{\prime}}(c^{\prime},[M,N])\rangle\label{eq:termB}%
\\
& :=\tfrac{1}{4}\left(  \frac{\hbar}{\mathrm{i}}\right)  ^{2}\frac{1}%
{\delta\delta^{\prime}}M(v_{0})\lambda(\vec{n}_{v_{0}}^{c^{\prime}})^{2}%
\sum_{e\in E(c^{\prime})|b(e)=v_{0}}\left(  N(v_{0}+\delta^{\prime}\dot
{e}(0))-N(v_{0})\right)  \nonumber\\
& \times\left[  \left(  \langle\hat{E}_{1}(L_{e})\rangle\left(  |c_{1}%
^{\prime},c_{2}^{\prime}\cup\alpha_{v_{0}}^{\delta}(\langle\hat{E}_{2}%
\rangle(v_{0}),n_{c_{1}^{\prime}}),c_{3}^{\prime}\rangle-|c^{\prime}%
\rangle\right)  -\langle\hat{E}_{2}(L_{e})\rangle\left(  |c_{1}^{\prime}%
,c_{2}^{\prime}\cup\alpha_{v_{0}}^{\delta}(\langle\hat{E}_{1}\rangle
(v_{0}),n_{c_{1}^{\prime}}),c_{3}^{\prime}\rangle-|c^{\prime}\rangle\right)
\right)  \right.  \nonumber\\
& -\left.  \left(  \langle\hat{E}_{2}(L_{e})\rangle\left(  |c_{1}^{\prime
},c_{2}^{\prime}\cup\alpha_{v_{0}}^{\delta}(\langle\hat{E}_{3}\rangle
(v_{0}),n_{c_{3}^{\prime}}),c_{3}^{\prime}\rangle-|c^{\prime}\rangle\right)
-\langle\hat{E}_{3}(L_{e})\rangle\left(  |c_{1}^{\prime},c_{2}^{\prime}%
\cup\alpha_{v_{0}}^{\delta}(\langle\hat{E}_{2}\rangle(v_{0}),n_{c_{3}^{\prime
}}),c_{3}^{\prime}\rangle-|c^{\prime}\rangle\right)  \right)  \right]
-\left(  N\leftrightarrow M\right)  \nonumber
\end{align}%
\begin{align}
& |\psi_{3}^{\delta,\delta^{\prime}}(c^{\prime},[M,N])\rangle\label{eq:termC}%
\\
& :=\tfrac{1}{4}\left(  \frac{\hbar}{\mathrm{i}}\right)  ^{2}\frac{1}%
{\delta\delta^{\prime}}M(v_{0})\lambda(\vec{n}_{v_{0}}^{c^{\prime}})^{2}%
\sum_{e\in E(c^{\prime})|b(e)=v_{0}}\left(  N(v_{0}+\delta^{\prime}\dot
{e}(0))-N(v_{0})\right)  \nonumber\\
& \times\left[  \left(  \langle\hat{E}_{2}(L_{e})\rangle\left(  |c_{1}%
^{\prime},c_{2}^{\prime},c_{3}^{\prime}\cup\alpha_{v_{0}}^{\delta}(\langle
\hat{E}_{3}\rangle(v_{0}),n_{c_{3}^{\prime}})\rangle-|c^{\prime}%
\rangle\right)  -\langle\hat{E}_{3}(L_{e})\rangle\left(  |c_{1}^{\prime}%
,c_{2}^{\prime},c_{3}^{\prime}\cup\alpha_{v_{0}}^{\delta}(\langle\hat{E}%
_{2}\rangle(v_{0}),n_{c_{3}^{\prime}})\rangle-|c^{\prime}\rangle\right)
\right)  \right.  \nonumber\\
& -\left.  \left(  \langle\hat{E}_{3}(L_{e})\rangle\left(  |c_{1}^{\prime
},c_{2}^{\prime},c_{3}^{\prime}\cup\alpha_{v_{0}}^{\delta}(\langle\hat{E}%
_{1}\rangle(v_{0}),n_{c_{1}^{\prime}})\rangle-|c^{\prime}\rangle\right)
-\langle\hat{E}_{1}(L_{e})\rangle\left(  |c_{1}^{\prime},c_{2}^{\prime}%
,c_{3}^{\prime}\cup\alpha_{v_{0}}^{\delta}(\langle\hat{E}_{3}\rangle
(v_{0}),n_{c_{1}^{\prime}})\rangle-|c^{\prime}\rangle\right)  \right)
\right]  -\left(  N\leftrightarrow M\right)  \nonumber
\end{align}

\end{document}